\documentstyle[12pt,epsf]{article}

\advance \topmargin by -\headheight
\advance \topmargin by -\headsep

\evensidemargin \oddsidemargin
\def\ie{{\em i.e.}}
\def\ie{\hbox{\it i.e.}}

\def\CC{{\mathchoice
{\rm C\mkern-8mu\vrule height1.45ex depth-.05ex
width.05em\mkern9mu\kern-.05em}
{\rm C\mkern-8mu\vrule height1.45ex depth-.05ex
width.05em\mkern9mu\kern-.05em}
{\rm C\mkern-8mu\vrule height1ex depth-.07ex
width.035em\mkern9mu\kern-.035em}
{\rm C\mkern-8mu\vrule height.65ex depth-.1ex
width.025em\mkern8mu\kern-.025em}}}

\def\RR{{\rm I\kern-1.6pt {\rm R}}}

\def\ZZ{{\rm Z}\kern-3.8pt {\rm Z} \kern2pt}
\def\IB{\relax{\rm I\kern-.18em B}}
\def\ID{\relax{\rm I\kern-.18em D}}
\def\II{\relax{\rm I\kern-.18em I}}
\def\IP{\relax{\rm I\kern-.18em P}}

\def\np{Nucl. Phys.}
\def\pl{Phys. Lett.}
\def\prl{Phys. Rev. Lett.}
\def\pr{Phys. Rev.}

\def\atmp{Adv. Theor. Math. Phys. }
\def\jhep{J. High Energy Phys.}
\def\ptp{Prog. Theor. Phys.}

\def\atmp{Adv. Theor. Math. Phys.}

\newcommand{\beq}{\begin{equation}}
\newcommand{\eeq}{\end{equation}}
\newcommand{\rc}{\nonumber\\}
\newcommand{\bear}{\begin{eqnarray}}
\newcommand{\eear}{\end{eqnarray}}

\def\to{\rightarrow}

\def\to{\rightarrow}

\newfont{\namefont}{cmr10}
\newfont{\addfont}{cmti7 scaled 1440}
\newfont{\boldmathfont}{cmbx10}
\newfont{\headfontb}{cmbx10 scaled 1728}
\renewcommand{\theequation}{{\rm\thesection.\arabic{equation}}}
\begin{document}
\begin{titlepage}

\begin{center} \Large \bf Open string modes at brane intersections

\end{center}

\vskip 0.3truein
\begin{center}
Daniel Are\'an
\footnote{arean@fpaxp1.usc.es}
and
Alfonso V. Ramallo
\footnote{alfonso@fpaxp1.usc.es}

\vspace{0.3in}

Departamento de F\'\i sica de Part\'\i culas, Universidade de
Santiago de Compostela \\and\\
Instituto Galego de F\'\i sica de Altas Enerx\'\i as (IGFAE)\\
E-15782 Santiago de Compostela, Spain
\vspace{0.3in}

\end{center}
\vskip.5
truein

\begin{center}
\bf ABSTRACT
\end{center}
We study systematically the open string modes of a general class of BPS intersections of
branes. We work in the approximation in which one of the branes is considered as a probe
embedded in the near-horizon geometry generated by the other type of branes. We mostly
concentrate on the D3-D5 and D3-D3 intersections, which are dual to defect theories with
a massive hypermultiplet confined to the defect. In these cases we are able to obtain
analytical expressions for the fluctuation modes of the probe and to compute the
corresponding mass spectra of the dual operators in closed form. Other BPS intersections
are also studied and their fluctuation modes and spectra are found numerically.

\vskip3.6truecm
\leftline{hep-th/0602174 \hfill February  2006}
\smallskip
\end{titlepage}
\setcounter{footnote}{0}

\tableofcontents



\setcounter{equation}{0}
\section{Introduction}
\medskip
The origin of the gauge/gravity correspondence is the twofold description of D-branes
\cite{jm, MAGOO}. On one hand the D-branes have an open string description as
hypersurfaces on which open strings can end. The dynamics of these hypersurfaces can be
described by the (super) Yang-Mills theory in flat space. On the other hand, the
D-branes also appear  as solitons of the type II low energy closed string effective
action and are solutions of the classical equations of supergravity. By relating these
two descriptions one can get information of the quantum dynamics of gauge theories by
studying classical supergravity. 

In its original form, all matter fields of the gauge theory side of the correspondence
are in the adjoint representation. Clearly, if we want to apply this duality to more
realistic scenarios we should be able to obtain a holographic description of theories
with matter in the fundamental representation \ie\ quarks. This can be achieved by adding
open strings to the supergravity side of the correspondence. The simplest way to do this
is by considering fundamental strings whose ends are fixed at the UV, as was done in ref.
\cite{Wilson} to compute the expectation values of the Wilson loop operators. Notice
that the  fundamentals introduced in this way are external static quark sources.

Alternatively, one can try to generalize the gauge/gravity correspondence by
adding brane probes embedded in  supergravity backgrounds. The fluctuations of
the probe correspond to degrees of freedom of open strings  connecting the brane probe
and those that generated the background \cite{KR}. On the field theory side these open
strings are identified with  fundamental hypermultiplets of dynamical quarks whose masses
are proportional to the distance between the two types of branes. 

One can use this setup to add flavor to some supergravity duals \cite{KKW}. In
particular, for the $AdS_5\times S^5$ geometry the appropriate flavor branes are
D7-branes which fill the spacetime directions of the gauge theory and are extended along
the holographic direction \cite{D3D7}. The starting point in this construction are two
stacks of D3- and D7-branes which intersect along three common spatial directions. If
the number of D3-branes is large we can take the decoupling limit and substitute them by
the $AdS_5\times S^5$ geometry. Moreover, when the number of D7-branes is small compared
to the number of D3-branes, we can assume that the D7-branes do not backreact on the
geometry and treat them as a probe whose fluctuation modes are identified with the
mesons of the dual gauge theory.  Remarkably, the  mass spectrum of the complete set of
fluctuations can be obtained analytically and the identification between the different
modes and the dual operators can be carried out \cite{KMMW}. Different flavor branes and
their spectra for several backgrounds have been considered in the recent literature (see
\cite{Sonnen}-\cite{APR}).

In this paper we generalize these results for the D3-D7 system to a general class of BPS
intersections of two types of branes, both in type II theories and in M-theory. In our
approach the lower dimensional brane is substituted by the corresponding near-horizon
geometry, while the higher dimensional one will be treated as a probe. Generically, the
addition of the probes to the supergravity background creates a defect in the gauge
theory dual in which extra hypermultiplets are localized. In the decoupling limit one
sends the string scale $l_s$ to zero keeping the gauge coupling of the lower
dimensional brane fixed. It is straightforward to see that the gauge coupling of the
higher dimensional brane vanishes in this limit and, as a consequence, the corresponding
gauge theory decouples and the gauge group of the higher dimensional brane becomes the
flavor symmetry of the effective theory at the intersection.

The prototypical example of a defect theory is the one dual to the D3-D5 intersection.
This system was proposed in ref. \cite{KR} as a generalization of the usual AdS/CFT
correspondence in the $AdS_5\times S^5$ geometry. Indeed, if the D5-branes are at zero
distance of the D3-branes they wrap an $AdS_4\times S^2$ submanifold of the 
$AdS_5\times S^5$ background. It was argued in ref. \cite{KR} that the AdS/CFT
correspondence acts twice in this system and, apart from 
the holographic description of the four
dimensional field theory on the boundary of $AdS_5$, the fluctuations of the D5-brane
probe should be dual to the physics confined to the boundary  of $AdS_4$. 

The field theory dual of the D3-D5 intersection corresponds to  ${\cal N}=4$, $d=4$ super
Yang-Mills theory coupled to  ${\cal N}=4$, $d=3$ fundamental hypermultiplets localized
at the defect. In ref. \cite{WFO}  the action of this model in the conformal limit of
zero D3-D5 separation was constructed and a precise
dictionary between operators of the field theory and fluctuation modes of the probe was
obtained (see also refs. \cite{EGK,ST}). We will extend these results to the case in
which the distance between the D3- and D5-branes is non-zero. This non-vanishing
distance breaks  conformal invariance by giving mass to the fundamental
hypermultiplets.  Interestingly, the differential equations for the quadratic
fluctuations can be decoupled, solved analytically and the corresponding mass spectra
can be given in closed form. These masses satisfy the degeneracy conditions expected
from the structure of the supersymmetric multiplets found in ref. \cite{WFO}.

The D3-D5 intersection can be generalized to the case of a Dp-D(p+2) BPS intersection, in
which the D(p+2)-brane creates a codimension one defect in the (p+1)-dimensional gauge
theory of the Dp-brane. The differential equations of the fluctuations can also be
decoupled in this more general case. Even if we will  not be able to solve these
equations in analytic form for $p\not=3$ , we will disentangle the mode structure and we
shall find the corresponding mass spectra by numerical methods. We will see that the
numerical values of the masses  satisfy degeneracy relations which are very similar to
the ones found for the exactly solvable  D3-D5 system.

Another interesting case of defect theory arises from the D3-D3 BPS intersection, in
which the two D3-branes share one spatial dimension. In the conformal limit this
intersection gives rise to a two-dimensional defect in a four-dimensional CFT. 
In this case one
has, in the probe approximation, a D3-brane probe wrapping an  $AdS_3\times S^1$ submanifold of
the  $AdS_5\times S^5$ background. In ref. \cite{CEGK} the spectrum of fluctuations of the
D3-brane probe in the conformal limit was obtained and the corresponding dual fields were
identified (see also \cite{Kirsch}). Notice that in this intersection both types of
branes have the same dimensionality and the decoupling argument explained above does not
hold anymore.  Therefore, it is more natural to regard this system as describing two
${\cal N}=4$ four-dimensional theories coupled to each other through a bifundamental
hypermultiplet living on the two-dimensional defect. This fact is reflected in the
appearance of a Higgs branch in the system, in which the two types of D3-branes merge
along some holomorphic curve. We will study this system when a non-zero mass is given to
the hypermultiplet. Again, we will be able to solve analytically the differential
equations for the fluctuations and to get the exact mass spectrum of the model.  This
system generalizes to the case of a Dp-Dp intersection, in which the two Dp-branes have
$p-2$ common spatial directions. For
$p\not=3$ we will get the mass spectrum of the different modes from a numerical
integration of the differential equations of the fluctuations.

The D3-D7 intersection
described above corresponds to a codimension zero ``defect". This configuration is a
particular case of the Dp-D(p+4) BPS intersection  in which the D(p+4)-brane fills
completely the (p+1)-dimensional worldvolume of the Dp-brane and acts as a flavor brane
of the corresponding supersymmetric gauge theory in p+1 dimensions. Again, for $p\not=3$
one has to employ numerical methods to get the mass spectrum. 

This paper is organized as follows. In section \ref{general} we will consider a general
intersection of two branes of arbitrary dimensionalities. By placing these two branes at
a non-zero distance, and by imposing a no-force condition on the static configuration,
we get an equation which determines the BPS intersections. Next, we consider
fluctuations of the scalars transverse to both types of branes around the static BPS
configurations. For D-brane probes embedded in $AdS_5\times S^5$ the
corresponding differential equation can be reduced, after a change of variables, to the
hypergeometric differential equation. Thus, in these cases the form of the fluctuations
can be obtained analytically and the mass spectra of the transverse scalar
fluctuations can be found by imposing suitable boundary conditions at the UV. In the
general case the fluctuation equation can be transformed into the Schr\"odinger equation
for some potential, which allows to apply the WKB method to get an estimate of the mass
levels. 

In section \ref{D3D5} we study in detail the complete set of fluctuations, involving all
scalar and vector worldvolume fields, of the D3-D5 intersection.  In general, these
fluctuations are coupled to each other and one has to decouple them in order  to get a
system of independent equations. The decoupling procedure is actually the same as in
the more general Dp-D(p+2) intersection and is given in detail in appendix
\ref{DpD(p+2)}.  In section  \ref{D3D5} we use this procedure to get the exact mass
spectrum of the D3-D5 system. We also recall the fluctuation/operator dictionary found
in ref. \cite{WFO} and we check that the masses we find for the modes are consistent with
the arrangement of the dual operators in supersymmetric multiplets. 

In section \ref{D3D3} we perform a complete analysis of the exactly solvable D3-D3
intersection. In this case the differential equations can also be decoupled and solved in
terms of the hypergeometric function. As a consequence, the exact mass spectrum can be
found and matched with the fluctuation/operator dictionary established in ref.
\cite{CEGK}. We will also show the appearance of the Higgs branch and how it is
modified by the fact that the hypermultiplet is massive.

In the main text we will concentrate on the study of the exactly solvable intersections
and we have left other cases to the appendices. These cases include the 
Dp-D(p+2), Dp-Dp, Dp-D(p+4) and F1-Dp intersections of the type II theory, as well as
the M2-M2, M2-M5 and M5-M5 intersections of M-theory. In all of them we compute the
numerical mass spectra and their WKB estimates. Finally, in section \ref{Conclusion} we
summarize our results and point out some possible extensions of our work.

\setcounter{equation}{0}
\section{Fluctuations of intersecting branes}
\medskip
\label{general}
\begin{figure}
\centerline{\hskip -.8in \epsffile{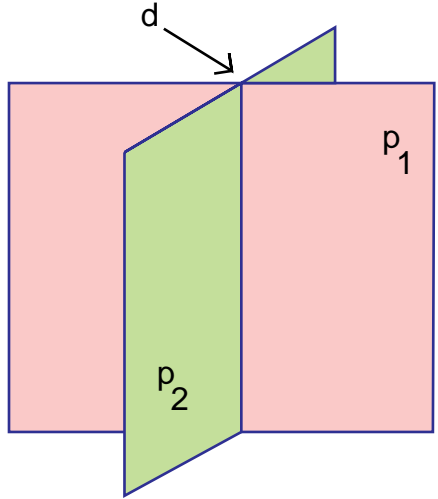}}
\caption{ A general orthogonal intersection of a $p_1$- and $p_2$-brane along $d$
spatial directions.}
\label{intersection}
\end{figure}

Let us consider an orthogonal intersection of a $p_1$-brane and a $p_2$-brane
along $d$ common spatial directions ($p_2\ge p_1$), as depicted in figure
\ref{intersection}.  We shall denote this intersection,
both in type II string theory and M-theory, as  $(d|p_1\perp p_2)$. We shall
treat the lower dimensional $p_1$-brane as a background, whereas the $p_2$-brane
will be considered as a probe. The background metric will be taken as:
\beq
ds^2\,=\,\Biggl[\,{r^2\over R^2}\,\Biggr]^{\gamma_1}\,\,
(\,-dt^2\,+\,(dx^1)^2\,+\cdots +(dx^{p_1})^2\,)\,+\,
\Biggl[\,{R^2\over r^2}\,\Biggr]^{\gamma_2}\,\,
d\vec y\cdot d\vec y\,\,,
\label{metric}
\eeq
where $R$, $\gamma_1$ and $\gamma_2$ are constants that depend on the case
considered, $\vec y=(y^1,\cdots,y^{D-1-p_1})$ with D=10,\,11 and $r^2=\vec y\cdot
\vec y$. In the type II theory the supergravity solution also contains a
dilaton $\phi$, which we will parametrize as:
\beq
e^{-\phi(r)}\,=\,\Biggl[\,{R^2\over r^2}\,
\Biggr]^{\gamma_3}\,\,,
\label{dilaton}
\eeq
with $\gamma_3$ being constant (in the case of a background of eleven
dimensional supergravity we just take $\gamma_3=0$).

Let us now place a $p_2$-brane in this background extended along the directions:
\beq
(\,t,x^1,\cdots,x^d,y^1,\cdots,y^{p_2-d}\,)\,\,.
\eeq
We shall denote by $\vec z$ the set of $y$ coordinates transverse to the probe:
\beq
\vec z\,=\,(z^1,\cdots, z^{D-p_1-p_2+d-1})\,\,,
\eeq
with $z^m=y^{p_2-d+m}$ for $m=1,\cdots, D-p_1-p_2+d-1$. Notice that the $\vec z$
coordinates are transverse to both background and probe branes. Moreover, we shall
choose spherical coordinates on the
$p_2$-brane worldvolume which is transverse to the $p_1$-brane. If we define:
\beq
\rho^2\,=\,(y^1)^2\,+\,\cdots+\,(y^{p_2-d})^2\,\,,
\label{rho}
\eeq
clearly, one has:
\beq
(dy^1)^2\,+\cdots (dy^{p_2-d})^2\,=\,d\rho^2\,+\,\rho^2 d\Omega^2_{p_2-d-1}\,\,,
\label{spherical}
\eeq
where $d\Omega^2_{p_2-d-1}$ is the line element of a unit $(p_2-d-1)$-sphere.
Obviously we are assuming that $p_2-d\ge 2$.

\subsection{BPS intersections}
\label{BPS}

\begin{figure}
\centerline{\hskip -.8in \epsffile{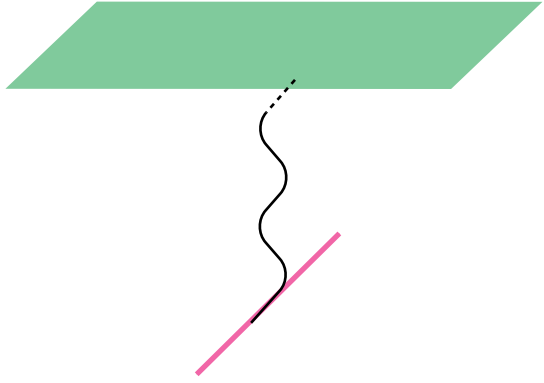}}
\caption{ The two branes of the intersection are separated a finite distance. In the
figure one of the branes is represented as a one-dimensional object. An open string can
be stretched between the two branes.}
\label{separation}
\end{figure}

Let us consider first a configuration in which the probe is located at a
constant value of $|\vec z|$, \ie\ at $|\vec z|=L$ (see figure \ref{separation}). If
$\xi^a$ are a set of worldvolume coordinates, the induced metric on the
probe worldvolume for such a static configuration  will be denoted by:
\beq
ds^2_{I}\,=\,\,{\cal G}_{ab}\,d\xi^a d\xi^b\,\,.
\eeq
In what follows we will use as worldvolume coordinates the cartesian ones
$x^0\cdots x^d$ and the radial and angular variables introduced in eqs. (\ref{rho}) and
(\ref{spherical}). Taking into account that, for an embedding with $|\vec z|=L$, one has 
$r^2=\rho^2\,+\,\vec z^{\,\,2}\,=\,\rho^2\,+\,L^2$, the induced metric can be
written as:
\beq
ds^2_{I}\,=\,\Biggl[\,{\rho^2+L^2\over
R^2}\,\Biggr]^{\gamma_1}\,\, (\,-dt^2\,+\,(dx^1)^2\,+\,\cdots + (dx^d)^2\,)\,+\,
\Biggl[\,{R^2\over \rho^2+L^2}\,\Biggr]^{\gamma_2}\,\,(\,
d\rho^2\,+\,\rho^2 d\Omega^2_{p_2-d-1}\,)\,\,.
\label{indmetric}
\eeq
The action of the probe is given by the Dirac-Born-Infeld action. In the 
configurations we  study in this section the worldvolume gauge field vanishes  and it
is easy to verify that the lagrangian density reduces to:
\beq
{\cal L}\,=\,-e^{-\phi}\,\sqrt{-\det {\cal G}}\,\,.
\eeq
For a static configuration such as the one with $|\vec z|=L$, the energy
density ${\cal H}$ is just ${\cal H}=-{\cal L}$. By using the explicit form of
${\cal G}$ in (\ref{indmetric}), one can verify that, for the $|\vec z|=L$ embedding,
${\cal H}$ is given by:
\beq
{\cal H}\,=\,
\Biggl[\,{\rho^2+L^2\over
R^2}\,\Biggr]^{{\gamma_1\over 2}\,(d+1)\,-\,{\gamma_2\over 2}\,(p_2-d)\,-
\gamma_3}\,\,
\rho^{p_2-d-1}\,\,\sqrt{\det \tilde g}\,\,,
\eeq
where $\tilde g$ is the metric of the unit $(p_2-d-1)$-sphere. In a BPS
configuration the no-force condition of a supersymmetric intersection requires
that ${\cal H}$ be independent of the distance $L$ between the branes. Clearly, this
can be achieved if the $\gamma_i$-coefficients are related as:
\beq
\gamma_3\,=\,{\gamma_1\over 2}\,(d+1)\,-\,{\gamma_2\over 2}\,(p_2-d)\,\,.
\label{BPScon}
\eeq
Let us rewrite this last equation as:
\beq
d\,=\,{\gamma_2\over \gamma_1+\gamma_2}\,p_2\,+\,
{2\gamma_3-\gamma_1\over \gamma_1+\gamma_2}\,\,,
\label{BPSrule}
\eeq
which gives the number $d$ of common dimensions of the intersection in terms of the 
parameters $\gamma_i$ of the background and of the dimension $p_2$ of the probe brane.
In the following subsections we shall  consider some particular examples.
\subsubsection{Dp-brane background}
In the string frame, 
the supergravity solution corresponding to a Dp-brane with $p<7$ has the form displayed
in eqs. (\ref{metric}) and (\ref{dilaton}) with $p_1=p$, $R$ given by
\beq
R^{7-p}\,=\,2^{5-p}\,\pi^{{5-p\over 2}}\,\Gamma\Big({7-p\over 2}\Big)\,g_s\,N\,
(\alpha')^{{7-p\over 2}}\,\,,
\label{RDp}
\eeq
and with the following  values for the exponents $\gamma_i$:
\beq
\gamma_1\,=\,\gamma_2\,=\,{7-p\over 4}\,\,,\qquad\qquad
\gamma_3\,=\,{(7-p)(p-3)\over 8}\,\,.\qquad\qquad
\label{Dpgammas}
\eeq
Moreover, the Dp-brane solution is endowed with a Ramond-Ramond $(p+1)$-form potential,
whose component along the Minkowski coordinates $x^0\cdots x^p$ can be taken as:
\beq
\Big[\,C^{(p+1)}\,\Big]_{x^0\cdots x^p}\,=\,
\Biggl[\,{r^2\over R^2}\,\Biggr]^{{7-p\over 2}}\,\,.
\label{CRR}
\eeq

Applying eq. (\ref{BPSrule}) to this background, we get the following relation 
between $d$ and $p_2$:
\beq
d={p_2+p-4\over 2}\,\,.
\eeq
Let us now consider the case in which the probe brane is another D-brane.
As the brane of the background and the probe should live in the same type II theory, 
$p_2-p$ should be even. Since $d\le p$, we are left with the following
three possibilities:
\beq
(p|Dp\perp D(p+4))\,\,,\qquad
(p-1|Dp\perp D(p+2))\,\,,\qquad
(p-2|Dp\perp Dp)\,\,.
\eeq

\subsubsection{Fundamental string background}
In the string frame, 
the metric and dilaton for
the background created by a fundamental string are of the form of eqs.
(\ref{metric}) and (\ref{dilaton}) for:
\beq
\gamma_1\,=\,3\,\,,\qquad
\gamma_2\,=\,0\,\,,\qquad
\gamma_3\,=\,{3\over 2}\,\,,\qquad
R^6\,=\,32\pi^2 (\alpha')^3\,g_s^2\,N\,\,. 
\label{F1back-parameters}
\eeq
In this case one gets from (\ref{BPSrule}) that $d=0$, which corresponds to the following
intersection:
\beq
(0|F1\perp Dp)\,\,.
\eeq

\subsubsection{M2-brane background}
Our next example is the  geometry created by an M2-brane  in M-theory. In this case
one has:
\beq
\gamma_1\,=\,2\,\,,\qquad
\gamma_2\,=\,1\,\,,\qquad
\gamma_3\,=\,0\,\,,\qquad
R^6\,=\,32\pi^2 l_P^{\,6}\,\,N\,\,,
\label{M2back-parameters}
\eeq
where $l_P$ is the Planck length in eleven dimensions. In this case
eq. (\ref{BPSrule}) becomes
\beq
d={p_2-2\over 3}\,\,.
\eeq
Taking $p_2=2,5$ we get the following intersections:
\beq
(0|M2\perp M2)\,\,\,,\qquad\qquad
(1|M2\perp M5)\,\,.
\eeq

\subsubsection{M5-brane background}
The background corresponding to an M5-brane in eleven dimensional supergravity has
\beq
\gamma_1\,=\,{1\over 2}\,\,,\qquad
\gamma_2\,=\,1\,\,,\qquad
\gamma_3\,=\,0\,\,,\qquad
R^3\,=\,\pi\,l_P^{\,3}\,N\,\,,
\label{M5back-parameters}
\eeq
which leads to
\beq
d={2p_2-1\over 3}\,\,.
\eeq
For $p_2=5$ in the previous expression we get the intersection:
\beq
(3|M5\perp M5)\,\,.
\eeq

\subsection{Fluctuations}
\label{generalfluctuations}

In what follows we will assume that the condition  (\ref{BPScon}) holds. This fact can be
checked for all the particular supersymmetric intersections that will be
analyzed below. 

Let us now study the fluctuations around the $|\vec z|=L$ embedding. Without
loss of generality we can take $z^1=L$, $z^m=0$ ($m>1$) as the unperturbed
configuration and consider a fluctuation of the type:
\beq
z^1=L+\chi^1\,\,,
\,\,\,\,\,\,\,\,\,\,\,\,\,
z^m=\chi^m\,\,(m>1)\,\,,
\eeq
where the $\chi$'s are small. The dynamics of the fluctuations is determined by
the Dirac-Born-Infeld lagrangian which, for the fluctuations of the transverse scalars
we study in this section, reduces to 
${\cal L}\,=\,-e^{-\phi}\sqrt{-\det g}$, where $g$ is the induced metric on the
worldvolume. By expanding this lagrangian and keeping up to
second order terms, one can prove that:
\beq
{\cal L}\,=\,-\,{1\over 2}\,\,\rho^{p_2-d-1}\,\,\sqrt{\det \tilde g}\,\,
\Biggl[\,{R^2\over \rho^2+L^2}\,\Biggr]^{\gamma_2}\,\,
{\cal G}^{ab}\,\partial_{a}\chi^m\,\partial_{b}\chi^m\,\,,
\label{fluct-lag-general}
\eeq
where ${\cal G}^{ab}$ is the (inverse) of the metric (\ref{indmetric}). The
equations of motion derived from this lagrangian are:
\beq
\partial_{a}\,\Bigg[\,{\rho^{p_2-d-1}\sqrt{\det \tilde g}\over 
(\rho^2+L^2)^{\gamma_2}}\,{\cal G}^{ab}\,\partial_{b}\chi
\,\,\Bigg]\,=\,0\,\,,
\label{eom-general}
\eeq
where we have dropped the index $m$ in the $\chi$'s. Using the explicit form of
the metric elements ${\cal G}^{ab}$, the above equation can be written as:
\beq
{R^{2\gamma_1+2\gamma_2}\over (\rho^2+L^2)^{\gamma_1+\gamma_2}
}\,\,\partial^{\mu}\partial_{\mu}\,\chi\,+\, {1\over
\rho^{p_2-d-1}}\,\partial_{\rho}\,(\rho^{p_2-d-1}\partial_{\rho}\chi)\,+\,
{1\over
\rho^2}\,\nabla^i\nabla_i\,\chi=0\,\,,
\eeq
where the index $\mu$ corresponds to the directions $x^{\mu}=(t, x^1,\cdots,
x^d)$ and $\nabla_i$ is the covariant derivative on the $(p_2-d-1)$-sphere. In
order to analyze this equation, let us separate variables as:
\beq
\chi\,=\,\xi(\rho)\,e^{ikx}\,Y^l(S^{p_2-d-1})\,\,,
\label{sepvar}
\eeq
where the product $kx$ is performed with the flat Minkowski metric and
$Y^l(S^{p_2-d-1})$ are scalar spherical harmonics which satisfy:
\beq
\nabla^i\nabla_i\,Y^l(S^{p_2-d-1})\,=\,-l(l+p_2-d-2)\,\,Y^l(S^{p_2-d-1})\,\,.
\label{casimir}
\eeq
If we redefine the variables as:
\beq
\varrho\,=\,{\rho\over L}\,\,,
\,\,\,\,\,\,\,\,\,\,\,\,\,\,
\bar M^2\,=\,-R^{2\gamma_1+2\gamma_2}\,L^{2-2\gamma_1-2\gamma_2}\,k^2\,\,,
\label{newvariables}
\eeq
the differential equation becomes:
\beq
\partial_{\varrho}\,(\varrho^{p_2-d-1}\partial_{\varrho}\,\xi)\,+\,\Big[\,
\bar M^2\,{\varrho^{p_2-d-1}\over (1+\varrho^2)^{\gamma_1+\gamma_2}}\,-\,
l(l+p_2-d-2)\varrho^{p_2-d-3}\,\Big]\,\xi\,=\,0\,\,.
\label{fluc}
\eeq

In order to study the solutions of eq. (\ref{fluc}), let us change
variables in such a way that this equation can be written as a Schr\"odinger
equation:
\beq
\partial^2_y\,\psi\,-\,V(y)\,\psi\,=\,0\,\,,
\label{Sch}
\eeq
where $V$ is some potential. The change of variables needed to convert eq.
(\ref{fluc}) into (\ref{Sch}) is:
\beq
e^y\,=\,\varrho\,\,,
\,\,\,\,\,\,\,\,\,\,\,\,\,
\psi\,=\,\varrho^{{p_2-d-2\over 2}}\,\,\xi\,\,.
\label{Sch-variables}
\eeq
Notice that in this change of variables $\varrho\to\infty$ corresponds to $y\to\infty$,
while the point $\varrho=0$ is mapped into $y=-\infty$. Moreover, 
the resulting potential $V(y)$ takes the form:
\beq
V(y)\,=\,\Biggl(l-1+{p_2-d\over 2}\Biggr)^2\,-\,\bar M^2\,\,{e^{2y}\over
(e^{2y}\,+\,1)^{\gamma_1+\gamma_2}}\,\,.
\label{potential}
\eeq

\begin{figure}
\centerline{\hskip -.8in \epsffile{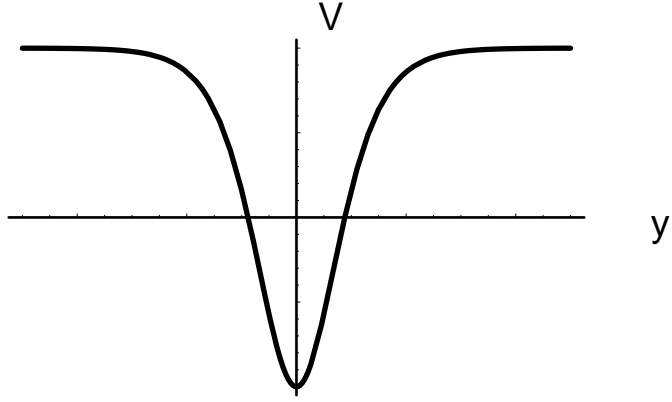}}
\caption{ The Schr\"odinger potential $V(y)$ of eq. (\ref{potential}).}
\label{poten}
\end{figure}

In figure \ref{poten} we have plotted the function $V(y)$. It is  interesting to notice
that, in these new variables, the problem of finding the mass spectrum can be rephrased
as that of finding the values of $\bar M$ such that a zero-energy level for the
potential  (\ref{potential}) exists. Notice that the classically allowed region in the
Schr\"odinger equation (\ref{Sch}) corresponds to the values of $y$ such that $V(y)\le
0$. We would have a discrete spectrum if this region is of finite size or, equivalently,
if the points
$y=\pm \infty$ are not allowed classically.  As, when $\gamma_1+\gamma_2>1$, one has:
\beq
\lim_{y\to\pm\infty}\,V(y)\,=\,\Biggl(l-1+{p_2-d\over 2}\Biggr)^2\,\,,
\label{generalV-limit}
\eeq
we will have a discrete spectrum for all values of $l\in\ZZ_+$ if $p_2-d>2$.
Notice that $p_2-d\ge 2$ and when $p_2-d=2$ and $l=0$ the turning points of the
potential $V(y)$ are at $y=\pm \infty$. Moreover, $V(y)$
has a unique minimum at a value of $y$ given by:
\beq
e^{2y_0}\,=\,{1\over\gamma_1+\gamma_2-1}\,\,.
\eeq

\subsection{The exactly solvable case}
\label{exactly-solvable}
When $\gamma_1+\gamma_2=2$, the differential equation for the fluctuation can
be solved exactly in terms of a hypergeometric function (see appendix
\ref{hyperchange}). To prove this statement,  let us change variables in eq.
(\ref{fluc}) as follows:
\beq
z=-\varrho^2\,\,.
\eeq
One can check that  eq. (\ref{fluc}) for $\gamma_1+\gamma_2=2$ is converted into:
\bear
&&z(1-z)\,{\partial^2\xi\over \partial z^2}\,+\,
{p_2-d\over 2}\,(1-z)\,{\partial\xi\over \partial z}
+\,\Bigg[\,\,{l(l+p_2-d-2)\over 4}\,\,(1-z^{-1})\,-\,
{\bar M^2\over 4}\,(1-z)^{-1}\,\,\Bigg]\,\xi\,=\,0\,\,,\rc\rc
\label{hyper}
\eear
which can be reduced to the hypergeometric differential equation. Indeed, 
let us  define $\lambda$ as:
\beq
\lambda\,\equiv\, {-1+\sqrt{\,\, 1+\phantom{}\bar M^2\phantom{\big)}}\over 2}\,\,.
\label{lambda}
\eeq
Notice that eq. (\ref{lambda}) can be easily inverted, namely:
\beq
\bar M^2=4\lambda (\lambda+1)\,\,.
\label{Mlambda}
\eeq
Then, in terms of the original variable $\varrho$, the solution of eq.
(\ref{hyper}) that is regular when
$\varrho\to 0$ is:
\beq
\xi(\varrho)\,=\,\varrho^l\,(\varrho^2+1)^{-\lambda}\,
F(-\lambda,  -\lambda+l-1+{p_2-d\over 2}; l+{p_2-d\over 2};-\varrho^2\,)\,\,.
\label{scalarhyper}
\eeq
We also want that $\xi$ vanishes when $\varrho\to\infty$. A way to ensure
this is by imposing that
\beq
-\lambda+l-1+{p_2-d\over 2}\,=\,-n\,\,,
\,\,\,\,\,\,\,\,\,\,\,\,\,\,\,\,\,\,\,\,
n=0,1,2,\cdots\,.
\label{quant}
\eeq
When this condition is satisfied the hypergeometric function behaves as $(\varrho^2)^n$
when $\varrho\to\infty$ and $\xi\sim \varrho^{-(l+p_2-d-2)}$ in this limit.
Notice that when $p_2-d=2$ the $l=0$ mode does not vanish at large $\varrho$, in
agreement with our general analysis based on the potential (\ref{potential}). By using
the condition (\ref{quant}) in eq. (\ref{Mlambda}), one gets:
\beq
\bar M^2\,=\,4\Bigg(n+l-1+{p_2-d\over 2}\,\Bigg)
\Bigg(n+l+{p_2-d\over 2}\,\Bigg)\,\,.
\label{exactM}
\eeq
Since in this case $\bar M^2=-R^4L^{-2}k^2$, one gets the following spectrum of
possible values of $k^2=-M^2$:
\beq
M^2\,=\,{4L^2\over R^4}\,\,\Bigg(n+l+{p_2-d-2\over 2}\,\Bigg)
\Bigg(n+l+{p_2-d\over 2}\,\Bigg)\,\,.
\label{exactMass}
\eeq

Let us look in detail which intersections satisfy the condition 
$\gamma_1+\gamma_2=2$, needed to reduce the fluctuation equation to the hypergeometric
one. From the list of intersections worked out in subsection \ref{BPS}, it is clear that
this exactly solvable cases can only occur if the background is a Dp-brane. Actually, in
this case one must have $\gamma_1=\gamma_2=1$ (see eq. (\ref{Dpgammas})), which only
happens if $p=3$. Thus, the list of exactly solvable intersections reduces to the
following cases:
\beq
(3|D3\perp D7)\,\,,\qquad
(2|D3\perp D5)\,\,,\qquad
(1|D3\perp D3)\,\,.
\label{AdSdefects}
\eeq
Notice that, for the three cases in (\ref{AdSdefects}), $p_2=2d+1$ for $d=3,2,1$.
Therefore, if $dx_{1,d}^2$ denotes the line element for the flat Minkowski space in
$d+1$ dimensions, one can write the induced metric (\ref{indmetric}) as:
\beq
ds^2_{I}\,=\,{\rho^2+L^2\over R^2}\,\,dx_{1,d}^2\,+\,
{R^2\over \rho^2+L^2}\,d\rho^2\,+\,R^2\,{\rho^2\over \rho^2+L^2}\,\,d\Omega^2_d\,\,.
\label{AdSindmetric}
\eeq
Moreover, by using the relation between
$p_2$ and $d$, one can rewrite the mass spectra (\ref{exactMass}) of scalar fluctuations
for the  intersections (\ref{AdSdefects}) as:
\beq
M_S^2\,=\,{4L^2\over R^4}\,\,\Bigg(\,n\,+\,l\,+\,{d-1\over 2}\,\Bigg)
\Bigg(\,n\,+\,l\,+\,{d+1\over 2}\,\Bigg)\,\,.
\label{generalMS}
\eeq

In the three cases in (\ref{AdSdefects}) the background geometry is $AdS_5\times S^5$. 
Moreover, one can see from (\ref{AdSindmetric}) that
the induced metric reduces in the UV limit $\rho\to\infty$ to that of a product space of
the form  $AdS_{d+2}\times S^{d}$. Indeed, the $(3|D3\perp D7)$ intersection is the case
extensively studied in ref. \cite{KMMW} and corresponds in the UV to an 
$AdS_5\times S^3\subset AdS_5\times S^5$ embedding. In this case the D7-brane is a flavor
brane for the ${\cal N}=4$ gauge theory.  The $(2|D3\perp D5)$ intersection represents
in the UV an $AdS_4\times S^2$ defect in $AdS_5\times S^5$. In the conformal limit $L=0$
the corresponding defect CFT has been studied in detail in ref. \cite{WFO} where, in
particular, the fluctuation/operator dictionary was found. In section \ref{D3D5} we will
extend these results to the case in which the brane separation $L$ is different from
zero and we will be able to find analytical expressions for the complete set of
fluctuations. The $(1|D3\perp D3)$ intersection corresponds in the UV to an
$AdS_3\times S^1$ defect in $AdS_5\times S^5$ which is of codimension two in the gauge
theory directions. The fluctuation spectra and the corresponding field theory dual for
$L=0$ have been analyzed in ref.
\cite{CEGK}. In section \ref{D3D3} we will integrate analytically the differential
equations for all the fluctuations of this $(1|D3\perp D3)$ intersection when the
D3-brane separation $L$ is non-vanishing.

The $\varrho\to\infty$ limit is simply the high energy regime of the theory, where the
mass of the quarks can be ignored and the theory becomes conformal. Therefore, the
$\varrho\to\infty$ behaviour of the fluctuations should provide us  information about
the the conformal dimension $\Delta$ of the corresponding dual operators. Indeed, in the
context of the AdS/CFT correspondence in $d+1$ dimensions, it is well known that, if the
fields are canonically normalized, the normalizable modes behave at infinity as
$\rho^{-\Delta}$, whereas the non-normalizable ones should behave as
$\rho^{\Delta-d-1}$. In the case in which the modes are not canonically normalized the
behaviours of both types of modes are of the form 
$\rho^{-\Delta+\gamma}$ and $\rho^{\Delta-d-1+\gamma}$ for some $\gamma$. Clearly we can
obtain the conformal dimension from the difference between the exponents. Let us apply
this method to  fluctuations which are given in terms of hypergeometric functions, as
in eq. (\ref{scalarhyper}). Since for large $\varrho$ the hypergeometric function behaves
as:
\beq
F(a_1, a_2;b;-\varrho^2)\approx c_1\, \varrho^{-2a_1}\,+\,
c_2\, \varrho^{-2a_2}\,\,,\qquad\qquad (\varrho\to \infty)\,\,,
\label{asymptotic-hyper}
\eeq
one immediately gets:
\beq
\Delta\,=\,{d+1\over 2}\,+\,a_2-a_1\,\,.
\label{Delta-hyper}
\eeq
For the scalar fluctuations studied above, one has from the solution (\ref{scalarhyper})
that $a_1=-\lambda$ and $a_2=-\lambda+l+{d-1\over 2}$. By applying eq.
(\ref{Delta-hyper}) to this case, we get the following value for the dimension of the
operator associated to the scalar fluctuations:
\beq
\Delta_S\,=\,l+d\,\,.
\label{generalDeltaS}
\eeq
Notice that the quantization condition (\ref{quant}) selects precisely the normalizable
modes, which behave at large $\varrho$  as $\xi\sim\varrho^{-\Delta_S+1}$. Notice also
that the modes that become constant at $\rho\to\infty$ correspond to operators with
$\Delta_S=1$ and, therefore, they should not be discarded. A glance at eq.
(\ref{generalDeltaS}) reveals that this situation only occurs when $d=1$ (\ie\ for the
$AdS_3\times S^1$ defect) and
$l=0$. This case will be studied in detail in section \ref{D3D3}.

As the hypergeometric function $F(a_1, a_2;b;-\varrho^2)$ is symmetric under
$a_1\leftrightarrow a_2$, it is clear that the roles of $a_1$ and $a_2$ can also be
exchanged in (\ref{Delta-hyper}). If the resulting conformal dimension lies in the
unitarity range $\Delta>0$ (or $\Delta\ge 0$ if $d=1$) we have a second branch of
fluctuations. For the case at hand $\Delta=1-l$ and the unitarity condition requires
generically that $l=0$. This second branch  is selected by
imposing  to the hypergeometric function (\ref{scalarhyper})
the truncation condition $\lambda=n$  for $n=1,2,\cdots$. The resulting spectrum is just
$\bar M^2=4n(n+1)$, $n=1,2,\cdots$. In the rest of this paper this second branch of  the
fluctuations of the transverse scalars will not be considered further.

\subsection{WKB quantization}
The mapping to the Schr\"odinger equation we have performed in section
\ref{generalfluctuations} allows us to apply the semiclassical WKB approximation to
compute the fluctuation spectrum. The WKB method has been very successful
\cite{MInahan,RS} in  the calculation of the glueball spectrum in the context of the
gauge/gravity correspondence \cite{glueball}. The starting point in this calculation is
the WKB quantization rule:
\beq
(n+{1\over 2})\pi\,=\,\int_{y_1}^{y_2}\,dy\,\sqrt{-V(y)}\,\,,
\,\,\,\,\,\,\,\,\,\,\,\,\,\,
n\ge 0\,\,,
\label{WKBquantization}
\eeq
where $n\in\ZZ$ and $y_1$, $y_2$ are the turning points of the potential
($V(y_1)=V(y_2)=0$). To evaluate the right-hand side of (\ref{WKBquantization}) we
expand it as a power series in $1/\bar M$ and keep the leading and subleading terms of
this expansion. We obtain  in this way the expression of $\bar M$ as a function of the
principal quantum number $n$ which is, in principle, reliable for large $n$, although in
some cases it happens to give the exact result. Let us recall the outcome of this
analysis for a general case, following \cite{RS}. With this purpose, 
let us come back to the original variable $\varrho$ and suppose that we have a
differential equation of the type:
\beq
\partial_{\varrho}\,\big(\,g(\varrho)\,\partial_{\varrho}\,\phi\,\big)\,+\,
\big(\,\bar M^2\,q(\varrho)\,+\,p(\varrho)\,\big)\,\phi\,=\,0\,\,,
\eeq
where the functions $g$, $h$ and $p$ behave near $\varrho\approx 0,\infty$ as:
\bear
&&g\approx g_1\varrho^{s_1}\,\,,
\,\,\,\,\,\,\,\,\,\,\,\,\,\,
q\approx q_1\varrho^{s_2}\,\,,
\,\,\,\,\,\,\,\,\,\,\,\,\,\,
p\approx p_1\varrho^{s_3}\,\,,
\,\,\,\,\,\,\,\,\,\,\,\,\,\,{\rm as}\,\,\varrho\to 0\,\,,\rc\rc
&&g\approx g_2\varrho^{r_1}\,\,,
\,\,\,\,\,\,\,\,\,\,\,\,\,\,
q\approx q_2\varrho^{r_2}\,\,,
\,\,\,\,\,\,\,\,\,\,\,\,\,\,
p\approx p_2\varrho^{r_3}\,\,,
\,\,\,\,\,\,\,\,\,\,\,\,\,\,{\rm as}\,\,\varrho\to \infty\,\,.
\eear
The consistency of the WKB approximation requires that $s_2-s_1+2$ and $r_1-r_2-2$ be
stricly positive numbers, whereas $s_3-s_1+2$ and $r_1-r_3-2$ can be either positive or
zero \cite{RS}. In our case (eq. (\ref{fluc})), the functions $g(\varrho)$, $q(\varrho)$
and
$p(\varrho)$ are:
\beq
g(\varrho)\,=\,\varrho^{p_2-d-1}\,\,,
\,\,\,\,\,\,\,\,\,\,
q(\varrho)\,=\,{\varrho^{p_2-d-1}\over (1+\varrho^2)^{\gamma_1+\gamma_2}}\,\,,
\,\,\,\,\,\,\,\,\,\,
p(\varrho)\,=\,-l(l+p_2-d-2)\varrho^{p_2-d-3}\,\,.
\label{gqp}
\eeq
From the behavior at $\rho\approx 0$ of the functions written above, we obtain:
\bear
&&g_1=1\,\,,
\,\,\,\,\,\,\,\,\,\,\,\,\,\,\,\,\,\,\,\,\,\,\,\,
\,\,\,\,\,\,\,\,\,\,\,\,\,\,\,\,\,\,\,\,\,\,\,\,
\,\,\,\,\,\,\,\,\,\,\,\,\,\,\,\,\,\,\,\,\,\,\,\,\,\,\,\,
s_1=p_2-d-1\,\,,\rc\rc
&&q_1=1\,\,,
\,\,\,\,\,\,\,\,\,\,\,\,\,\,\,\,\,\,\,\,\,\,\,\,
\,\,\,\,\,\,\,\,\,\,\,\,\,\,\,\,\,\,\,\,\,\,\,\,
\,\,\,\,\,\,\,\,\,\,\,\,\,\,\,\,\,\,\,\,\,\,\,\,\,\,\,\,\,
s_2=p_2-d-1\,\,,\rc\rc
&&p_1=-l(l+p_2-d-2)\,
\,\,,
\,\,\,\,\,\,\,\,\,\,\,\,\,\,\,\,\,\,\,\,\,\,\,\,\,\,\,\,
\,\,\,\,\,\,
s_3=p_2-d-3\,\,.
\label{IR}
\eear
Notice that $s_2-s_1+2=2$ and $s_3-s_1+2=0$ and, thus, we are within the range of
applicability of the WKB approximation. Moreover, from the behavior at $\rho\to\infty$ 
of the functions written in (\ref{gqp}) we obtain:
\bear
&&g_2=1\,\,,
\,\,\,\,\,\,\,\,\,\,\,\,\,\,\,\,\,\,\,\,\,\,\,\,
\,\,\,\,\,\,\,\,\,\,\,\,\,\,\,\,\,\,\,\,\,\,\,\,
\,\,\,\,\,\,\,\,\,\,\,\,\,\,\,\,\,\,\,\,\,\,\,\,\,\,\,\,
r_1=p_2-d-1\,\,,\rc\rc
&&q_2=1\,\,,
\,\,\,\,\,\,\,\,\,\,\,\,\,\,\,\,\,\,\,\,\,\,\,\,
\,\,\,\,\,\,\,\,\,\,\,\,\,\,\,\,\,\,\,\,\,\,\,\,
\,\,\,\,\,\,\,\,\,\,\,\,\,\,\,\,\,\,\,\,\,\,\,\,\,\,\,\,
r_2=p_2-d-1-2\gamma_1-2\gamma_2\,\,,\rc\rc
&&p_2=-l(l+p_2-d-2)
\,\,,
\,\,\,\,\,\,\,\,\,\,\,\,\,\,\,\,\,\,\,\,\,\,\,\,\,\,\,\,
\,\,\,\,\,
r_3=p_2-d-3\,\,.
\label{UV}
\eear
Now $r_1-r_2-2=2(\gamma_1+\gamma_2-1)$ and $r_1-r_3-2=0$ and we are also in the
range of applicability of the WKB method if $\gamma_1+\gamma_2>1$. Coming back to the
general case, let us define \cite{RS} the quantities:
\beq
\alpha_1\,=\,s_2-s_1+2\,\,,
\,\,\,\,\,\,\,\,\,\,\,\,\,\,
\beta_1\,=\,r_1-r_2-2\,\,,
\label{WKBalphabetauno}
\eeq
and (as $s_3-s_1+2=r_1-r_3-2=0$, see \cite{RS}):
\beq
\alpha_2\,=\,\sqrt{(s_1-1)^2\,-\,4\,{p_1\over g_1}}\,\,,
\,\,\,\,\,\,\,\,\,\,\,\,\,\,\,\,\,\,\,\,\,\,\,\,\,\,\,\,
\beta_2\,=\,\sqrt{(r_1-1)^2\,-\,4\,{p_2\over g_2}}\,\,.
\label{alphabeta2}
\eeq
Then, the mass levels for large quantum number $n$ can be written in terms of
$\alpha_{1,2}$ and $\beta_{1,2}$ as \cite{RS}:
\beq
\bar M^2_{WKB}\,=\,{\pi^2\over \zeta^2}\,(n+1)\,\bigg(n\,+\,{\alpha_2\over
\alpha_1}\,+\, {\beta_2\over \beta_1}\bigg)\,\,,
\label{generalWKBlevels}
\eeq
where $\zeta$ is the following integral:
\beq
\zeta\,=\,\int_0^{\infty} d\varrho\,\,\sqrt{{q(\varrho)\over g(\varrho)}}\,\,.
\eeq
In our case $\alpha_{1,2}$ and $\beta_{1,2}$ are easily obtained from the coefficients
written in (\ref{IR}) and (\ref{UV}), namely:
\beq
\alpha_1\,=\,2\,\,,
\,\,\,\,\,\,\,\,\,\,\,\,\,\,
\beta_1\,=\,2(\gamma_1+\gamma_2-1)\,\,,
\,\,\,\,\,\,\,\,\,\,\,\,\,\,
\alpha_2\,=\,\beta_2\,=\,2l+p_2-d-2\,\,.
\label{alphabeta}
\eeq
Moreover, the integral $\zeta$ for our system is given by:
\beq
\zeta\,=\,\int_{0}^{\infty}{d\varrho\over 
(1+\varrho^2)^{{\gamma_1+\gamma_2\over 2}}}\,=\, 
{\sqrt{\pi}\over 2}\,\,
{\Gamma\Big({\gamma_1+\gamma_2-1\over 2}\Big)\over 
\Gamma\Big({\gamma_1+\gamma_2\over 2}\Big)}\,\,,
\label{zeta}
\eeq
and we get the following WKB formula for the masses:
\beq
\bar M^{WKB}_S\,=\,2\sqrt{\pi}\,
{\Gamma\Big({\gamma_1+\gamma_2\over 2}\Big)\over
 \Gamma\Big({\gamma_1+\gamma_2-1\over 2}\Big)}\,
\sqrt{(n+1)\,\Bigg(n\,+\,
{\gamma_1+\gamma_2\over \gamma_1+\gamma_2-1}\,\Big(\,l-1
+{p_2-d\over 2}\,\,\Big)\,\Bigg)}\,\,.
\label{WKBM}
\eeq

\subsection{Numerical computation}
\label{generalnumeric}

The  formula (\ref{WKBM}) for the masses can be checked numerically by
means of the shooting technique. Notice that the behaviour for small $\varrho$ of the
fluctuation $\xi$ (needed when this technique is applied) can be easily obtained. Indeed,
let us try to find a solution of eq. (\ref{fluc}) of the form:
\beq
\xi\sim \varrho^{\gamma}\,\,,
\eeq
and let us neglect the term containing $\bar M^2$ of eq. (\ref{fluc}). It is
immediate to see that $\gamma$ satisfies the equation:
\beq
\gamma^2+(p_2-d-2)\gamma-l(l+p_2-d-2)=0\,\,,
\eeq
whose roots are: 
\beq
\gamma=l,-(l+p_2-d-2)\,\,.
\label{IRroots}
\eeq
Clearly, the solution regular at $\varrho=0$ should correspond to the root $\gamma=l$.
Then, we conclude that near $\varrho=0$ one has:
\beq
\xi\sim \varrho^{l}\,\,,\qquad\qquad (\varrho\approx 0)
\,\,.
\label{generalIR}
\eeq
In order to get the mass levels in the numerical calculation we have to match the 
$\varrho\approx 0$ behaviour (\ref{generalIR}) with the behaviour for large $\varrho$.
The latter can be easily obtained by using the mapping written in (\ref{Sch-variables})
to the Schr\"odinger equation  (\ref{Sch}). Indeed, for $\varrho\to\infty$, or
equivalently for $y\to\infty$, the potential $V(y)$ becomes asymptotically constant and
the wave equation (\ref{Sch}) can be  trivially integrated. Let us call
$V_*\equiv \lim_{y\to \infty} V(y)$, with $V_*>0$. Then, the solutions of (\ref{Sch})
are of the form $\psi\sim e^{\pm \sqrt{V_*}y}$ which, in terms of the original variable 
$\varrho$ are simply $\psi\sim \varrho^{\pm \sqrt{V_*}}$. The actual value of $V_*$ is
given in eq. (\ref{generalV-limit}). Taking into account the relation 
(\ref{Sch-variables}) between $\psi$ and $\xi$, we get that 
$\xi\sim \varrho^{\gamma}$ for $\varrho\to\infty$, where $\gamma$ are exactly the two
values written in eq. (\ref{IRroots}). The so-called $S^l$ modes are characterized by
the following UV behaviour:
\beq
\xi\sim \varrho^{-(l+p_2-d-2)}\,\,,\qquad\qquad (\varrho\to\infty)\,\,.
\label{generalUV}
\eeq
In the shooting technique one solves the differential equation for the fluctuations by
imposing the behaviour (\ref{generalIR}) at $\varrho\approx 0$ and then one scans the
values of $\bar M$ until the UV behaviour (\ref{generalUV}) is fine tuned. This occurs
only for a discrete set of values of $\bar M$, which determines the mass spectrum we are
looking for. The numerical values obtained in the different intersections  and their
comparison with the WKB mass formula (\ref{WKBM}) are studied in the appendices.

\setcounter{equation}{0}
\section{Fluctuations of the D3-D5 system}
\medskip
\label{D3D5}
In this section we study in detail the complete set of fluctuations corresponding to the 
$(2|D3\perp D5)$ intersection. The dynamics of the D5-brane probe in the 
$AdS_5\times S^5$ background is governed by the Dirac-Born-Infeld action, which in this
case reduces to
\beq
S\,=\,-\,\int d^6\xi\,\sqrt{-\det (g+F)}\,+\,
\int d^6\xi \,\,\,P\big[\,C^{(4)}\,\big]\wedge F\,\,,
\label{DBI-D5}
\eeq
where $g$ is the induced metric on the worldvolume, $P\big[\cdots]$ denotes the pullback
of the form inside the brackets and, for convenience, we are taking the D5-brane tension
equal to one. In (\ref{DBI-D5}) $F$ is the two-form corresponding to the worldvolume
field strength, whose one-form potential will be denoted by $A$ ($F=dA$).

Let us now find the action for the complete set of quadratic fluctuations around the
static configuration in which the two branes are at a distance $L$. Recall that in this
embedding the worldvolume metric in the UV is $AdS_4\times S^2$. As in section 
\ref{general}, let us denote by $\chi$ the scalars transverse to both types of branes
and let us assume that the probe is extended along  $x^1$ and $x^2$. We will 
call simply $X$ to the coordinate  $x^3$. By expanding up to second order the
action (\ref{DBI-D5}), one gets the following lagrangian for the fluctuations:
\bear
&&{\cal L}\,=\,-\rho^2\,\sqrt{\tilde g}\,\Bigg[\,{1\over 2}\,
{R^2\over \rho^2+L^2}\,
{\cal G}^{ab}\partial_a\chi\partial_b\chi\,+\,{1\over 2}\,{\rho^2+L^2\over R^2}\,
{\cal G}^{ab}\partial_a X\partial_b X\,+\,{1\over 4}\,F_{ab}F^{ab}\,\Bigg]\,-\,\rc\rc
&&\qquad\qquad\qquad\qquad\qquad\qquad-\,
2\,{\rho\over R^4}\,(\rho^2+L^2)X\epsilon^{ij}F_{ij}\,\,,
\label{D3D5-quad-action}
\eear
where $i,j$ are indices of the two-sphere of the worldvolume, $\epsilon^{ij}=\pm 1$
and ${\cal G}_{ab}$ is the induced metric for the static configuration, \ie\ the metric
displayed in eq. (\ref{AdSindmetric}) for $d=2$.

The equation of motion of the scalar $\chi$ is just (\ref{eom-general}) for $p_2=5$,
$d=2$ and $\gamma_2=1$. As shown in subsection \ref{exactly-solvable} this equation can
be solved exactly in terms of the hypergeometric  function (\ref{scalarhyper}). Upon
imposing the quantization condition (\ref{quant}) we obtain a tower of normalizable
modes, which we will denote by $S^l$,  given by:
\beq
\xi_S(\varrho)\,=\,\,=\,\varrho^l\,(\varrho^2+1)^{-n-l-{1\over 2}}\,
F(-n-l-{1\over 2},  -n; l+{3\over 2};-\varrho^2\,)\,\,,
\qquad (l,n\ge 0)\,\,.
\eeq
Recall from (\ref{generalMS}) and (\ref{generalDeltaS}) that the associated mass
$M_S(n,l)$ and conformal dimension $\Delta_S$ for these scalar modes are given by:
\beq
M_S(n,l)\,=\,{2L\over R^2}\,
\sqrt{\bigg(n+l+{1\over 2}\bigg)\bigg(n+l+{3\over 2}\bigg)}\,\,,
\qquad\qquad
\Delta_S\,=\,l+2\,\,.
\label{MSd3d5}
\eeq
Notice that the value found here for $\Delta_S$ is in agreement with the result 
of ref. \cite{WFO}.

As shown in (\ref{D3D5-quad-action}), the scalar $X$ is coupled to the components
$F_{ij}$ of the gauge field strength along the two-sphere. The equation of
motion of $X$ derived from the lagrangian (\ref{D3D5-quad-action}) is:
\beq
R^2\,
\partial_a\,\Bigg[\,\rho^2\sqrt{\tilde g}\,(\rho^2+L^2)\,
{\cal G}^{ab}\partial_b X\,\Bigg]\,-\,2\rho\,( \rho^2+L^2)\,
\epsilon^{ij}F_{ij}\,=\,0\,\,.
\label{D3D5Xeq}
\eeq
Moreover, the equation of motion of the gauge field is:
\beq
R^4\,
\partial_a\,\Big[\,\rho^2\sqrt{\tilde g}\,F^{ab}\,\Big]\,-\,4\rho\,
(\rho^2+L^2)\,\epsilon^{bj}\,\partial_j X\,=\,0\,\,,
\label{D3D5Aeq}
\eeq
where $\epsilon^{bj}$ is zero unless $b$ is an index along the two-sphere.

Let us now see how the equations of motion (\ref{D3D5Xeq}) and (\ref{D3D5Aeq}) can be
decoupled and, subsequently, integrated in  analytic form. 
With this purpose in mind, let us see
how one can obtain vector spherical harmonics for the two-sphere from the scalar
harmonics
$Y^l$. Clearly, from a scalar harmonic in $S^2$ we can can construct a vector  by simply
taking the derivative  with respect to the coordinates of the
two-sphere, namely:
\beq
Y_i^l(S^2)\equiv \nabla_i\,Y^l(S^2)\,\,.
\label{S2har}
\eeq
One can check from (\ref{casimir}) that these functions satisfy:
\bear
&&\nabla^i\,Y_i^l\,=\,{1\over \sqrt{\tilde g}}\,
\partial_i\,\Big[\,\sqrt{\tilde g}\, \tilde g^{ij}\,Y_j^l\,\Big]\,=\,-l(l+1)\,Y^l\,\,,\rc
&&\epsilon^{ij}\,\partial_i\,Y_j^l\,=\,0\,\,.
\eear
Alternatively, we can take the Hogde dual in the sphere and define a new vector
harmonic function $\hat Y_i^l(S^2)$ as:
\beq
\hat Y_i^l(S^2)\,\equiv\,{1\over \sqrt{\tilde g}}\,\tilde g_{ij}\,\epsilon^{jk}\,
\nabla_k\,Y^l(S^2)\,\,.
\label{hatS2har}
\eeq
The $\hat Y_i^l$ vector harmonics satisfy:
\bear
&&\nabla^i\,\hat Y_i^l\,=\,0\,\,,\rc
&& \epsilon^{ij}\,\partial_i\,\hat Y_j^l\,=\,\,l(l+1)\,\sqrt{\tilde g}\,\,Y^l\,\,.
\label{hatYprop}
\eear

Let us analyze the different types of modes, in analogy with the D3-D7 case in
\cite{KMMW}.

\subsection{Type I modes}

We are going to study first the modes which involve the scalar field $X$ and the
components $A_i$ of the gauge field along the two-sphere directions. Generically, the
equations of motion couple   $A_i$ to the other gauge field components $A_{\mu}$ and
$A_{\rho}$. However,  due to the property $\nabla^i\,\hat Y_i^l\,=\,0$ (see eq.
(\ref{hatYprop})), if $A_i$ is proportional
to $\hat Y_i^l$ it does not mix with other components of the gauge field, although it
mixes with the scalar $X$. Accordingly, let us take the ansatz:
\beq
A_{\mu}=0\,\,,\qquad\qquad
A_{\rho}=0\,\,,\qquad\qquad
A_i\,=\,\phi(x,\rho)\,\hat Y_i^l(S^2)\,\,,
\label{typeIA}
\eeq
while we represent $X$ as:
\beq
X\,=\,\Lambda(x,\rho)\,Y^l(S^2)\,\,.
\label{typeIX}
\eeq
Taking into account that
\beq
{1\over 2}\,\epsilon^{ij}\,F_{ij}\,=\,l(l+1)\,\sqrt{\tilde g}\,\phi\,Y^l\,\,,
\eeq
one can prove that the equation of motion  of $X$ (eq. (\ref{D3D5Xeq})) becomes:
\bear
&&R^4\,\rho^2\,\partial_{\mu}\partial^{\mu}\Lambda\,+\,\partial_{\rho}\Big[\,
\rho^2\,(\,\rho^2\,+\,L^2\,)^2\,\partial_{\rho}\,\Lambda\,\Big]\,-\,\rc\rc
&&-
l(l+1)\,(\,\rho^2+L^2\,)^2\,\Lambda\,-\,4l(l+1)\,\rho\,(\,\rho^2+L^2\,)\phi\,=\,0\,\,.
\label{D3D5Lambdaeq}
\eear
It can be easily verified that the equations for $A_{\mu}$ and $A_{\rho}$ are
automatically satisfied as a consequence of the relation 
$\nabla^i\,\hat Y_i^l\,=\,0$. Moreover, for $l\not=0$ the equation of motion 
(\ref{D3D5Aeq}) for the gauge field
components along $S^2$ reduces to:
\beq
R^4\,
\partial_{\mu}\partial^{\mu}\phi\,+\,
\partial_{\rho}\Big[\,
\,(\,\rho^2\,+\,L^2\,)^2\,\partial_{\rho}\,\phi\,\Big]\,-\,
l(l+1)\,{(\,\rho^2\,+\,L^2\,)^2\over
\rho^2}\,\phi\,-4\rho\,(\,\rho^2\,+\,L^2\,)\Lambda\,=\,0\,\,.
\label{D3D5phieq}
\eeq
In order to decouple this system of equations, let us follow closely the steps of ref.
\cite{WFO}. First, we redefine the scalar field $\Lambda$ as follows:
\beq
V\,=\,\rho\,\Lambda\,\,.
\label{VLambda}
\eeq
The system of equations which results after this redefinition can be decoupled by simply
taking suitable linear combinations of the unknown functions $V$ and $\phi$. This
decoupling procedure was used in ref. \cite{WFO} for the conformal case $L=0$ and,
remarkably, it also works for the case in which the brane separation does not vanish. In
appendix \ref{DpD(p+2)} we will apply this method to decouple the fluctuations of the
type written in eqs. (\ref{typeIA}) and (\ref{typeIX}) for the more general 
$(p-1|Dp\perp D(p+2))$ intersection. Here we just need the decoupled functions, which
are:
\bear
&&Z^+\,=\,V\,+\, l\phi\,\,,\rc\rc
&&Z^-\,=\,V\,-\,(l+1)\,\phi\,\,.
\label{Zpm}
\eear
It is interesting at this point to notice that the $Z^-$ modes only exist for $l\ge 1$
while the $Z^+$ modes make sense for $l\ge 0$. Indeed, the vector harmonic $\hat Y_i^l$
vanishes for $l=0$, since it is the derivative of a constant function (see eq.
(\ref{S2har})). Then, it follows from (\ref{typeIA}) that the vector field vanishes for
these $l=0$ modes and, as a result, the $l=0$ mode of the scalar field is uncoupled. It
is clear from (\ref{Zpm}) that this $l=0$ mode of the field $X$ is just $ Z^+$ for
vanishing $l$. As we have just mentioned, for $l>0$ the equations for $ Z^+$ and 
$ Z^-$ are decoupled. These equations can be obtained by substituting the definition
(\ref{Zpm}) into eqs. (\ref{D3D5Lambdaeq}) and (\ref{D3D5phieq}). In order to
get the corresponding spectra, let us adopt  a plane wave ansatz for $Z^{\pm}$, namely:
\beq
Z^{\pm}\,=\,e^{ikx}\,\xi^{\pm}(\rho)\,\,.
\eeq
Moreover, let us define the reduced variables $\varrho$ and $\bar M$ as in eq.
(\ref{newvariables}), namely $\varrho=\rho/L$ and $\bar M=-R^4L^{-2}k^2$. Furthermore,
we define $\lambda$ as in eq. (\ref{lambda}). We will consider separately the $Z^+$ and
$Z^-$ equations.

\subsubsection{$Z^+$ spectra}
By combining appropriately eqs. (\ref{D3D5Lambdaeq}) and (\ref{D3D5phieq}) one can show
that the  equation for $\xi^+$ is indeed decoupled and given by:
\beq
{1\over 1+\varrho^2}\,\partial_{\varrho}\,\Big[\,
(1+\varrho^2)^2\,\partial_{\varrho}\,\xi^+\,\Big]\,+\,
\Bigg[\,{\bar M^2\over 1+\varrho^2}\,\,-\,
(l+1)\,\big(l+4+{l\over \varrho^2}\big)\,
\Bigg]\,\xi^+\,=\,0\,\,.
\label{xi+eq}
\eeq
Remarkably, eq. (\ref{xi+eq}) can be analytically solved in terms of a hypergeometric
function. Actually, by using the change of variables of appendix \ref{hyperchange}, one
can show that the solution of (\ref{xi+eq}) which is 
regular at $\varrho=0$ is:
\beq
\xi^+(\varrho)\,=\,\varrho^{1+l}\,(1+\varrho^2)\,^{-1-\lambda}\,\,
F(-\lambda-1, l+{3\over 2}-\lambda;l+{3\over 2}; -\varrho^2)\,\,.
\label{xi+expression}
\eeq
By using eq. (\ref{asymptotic-hyper}) one can show that, for large values of the
$\varrho$ coordinate, the function $\xi^+$ written in (\ref{xi+expression}) behaves as:
\beq
\xi^+(\varrho)\,\sim\,c_1\varrho^{l+1}\,+\,c_2\varrho^{-l-4}\,\,,
\qquad\qquad (\varrho\to\infty)\,\,.
\eeq
Clearly, the only normalizable solutions are those for which $c_1=0$. 
This regularity condition at $\varrho=\infty$ can be enforced by means of 
the following quantization condition:
\beq
l+{3\over 2}-\lambda\,=\,-n\,\,,\qquad\qquad
n=0,1,2,\cdots\,\,.
\label{I+quantization}
\eeq
The $\xi^+$ fluctuations for which (\ref{I+quantization}) holds will be referred to as
$I_+^l$ modes. Their analytical expression is given by:
\beq
\xi_{I_+}(\varrho)\,=\,\,=\,\varrho^{1+l}\,(\varrho^2+1)^{-n-l-{5\over 2}}\,
F(-n-l-{5\over 2},  -n; l+{3\over 2};-\varrho^2\,)\,\,,
\qquad (l,n\ge 0)\,\,,
\eeq
and the corresponding energy levels are:
\beq
M_{I_+}(n,l)\,=\,{2L\over R^2}\,
\sqrt{\bigg(n+l+{3\over 2}\bigg)\bigg(n+l+{5\over 2}\bigg)}\,\,. 
\eeq
By using the general expression (\ref{Delta-hyper}) one reaches the conclusion that the
conformal dimensions of the operators dual to the $I_+^l$ modes are:
\beq
\Delta_{I_+}=l+4\,\,,
\eeq
which  agrees with the values found in ref. \cite{WFO}.

\subsubsection{$Z^-$ spectra}
The equation for $\xi^-$ can be shown to be:
\beq
{1\over 1+\varrho^2}\,\partial_{\varrho}\,\Big[\,
(\,1+\varrho^2\,)^2\,\partial_{\varrho}\,\xi^-\,\Big]\,+\,
\Bigg[\,{\bar M^2\over 1+\varrho^2}\,\,-\,
l\,\big(l-3+{l+1\over \varrho^2}\big)\,
\Bigg]\,\xi^-\,=\,0\,\,,
\eeq
which again can be solved in terms of the hypergeometric function. The solution
regular at $\varrho=0$ is:
\beq
\xi^-(\varrho)\,=\,\varrho^{1+l}\,(\varrho^2+1)\,^{-1-\lambda}\,\,
F(-\lambda+1, l-{1\over 2}-\lambda;l+{3\over 2}; -\varrho^2)\,\,.
\eeq
It is easy to verify by using eq. (\ref{asymptotic-hyper}) that $\xi^-(\varrho)$ has two
possible behaviours at
$\varrho\to
\infty$, namely $\varrho^{-l}$, $\varrho^{l-3}$, where $l\ge 1$.  The former corresponds
to a normalizable mode with  conformal dimension $\Delta=l$, while the latter is
associated to operators with
$\Delta=3-l$. Notice that the existence of these two branches is in agreement with
the results of ref. \cite{WFO}. 

Let us consider first the branch with  $\Delta=l$, which we will refer to as $I_-^l$
fluctuations. One can select these fluctuations by imposing the following
quantization condition:
\beq
l-{1\over 2}-\lambda\,=\,-n\,\,,\qquad\qquad
n=0,1,2,\cdots\,\,.
\eeq
The corresponding functions $\xi^-(\varrho)$ are:
\beq
\xi_{I_-}(\varrho)\,=\,\varrho^{1+l}\,(\varrho^2+1)\,^{-n-l-{1\over 2}}\,\,
F(-n-l+{3\over 2}, -n;l+{3\over 2}; -\varrho^2)\,\,,
\qquad (l\ge 1, n\ge 0)\,\,,
\label{I-sol}
\eeq
and the mass spectrum and conformal dimension are:
\beq
M_{I_-}(n,l)\,=\,{2L\over R^2}\,
\sqrt{\bigg(n+l-{1\over 2}\bigg)\bigg(n+l+{1\over 2}\bigg)}\,\,,
\qquad\qquad \Delta_{I_-}=l\,\,.
\eeq

One can check that, indeed, the solution (\ref{I-sol}) behaves as $\varrho^{-l}$
at $\varrho\to\infty$ and, therefore, the associated operator in the conformal limit has
$\Delta=l$ as it should. 

One can select the  branch with $\Delta=3-l$ by requiring that:
\beq
-\lambda+1\,=\,-n\,\,,\qquad\qquad
n=0,1,2,\cdots\,\,.
\label{Itildequant}
\eeq
The corresponding functions are:
\beq
\xi_{\tilde I_-}(\varrho)\,=\,\varrho^{1+l}\,(\varrho^2+1)\,^{-2-n}\,\,
F(-n,l -n-{3\over 2};l+{3\over 2}; -\varrho^2)\,\,.
\eeq
Notice that the condition $\Delta=3-l>0$ is only fulfilled for two possible values of
$l$, namely $l=1,2$. We will refer to this branch of solutions as $\tilde I_-^l$
fluctuations. Their mass spectrum is independent of $l$, as follows from eq.
(\ref{Itildequant}).  Thus, one has:
\beq
M_{\tilde I_-}(n,l)\,=\,{2L\over R^2}\,
\sqrt{\bigg(n+1\bigg)\bigg(n+2\bigg)}\,\,,
\qquad \qquad 
\Delta_{\tilde I_-}=3-l\,\,,
\qquad \qquad 
(l=1,2)\,\,.
\label{Itildelevels}
\eeq

\subsection{Type II modes}

Let us consider a configuration with $X=0$ and with the following ansatz for the gauge
fields:
\beq
A_{\mu}\,=\,\phi_{\mu}(x,\rho)\,Y^l(S^2)\,\,,\qquad\qquad
A_{\rho}\,=\,0\,\,,\qquad\qquad
A_i\,=\,0\,\,,
\label{D3D5II}
\eeq
with 
\beq
\partial^{\mu}\phi_{\mu}\,=\,0\,\,.
\eeq
Due to this last condition one can check that the equations for $A_{\rho}$ and 
$A_i$ are satisfied, while the equation for $A_{\mu}$ yields:
\beq
\partial_{\rho}\big(\rho^2\partial_\rho\phi_{\nu}\big)\,+\,
R^4\,\,{\rho^2\over (\rho^2+L^2)^2}\,\partial_{\mu}\,\partial^{\mu}\phi_{\nu}\,-\,
l(l+1)\,\phi_{\nu}\,=\,0\,\,.
\eeq
Expanding in a plane wave basis we get exactly the same equation as in the scalar
fluctuations. Actually, let us represent $\phi_{\mu}$ as:
\beq
\phi_{\mu}\,=\,\xi_{\mu}\,e^{ikx}\,\chi(\rho)\,\,,
\label{D3D5IIbis}
\eeq
where $k^{\mu}\,\xi_{\mu}=0$. The equation for $\chi$ is:
\beq
{1\over \varrho^2}\,\partial_{\varrho}\Big(\,\varrho^2\partial_{\varrho}\chi\Big)\,+\,
\Bigg[\,{\bar M^2\over (1+\varrho^2)^2}\,-\,{l(l+1)\over
\varrho^2}\,\Bigg]\chi\,=\,0\,\,,
\eeq
where we have already introduced the reduced quantities $\varrho$ and $\bar M$. 
This equation is identical to the one corresponding to the transverse scalars. Thus, the
spectra are the same in both cases. These fluctuations correspond to an operator with
conformal weight $\Delta_{II}=l+2$, in agreement with ref. \cite{WFO}.

\subsection{Type III modes}

These modes have $X=0$ and the following form for the gauge field:
\beq
A_{\mu}\,=\,0\,\,,\qquad\qquad
A_{\rho}\,=\,\phi(x,\rho)\,Y^l(S^2)\,\,,\qquad\qquad
A_i\,=\,\tilde\phi(x,\rho)\,Y^l_i(S^2)\,\,.
\label{D3D5IIIansatz}
\eeq
Notice that for the gauge field potential written above   the field strength
components $F_{ij}$ along the two-sphere vanish, which ensures that the equation of
motion of $X$ is satisfied for $X=0$. The non-vanishing components of the gauge field
strenght are:
\beq
F_{\mu i}\,=\,\partial_{\mu}\,\tilde\phi\,Y_i^l\,\,,\qquad\qquad
F_{\mu \rho}\,=\,\partial_{\mu}\,\phi\,Y^l\,\,,\qquad\qquad
F_{\rho i}\,=\,(\,\partial_{\rho}\,\tilde\phi\,-\,\phi\,)\,Y_i^l\,\,.
\eeq
The equation for $A_{\rho}$ becomes:
\beq
R^4\,\rho^2\,\partial_{\mu}\,\partial^{\mu}\phi\,-\,l(l+1)\,(\rho^2+L^2)^2\,
(\phi-\partial_{\rho}\tilde\phi\,)\,=\,0\,\,,
\eeq
while the equation for $A_i$ is:
\beq
R^4\,\partial_{\mu}\,\partial^{\mu}\tilde\phi\,+\,
\partial_{\rho}\,\Big[\,(\rho^2+L^2)^2\,(\partial_{\rho}\tilde\phi\,-\,\phi)\,\Big]
\,=\,0\,\,.
\eeq
Moreover, the equation for $A_{\mu}$ can be written as:
\beq
\partial_{\mu}\,\Big[\,l(l+1)\tilde\phi\,-\,\partial_{\rho}(\rho^2\phi)\,\Big]
\,=\,0\,\,.
\label{D3D5IIIAmu}
\eeq

Expanding $\phi$ and $\tilde\phi$ in a plane wave basis we can get rid of the $x^{\mu}$
derivative and we can write the following relation between  $\tilde\phi$ and $\phi$:
\beq
l(l+1)\tilde\phi\,=\,\partial_{\rho}(\rho^2\phi)\,\,.
\eeq
For $l\not= 0$, one can use this relation to eliminate $\tilde\phi$ in favor of $\phi$.
The equation of motion of $A_{\rho}$ becomes:
\beq
\partial^2_{\rho}\,(\rho^2\,\phi)\,-\,l(l+1)\,\phi\,+\,{R^4\,\rho^2\over
(\rho^2+L^2)^2}\,
\partial_{\mu}\,\partial^{\mu}\phi\,=\,0\,\,,
\eeq
which, again,  can be solved in terms of hypergeometric functions. The equation of motion
of
$A_i$ is just equivalent to the above equation. To solve this equation, let us write
\beq
\phi(x,\rho)\,=\,e^{ikx}\,\zeta(\rho)\,\,.
\eeq
Then, in terms of the reduced variable $\varrho$ the equation for $\zeta$ becomes
\beq
{1\over \varrho^2}\,\partial^2_{\varrho}\,\Big(\varrho^2\zeta\Big)\,+\,
\Bigg[\,{\bar M^2\over (1+\varrho^2)^2}\,-\,{l(l+1)\over
\varrho^2}\,\Bigg]\zeta\,=\,0\,\,,
\label{D3D5IIIeq}
\eeq
where $\bar M$ is the same as for the type I modes. 
Eq. (\ref{D3D5IIIeq}) can be solved in terms of the hypergeometric function as:
\beq
\zeta(\varrho)\,=\,\varrho^{l-1}\,(\varrho^2+1)\,^{-\lambda}\,\,
F(-\lambda, l+{1\over 2}-\lambda;l+{3\over 2}; -\varrho^2)\,\,.
\eeq
The quantization condition and the energy levels are just the same as for the transverse
scalars and the type II modes. At $\rho\to\infty$, $\zeta\sim\rho^{-l-2}$. These
fluctuations correspond to a field with conformal weight $\Delta_{III}=l+2$, as predicted
in \cite{WFO}. 

\subsection{Fluctuation/operator correspondence}
\medskip
Let us recall the array corresponding to the D3-D5 intersection:
\beq
\begin{array}{ccccccccccl}
 &1&2&3& 4& 5&6 &7&8&9 & \nonumber \\
D3: & \times &\times &\times &\_ &\_ & \_&\_ &\_ &\_ &     \nonumber \\
D5: &\times&\times&\_&\times&\times&\times&\_&\_&\_ &
\end{array}
\label{D3D5intersection}
\eeq
Before adding the D5-brane we have a $SO(6)$ R-symmetry which corresponds to the rotation 
in the $456789$ directions. The D5-brane breaks this $SO(6)$  to
$SU(2)_H\times SU(2)_V$, where the $SU(2)_H$  corresponds to  rotations in the 
$456$ directions (which are along the D5-brane worldvolume) and the $SU(2)_V$ is
generated by rotations in the $789$ subspace (which are the directions orthogonal to
both types of branes). 

Let us recall how the ${\cal N}=4$, $d=4$ gauge multiplet decomposes under the 
${\cal N}=4$, $d=3$ supersymmetry. As it is well-known, the ${\cal N}=4$ gauge multiplet
in four dimensions contains a vector $A_{\mu}$ (which has components along the four
coordinates of the D3-brane worldvolume), six real scalars $X^i$ (corresponding to the
directions orthogonal to the D3-brane worldvolume) and four complex Weyl spinors
$\lambda^a$. All these fields are in the adjoint representation of the gauge group. The
$d=4$ vector field $A_{\mu}$ gives rise to a $d=3$ gauge field  $A_{k}$ and to a
scalar field $A_3$. Both types of fields are singlets with respect to the 
$SU(2)_H\times SU(2)_V$ symmetry. Moreover, the adjoint scalars can be arranged in two sets
as:
\beq
X_H\,=\,(X^4,X^5,X^6)\,\,,
\qquad
X_V\,=\,(X^7,X^8,X^9)\,\,.
\eeq
Clearly, $X_H$ transforms in the $({\bf 3},{\bf 1})$ representation of 
$SU(2)_H\times SU(2)_V$ while $X_V$ does it in the $({\bf 1}, {\bf 3})$. Finally, the
spinors transform in the  $({\bf 2},{\bf 2})$ representation and will be denoted by
$\lambda^{im}$. In addition to the bulk fields, the 3-5 strings introduce a $d=3$ complex
hypermultiplet in the fundamental representation of the gauge field, whose components will
be denoted by $(q^m, \psi^i)$. The bosonic components $q^m$ of this hypermultiplet 
transform in the $({\bf 2},{\bf 1})$ representation, whereas the fermionic ones
$\psi^i$ are in the $({\bf 1}, {\bf 2})$ of $SU(2)_H\times SU(2)_V$. 
The dimensions and quantum numbers of the different fields just discussed are summarized 
in table \ref{tableFieldsD3D5}.

\begin{table}[!h]
\centerline{
\begin{tabular}[b]{|c|c|c|c|}   
 \hline
 Field  & $\Delta$  & $SU(2)_H$ & $SU(2)_V$ \\ 
\hline 
\ \ $A_k$ & $1$  & $0$ &0 \\  \hline  
\ \ $A_3$ & $1$  & $ 0$ & 0 \\ \hline  
\ \ $X_H$ & $1$  & $ 1$ &0 \\  \hline   
\ \ $X_V$ & $1$  & $ 0 $ &1 \\ \hline  
\ \ $\lambda^{im}$ & $3/2$  & $ 1/2 $ &1/2 \\ \hline  
\ \ $q^m$ & $1/2$  & $ 1/2 $ &0 \\ \hline  
\ \ $\psi^i$ & $1$  & $ 0 $ &1/2 \\ \hline 
\end{tabular}
}
\caption{Quantum numbers and dimensions of the fields of the D3-D5 intersection.}
\label{tableFieldsD3D5} 
\end{table}

Let us now determine the quantum numbers of the different fluctuations of the D3-D5
system. We will denote by $S^l$ the scalar fluctuations and by $I_+^l$ and $I_-^l$  the
two types of vector fluctuations of type I. The $I_-^l$ fluctuations which we will
consider from now on are those corresponding to the operator of dimension $\Delta=l$.
Moreover, the modes of type II and III correspond to fluctuations of the vector gauge field
in $AdS_4$ and will be denoted collectively by $V^l$. In all cases, $l$ corresponds to the
quantum number of the spherical harmonics in the $456$ directions and, thus, it can be
identified with the isospin of the $SU(2)_H$ representation. Moreover, it is clear that
the scalar modes are fluctuations in the $789$ directions and therefore are in the vector
representation of $SU(2)_V$, while the other fluctuations are singlets under $SU(2)_V$.
With all this data and with the values of the dimensions determined previously, we can
fill the values displayed in  table \ref{tableModesD3D5}.

\begin{table}[!h]
\centerline{
\begin{tabular}[b]{|c|c|c|c|}   
 \hline
 Mode  & $\Delta$  & $SU(2)_H$ & $SU(2)_V$ \\ 
\hline 
\ \ $S^l$ & $l+2$  & $l\ge 0$ &1 \\  \hline  
\ \ $I_+^l$ & $l+4$  & $l\ge 0$ & 0 \\ \hline  
\ \ $I_-^l$ & $l$  & $l\ge 1$ &0 \\  \hline   
\ \ $V^l$ & $l+2$  & $l\ge 0 $ &0 \\ \hline  
\end{tabular}
}
\caption{Quantum numbers and dimensions of the modes of the D3-D5 intersection.}
\label{tableModesD3D5} 
\end{table}

Let us now recall our results for the mass spectra. 
The mass of the scalar fluctuations $M_S(n,l)$ is given in eq. (\ref{MSd3d5}).
The masses of the other modes are given in terms of $M_S(n,l)$ as:
\beq
M_{I_+}(n,l)\,=\,M_S(n,l+1)\,\,,\,\,\,\,
M_{I_-}(n,l)\,=\,M_S(n,l-1)\,\,,\,\,\,\,
M_{V}(n,l)\,=\,M_S(n,l)\,\,.
\label{massrelations}
\eeq

Let us now match, following ref. \cite{WFO}, the fluctuation modes with composite
operators  of the ${\cal N}=4$, $d=3$ defect theory by looking at the dimensions 
and $SU(2)_H\times SU(2)_V$ quantum numbers of these two types of objects. Let us consider
first the fluctuation mode with the lowest dimension, which according to our previous
results is $I_-^1$. This mode is a triplet of $SU(2)_H$ and a singlet of $SU(2)_V$ and has
$\Delta=1$. There is only one operator with these characteristics. Indeed, let us define
the following operator:
\beq
{\cal C}^I\equiv \bar q^m \sigma_{mn}^I\, q^n\,\,,
\eeq
where the $\sigma^I$ are Pauli matrices. This operator has clearly the same dimension and
$SU(2)_H\times SU(2)_V$ quantum numbers as the mode $I_-^1$. Therefore, we have the
identification \cite{WFO}:
\beq
I_-^1\sim {\cal C}^I\,\,.
\eeq
Moreover, by acting with the supersymmetry generators we can obtain the other operators
in the same multiplet as ${\cal C}^I$. The bosonic ones are \cite{WFO}:
\bear
&&{\cal E}^A\,=\,\bar\psi^i\,\sigma_{ij}^A\,\psi^j\,+\,2\bar q^m
X_V^{Aa}T^a q^m\,\,,
\rc\rc  &&J_B^k\,=\,i\bar q^m D^k
q^m\,-\,i(D^kq^m)^{\dagger}\,q^m\,+\,
\bar\psi^i\,\rho^k \psi^i\,\,,
\eear
where the $T^a$ are the matrices of the gauge group and the $\rho^k$ are Dirac matrices in
$d=3$. Notice that these two operators have dimension $\Delta=2$. Moreover, 
${\cal E}^A$ transforms in the $({\bf 1}, {\bf 3})$ representation of $SU(2)_H\times
SU(2)_V$, whereas $J_B^k$ is a $SU(2)_H\times SU(2)_V$ singlet. It is straightforward to
find the modes that have these same quantum numbers and dimension. Indeed, one gets that:
\beq
S^0\sim {\cal E}^A\,\,,\qquad\qquad
V^0\sim J_B^k\,\,.
\eeq
Notice that our mass spectrum is consistent with these identifications, since according to
eq. (\ref{massrelations}),  the fluctuations $I_-^1$, $S^0$ and $V^0$ have all the same
mass spectrum, namely $M_s(n,0)$. 

Let us next consider the modes corresponding to higher values of $l$. Following ref.
\cite{WFO} we define the operator
\beq
{\cal C}_l^{I_0\cdots I_l}\,\equiv\,
{\cal C}^{(I_0}\,X_H^{I_1}\cdots X_H^{I_l)}\,\,,
\eeq
where the parentheses stand for the traceless symmetrization of the indices. The operator
${\cal C}_l$ has dimension $\Delta=l+1$, is a singlet of $SU(2)_V$ and transforms in the
spin
$l+1$ representation of $SU(2)_H$. It is thus a natural candidate to be identified with
the mode $I_-^{l+1}$, namely:
\beq
I_-^{l+1}\sim {\cal C}_l^{I_0\cdots I_l}\,\,.
\eeq
The mode $I_+^0$ has been identified in \cite{WFO} with a four-supercharge descendant of
the second-floor chiral primary ${\cal C}_1^{IJ}$. Notice that our mass spectra supports
this identification since $M_{I_+}(n,0)\,=\,M_{I_-}(n,2)$. Actually, our results are
consistent with having the modes $I_-^{l+1}$, $S^l$, $V^l$ and $I_+^{l-1}$ in the same
massive supermultiplet for $l\ge 1$ and with the identification of this supermultiplet
with the one obtained from the chiral primary ${\cal C}_l$, \ie:
\beq
(I_-^{l+1}, S^l, V^l, I_+^{l-1})\sim ({\cal C}_l,\cdots)\,\,,\qquad\qquad
(l\ge 1)\,\,.
\label{identf}
\eeq
As a check notice that the four modes on the left-hand side of eq. (\ref{identf}) have the
same mass spectrum, namely $M_S(n,l)$. Moreover, 
$\Delta(S^l)=\Delta(V^l)=\Delta(I_-^{l+1})+1$ and 
$\Delta(I_+^{l-1})=\Delta(I_-^{l+1})+2$, which is in agreement with the fact that the
supercharge has dimension $1/2$.

\setcounter{equation}{0}
\section{Fluctuations of the D3-D3 system}
\medskip
\label{D3D3}
Let us analyze in this section the modes of the $(1|D3\perp D3)$ intersection. In the
probe approximation we are considering the equations of motion of these fluctuation
modes are obtained from the Dirac-Born-Infeld action of a D3-brane in the $AdS_5\times
S^5$ background. This action is given by:
\beq
S\,=\,-\,\int d^4\xi\,\sqrt{-\det (g+F)}\,+\,
\int d^4\xi \,\,\,P\big[\,C^{(4)}\,\big]\,\,.
\label{DBID3-D3}
\eeq
We want to expand the action (\ref{DBID3-D3}) around the static configuration in which
the two branes are separated a distance $L$. 
Recall from section \ref{general} that the induced metric on the worldvolume of the
D3-brane probe of such a configuration is just the one written in eq.
(\ref{AdSindmetric}) for $d=1$, which reduces to $AdS_3\times S^1$ in the UV limit. As
in section \ref{general}, let us denote by $\chi$ the scalars transverse to both types
of branes. In this case the defect created by the probe has codimension two in the
Minkowski directions of $AdS_5\times S^5$. Let us assume that the D3-brane probe is
extended along the Minkowski coordinate $x^1$ and let us define:
\beq
\lambda_1=x^2\,\,,\qquad\qquad
\lambda_2=x^3\,\,.
\eeq 
With these notations, the lagrangian for the quadratic fluctuations can be readily
obtained from (\ref{DBID3-D3}):
\bear
&&{\cal L}\,=\,-\,{\rho\over 2}\,
{R^2\over \rho^2+L^2}\,
{\cal G}^{ab}\partial_a\chi\partial_b\chi\,-\,{\rho\over 2}\,{\rho^2+L^2\over R^2}\,
{\cal G}^{ab}\partial_a \lambda_i\partial_b \lambda_i\,-\,
{\rho \over 4}\,F_{ab}F^{ab}\,+\,\rc\rc
&&\qquad\qquad\qquad\qquad\qquad\qquad
+\,{(\rho^2+L^2)^2\over R^4}
 \,\epsilon^{ij}\partial_{\rho} \lambda_i\,\partial_{\varphi} \lambda_j\,\,,
\label{D3D3quadraticL}
\eear
where $i,j=1,2$ and ${\cal G}_{ab}$ is  the metric (\ref{AdSindmetric}) for
$d=1$. The equation of motion of $\chi$ derived from (\ref{D3D3quadraticL}) is just the
one studied in section \ref{general}, namely (\ref{eom-general}), for $p_2=3$, $d=1$ and
$\gamma_2=1$. Separating variables as in (\ref{sepvar}) we arrive at an equation which
can be analytically solved in terms of hypergeometric functions. The corresponding
solution has been written in eq. (\ref{scalarhyper}). After imposing the truncation
condition (\ref{quant})  we obtain the so-called $S^l$ modes, whose explicit expression
is:
\beq
\xi_S(\varrho)\,=\,\,=\,\varrho^l\,(\varrho^2+1)^{-n-l}\,
F(-n-l,  -n; l+1;-\varrho^2\,)\,\,.
\label{D3D3xiS}
\eeq
Notice that in this case the harmonics are just exponentials of the type
$e^{i\l\varphi}$, where $\varphi$ is just the angular coordinate of the $S^1$ circle.
Then, the quantum number $l$ can take also negative values. The modes written in
(\ref{D3D3xiS}) are those which are regular at $\rho=0$ for non-negative $l$. When $l<0$
one can get regular modes at the origin by using the second solution of the
hypergeometric function. The result is just (\ref{D3D3xiS}) with $l$ changed by $-l$.
However, since the scalar field $\chi$ whose fluctuation we are analyzing is real,
changing $l$ by $-l$ in $e^{i\l\varphi}$ makes no difference and we can restrict
ourselves to the case $l\ge 0$.  The mass spectrum and associated conformal dimensions of
the fluctuations  (\ref{D3D3xiS}) are:
\beq
M_S(n,l)\,=\,{2L\over R^2}\,
\sqrt{\bigg(n+l\bigg)\bigg(n+l+1\bigg)}\,\,,\qquad
\Delta_S\,=\,l+1\,\,,
\label{MSd3d3}
\eeq
where $n\ge 0$, except for the case $l=0$ where $n\ge 1$. Notice that for $n=l=0$ the
function $\xi_S(\varrho)$ is just constant. Moreover,  $M_S$ vanishes in this case and
thus we can take the solution $\chi$ to be also independent of the Minkowski
coordinates. This constant zero mode corresponds just to changing the value of the
distance $L$ and should not be considered as a true fluctuation. Therefore, we shall
understand that $n\ge 0$ in (\ref{D3D3xiS}) and (\ref{MSd3d3}), except in the case $l=0$
where $n\ge 1$.

\subsection{Scalar fluctuations}
Let us now study the fluctuations of the $\lambda_i$ scalars. Notice that these fields
are coupled through the Wess-Zumino term in (\ref{D3D3quadraticL}). Actually, 
the equations of motion for the $\lambda_i$'s derived from (\ref{D3D3quadraticL}) are:
\beq
R^2\,\partial_a\,\Bigg[\,\rho\,(\rho^2+L^2)\,
{\cal G}^{ab}\partial_b \lambda_i\,\Bigg]\,-\,4\rho\,( \rho^2+L^2)\,\epsilon^{ij}
\partial_{\varphi} \lambda_j\,=\,0\,\,.
\eeq
To solve these equations let us introduce the reduced variables 
$\bar M^2\,=\,-R^4L^{-2}k^2$, $\varrho=\rho/L$ and 
let us expand the $\lambda_i$'s in modes as:
\beq
\lambda_i\,=\,e^{ikx}\,e^{-il\varphi}\,\xi_i(\rho)\,\,,\qquad\qquad
i=1,2\,\,.
\label{D3D3separation}
\eeq
Then, the functions $\xi_i(\rho)$ satisfy the coupled equations:
\beq
{1\over \varrho (1+\varrho^2)}\,\partial_{\varrho}\,\Big[\,\varrho\,
(1+\varrho^2)^2\,\partial_{\varrho}\,\xi_i\,\Big]\,+\,
\Bigg[\,{\bar M^2\over 1+\varrho^2}\,\,-\,
l^2\,\big(1+{1\over \varrho^2}\big)\,
\Bigg]\,\xi_i\,+4il\,\epsilon^{ij}\xi_j\,=\,0\,\,.
\eeq
In order to diagonalize this system of equations, let us define the following complex
function:
\beq
w\,=\,\xi_1+ i\,\xi_2\,\,,
\label{D3D3w-def}
\eeq
which satisfies the differential equation:
\beq
{1\over \varrho (1+\varrho^2)}\,\partial_{\varrho}\,\Big[\,\varrho\,
(1+\varrho^2)^2\,\partial_{\varrho}\,w\,\Big]\,+\,
\Bigg[\,{\bar M^2\over 1+\varrho^2}\,\,-\,
l^2\,\big(1+{1\over \varrho^2}\big)\,
\Bigg]\,w\,+ 4l\,w\,=\,0\,\,.
\label{D3weqn}
\eeq
Equation  (\ref{D3weqn}) can be solved in terms of a hypergeometric function, namely:
\beq
w^{(1)}\,=\,
\varrho^{l}\,(1+\varrho^2)\,^{-\lambda-1}\,\,
F(-\lambda+1, l-\lambda-1;l+1; -\varrho^2)\,\,,
\label{w1}
\eeq
where $\lambda$ is related to $\bar M$ as in  (\ref{lambda}). 
Notice that $w^{(1)}$ is regular at $l=0$ for $l\ge 0$. Actually, since $l$ can be
negative in this case  one can also consider the second solution of the hypergeometric
equation, which is:
\beq
w^{(2)}\,=\,
\varrho^{-l}\,(1+\varrho^2)\,^{-\lambda-1}\,\,
F(-\lambda-1, -\lambda-l+1;1-l; -\varrho^2)\,\,.
\label{w2}
\eeq
By applying eq. (\ref{Delta-hyper}) to the present case, we obtain that 
the conformal dimension of a fluctuation of the type 
(\ref{hyper}) is just $\Delta=l-1$ or $\Delta=3-l$. Actually, 
it is straightforward to verify that the solutions of the differential equation 
(\ref{D3weqn}) present two different behaviours at $\varrho\to \infty$, namely 
$\varrho^{-l}$ and $\varrho^{l-4}$. The first behaviour corresponds to an operator with
$\Delta=l-1$, while $\Delta=3-l$ is the dimension of an operator whose dual fluctuation
behaves as $\varrho^{l-4}$ for large $\varrho$. In the following we will refer to the
fluctuations with
$\Delta=l-1$ as
$W_+^l$, while those with
$\Delta=3-l$ will be denoted by $W_-^l$. These two branches
\footnote{For $l=2$ both UV behaviours coincide and   there is no distinction between
the two branches. In this case there are solutions of the fluctuation equation which
behave as $\varrho^{-2}\log \varrho$ for large $\varrho$.  }
will be studied
separately in their unitarity range
$\Delta\ge 0$ by finding the truncations of the hypergeometric series of 
$w^{(1)}$ and $w^{(2)}$ with the appropriate behaviour at large $\varrho$.

\subsubsection{$W_+^l$ fluctuations}
Let us consider the solution $w^{(1)}$ in eq. (\ref{w1}) with the following truncation
condition:
\beq
l-\lambda-1\,=\,-n\,\,,\qquad\qquad
n=0,1,2,\cdots\,\,.
\eeq
The resulting solution is:
\beq
w\,=\,
\varrho^{l}\,(1+\varrho^2)\,^{-l-n}\,\,
F(2-l-n, -n;l+1; -\varrho^2)\,\,.
\label{W+sol}
\eeq
One can check easily that $w\sim\varrho^{-l}$ for large $\varrho$ if $l=1, n=0$ or $l\ge
2, n\ge 0$. Notice that $\Delta=l-1$ in this case and the unitarity range is $l\ge 1$.
The mass spectrum becomes:
\beq
M_{W_+}(n,l)\,=\,{2L\over R^2}\,
\sqrt{\Big(n+l-1\Big)\Big(n+l\Big)}\,\,,\qquad\qquad
\Delta_{W_+}\,=\,l-1\,\,,
\qquad\qquad (l\ge 1)\,\,.
\label{D3D3W+mass}
\eeq
The  $l=1$ fluctuation  is a special case. Indeed, in this case one has
$\Delta=0$ and, as $n$ must vanish,  $M$ is zero, \ie\ we
have a massless mode despite of the fact that we have introduced a mass scale by
separating the branes. As argued in ref. \cite{CEGK} this is related to the appearance of
the Higgs branch on the field theory side. Let us look closer at this $l=1$, $n=0$ mode.
In this case
$M=\lambda=0$ and  the hypergeometric function is just equal to one. Thus:
\beq
w\sim {\rho\over \rho^2+L^2}\,\,,
\label{higgsbranch}
\eeq
where we have reintroduced the constant $L$. In particular, for large $\rho$
\beq
w \approx {c\over \rho}\,\,,\qquad\qquad (\rho \to \infty)\,\,,
\eeq
where $c$ is a constant. Let us consider the solution in which $k=0$ (which certainly has 
$M=0$). This solution does not depend on the coordinates $(t,x^1)$. Let us introduce the
$\varphi$ dependence and define the following two complex variables:
\beq
\Lambda\equiv \lambda_1+i\lambda_2=x^2+ix^3\,\,,\qquad
Y=\rho \, e^{i\varphi}\,\,.
\eeq
Then the modes we are studying satisfy for large $\rho$:
\beq
\Lambda Y\approx c\,\,,\qquad\qquad (\rho \to \infty)\,\,,
\eeq
which is just the holomorphic curve of the Higgs branch found in ref. \cite{CEGK} by
looking at the vanishing of the F-terms of the susy theory. 

It is worth to stress here the difference between the $l=1$ solution (\ref{higgsbranch})
and the fluctuations (\ref{W+sol}) for $l>1$. Indeed, in the latter case we get a full
tower of solutions, depending on the excitation number $n$, whereas for $l=1$ we have
only the single function (\ref{higgsbranch}). Moreover, the mass spectra
(\ref{D3D3W+mass}) is simply related to the one corresponding to the transverse scalar
fluctuations $S^l$ only for $l>1$ (see section \ref{D3D3dictionary}). One can regard
(\ref{higgsbranch}) as a non-trivial solution in which the D3-brane probe is deformed at
no cost along the directions of the worldvolume of the D3-brane of the background.

It is also interesting to point out that the differential equation (\ref{D3weqn}) can be
solved by taking $w=\rho^{-l}$ and $M=0$. This fact can be checked directly from eq.
(\ref{D3weqn}) or by taking $\lambda\to 0$ in the solution (\ref{w2}). Notice, however
that this solution is not well-behaved at $\rho \to 0$ for $l\ge 1$, contrary to what
happens to the function written in eq. (\ref{higgsbranch}). A second solution with $M=0$
can be obtained by putting $\lambda=0$ in (\ref{w1}). This solution is regular at
$\rho\to 0$. However, for $l>1$ the hypergeometric function which results from taking 
$\lambda=0$ in (\ref{w1}) contains logarithms for $L\not=0$ and its interpretation in
terms of a holomorphic curve is unclear to us.

\subsubsection{$W_-^l$ fluctuations}
The allowed range of values of $l$ for the fluctuations $W_-^l$ is $l\le 3$. We have 
found a discrete tower of states only for $l\le 1$. As in the previous subsection, $l=1$ is
special. In this case the solutions regular at $\varrho=0$ which decrease as
$\varrho^{-3}$ for large 
$\varrho$ are:
\beq
w\,=\,
\varrho\,(1+\varrho^2)\,^{-n-1}\,\,
F(1-n, -n;2; -\varrho^2)\,\,,\qquad (l=1)\,\,,
\eeq
where $n\ge 1$. Indeed, this solution is just (\ref{W+sol}) for $l=1$ and $n\ge 1$.
Moreover, for $l\le 0$ the solutions which behave as $\varrho^{l-4}$ for large $\varrho$
can be obtained by putting $-\lambda-l+1\,=\,-n$, with $n\ge 0$, on the solution
$w^{(2)}$ of eq.  (\ref{w2}). One gets:
\beq
w\,=\,
\varrho^{-l}\,(1+\varrho^2)\,^{-2-n+l}\,\,
F(-2-n+l, -n;1-l; -\varrho^2)\,\,,\qquad (l\le 0)\,\,.
\eeq
The mass spectrum for $l\le 1$ can be written as:
\beq
M_{W_-}(n,l)\,=\,{2L\over R^2}\,
\sqrt{\Big(n+1-l\Big)\Big(n+2-l\Big)}\,\,,\qquad 
\Delta_{W_-}\,=\,3-l\,\,,\qquad
(l\le 1)\,\,,
\eeq
where it should be understood that $n\ge 1$ for $l=1$ and $n\ge 0$ otherwise.

\subsection{Vector fluctuations}
We will now study the fluctuations of the worldvolume gauge field. We will try to
imitate the discussion of section \ref{D3D5} for the D3-D5 system. Obviously, the
analogue of the type I modes does not exists for a one-dimensional sphere. Let us
analyze the spectra of the other two types of modes. 
\subsubsection{Type II modes}
Let us consider the ansatz
\beq
A_{\mu}\,=\,\xi_{\mu}\,\phi(\rho)\,e^{ikx}\,e^{-il\varphi}\,\,,\qquad
A_{\rho}\,=\,0\,\,,\qquad
A_{\varphi}\,=\,0\,\,,
\label{D3D3typeII}
\eeq
with $\xi_{\mu}$ being a constant vector such that $k^{\mu}\xi_{\mu}=0$. In terms of
the reduced variables $\varrho$ and $\bar M$, the equation for
$\phi(\rho)$ is:
\beq
\partial_{\varrho}\,\Big[\,\varrho\,\partial_{\varrho}\phi\,\Big]\,+\,
\Bigg[\,{\varrho\over (\varrho^2+1)^2}\,\bar M^2\,-\,{l^2\over
\varrho}\,\Bigg]\,\phi\,=\,0\,\,,
\eeq
which is the same as for the transverse scalars $\chi$. Therefore, the mass spectrum of
these type II vector modes is just the same as in (\ref{MSd3d3}).

\subsubsection{Type III modes}
We now adopt the ansatz:
\beq
A_{\mu}\,=\,0\,\,,\qquad
A_{\rho}\,=\,\phi(\rho)\,e^{ikx}\,e^{-il\varphi}\,\,,\qquad
A_{\varphi}\,=\,\tilde\phi(\rho)\,e^{ikx}\,e^{-il\varphi}\,\,. 
\label{D3D3typeIII}
\eeq
The equation for $A_{\rho}$ is:
\beq
il\partial_{\rho}\tilde\phi\,-\,l^2\phi\,+\,
M^2\,R^4\,{\rho^2\over (\rho^2+L^2)^2}\,\phi\,=\,0\,\,,
\eeq
while the equation for $A_{\varphi}$ yields:
\beq
\partial_{\rho}\Bigg[\,{(\rho^2+L^2)^2\over \rho}\,
\Big(\,\partial_{\rho}\tilde\phi\,+\,il\phi\,\Big)\,\Bigg]\,+\,
{M^2\,R^4\over \rho}\,\tilde\phi\,=\,0\,\,.
\eeq
Finally, the equation for $A_{\mu}$ gives a relation between $\phi$ and $\tilde\phi$,
namely:
\beq
\rho\,\partial_{\rho}\Big(\,\rho\phi\,\Big)\,=\,il\tilde\phi\,\,.
\label{D3D3phi-tildephi}
\eeq
For $l\not=0$ we can use (\ref{D3D3phi-tildephi}) to eliminate $\tilde\phi$ in favor of
$\phi$. The remaining equations reduce to the following equation for $\phi$:
\beq
\partial_{\varrho}\,\Big[\,\varrho\,\partial_{\varrho}(\,\varrho \phi\,)\,\Big]\,+\,
\Bigg[\,{\varrho^2\over (\varrho^2+1)^2}\,\bar M^2\,-\,l^2\,\Bigg]\,\phi\,=\,0\,\,,
\label{D3D3-typeIII-phi}
\eeq
where we have already introduced the reduced variables $\varrho$ and $\bar M$. 
The solution of (\ref{D3D3-typeIII-phi}) regular at $\rho=0$ is:
\beq
\phi\,=\,\varrho^{l-1}\,(1+\varrho^2)\,^{-\lambda}\,\,
F(-\lambda, l-\lambda;l+1; -\varrho^2)\,\,,
\eeq
with  $\lambda$ being the quantity defined in (\ref{lambda}). 
By imposing the quantization condition:
\beq
l-\lambda\,=\,-n\,\,,\qquad\qquad n=0,1,\cdots\,\,,
\eeq
we get a tower of fluctuation modes which behaves as $\rho^{-l-1}$ when $\rho\to\infty$.
The corresponding   mass levels and conformal dimensions are:
\beq
M_V(n,l)\,=\,{2L\over R^2}\,\sqrt{\Big(n+l\Big)\Big(n+l+1\Big)}\,\,,
\qquad\qquad \Delta_V\,=\,l+1\,\,,
\eeq
which again coincide with the results obtained for the scalar modes. 
\subsection{Fluctuation/operator correspondence}
\medskip
\label{D3D3dictionary}
The array corresponding to the D3-D3 intersection is:
\beq
\begin{array}{ccccccccccl}
 &1&2&3& 4& 5&6 &7&8&9 & \nonumber \\
D3: & \times &\times &\times &\_ &\_ & \_&\_ &\_ &\_ &     \nonumber \\
D3': &\times&\_&\_&\times&\times&\_&\_&\_&\_ &
\end{array}
\label{D3D3intersection}
\eeq
where the $D3'$ is the probe brane. First of all, let us discuss the isometries of this
configuration. Clearly, the addition of the brane probe breaks the $SO(6)$ symmetry
corresponding to rotations in the 456789 directions to $SO(4)\times U(1)_{45}$, where the
$SO(4)\approx SU(2)\times SU(2)$ factor is generated by rotations in the 6789 subspace and
the $U(1)_{45}$ corresponds to rotations in the plane spanned by coordinates $4$ and $5$.
In addition we have an extra $U(1)_{23}$ generated by the rotations in the $23$ plane. 

The field content of the defect theory can be obtained by reducing the ${\cal N}=4$,
$d=4$  gauge multiplet  down to two dimensions and by adding the corresponding $3-3'$
sector \cite{CEGK}. The resulting theory has $(4,4)$ supersymmetry in $d=2$. In
particular, two of the six $d=4$ adjoint chiral scalar superfields give rise to a field
$Q$ whose lowest component (which we will denote by $q$) describes the fluctuations of
the D3 in the directions $4$ and $5$. This field $q$ is a singlet of $SO(4)$ and 
$U(1)_{23}$ and is charged under $U(1)_{45}$. The strings stretched between the D3 and
the D3' give rise to two chiral multiplets $B$ and $\tilde B$ which are fundamental and
antifundamental with respect to the gauge group. The lowest components of $B$ and
$\tilde B$ are two scalar fields $b$ and $\tilde b$ which are singlets under $SO(4)$ and
are  charged under $U(1)_{23}$ and $U(1)_{45}$. Moreover, the fermionic components of
$B$ and $\tilde B$ can be arranged in two $SU(2)$ multiplets $\psi^+$ and $\psi^-$ which
are neutral with respect to the two $U(1)$'s and charged under one of the two $SU(2)$'s
of the decomposition $SO(4)\approx SU(2)\times SU(2)$. The dimensions and quantum
numbers of the different fields just discussed are summarized in table
\ref{tableFieldsD3D3}.
\begin{table}[!h]
\centerline{
\begin{tabular}[b]{|c|c|c|c|c|}   
 \hline
\rule{0mm}{4.5mm} Field  & $\Delta$  & $SO(4)$ & $U(1)_{23}$ &$U(1)_{45}$ \\ 
\hline 
  \rule{0mm}{5.5mm} $q$ & $1$  & $(0,0)$ & $0$ & $1$ \\  \hline  
\rule{0mm}{5.5mm}  $b, \tilde b$ &$0$  &  $(0,0)$  & $-1/2$ & $1/2$ \\ \hline  
\rule{0mm}{5.5mm} $\psi^+$ & $1/2$  & $(1/2, 0)$ &$0$&$0$ \\  \hline   
\rule{0mm}{5.5mm} $\psi^-$ & $1/2$  & $(0, 1/2)$ &$0$&$0$ \\ \hline  
\end{tabular}
}
\caption{Quantum numbers and dimensions of the fields of the D3-D3 intersection.}
\label{tableFieldsD3D3} 
\end{table}

In order to establish the fluctuation/operator dictionary in this case, let us determine
the quantum numbers of the different fluctuations. The fluctuations in the directions
$6789$ (which are transverse to both types of D3-branes) will be denoted by $S^l$. Clearly
they transform in the $(1/2,1/2)$ representation of $SO(4)$ and are neutral under 
$U(1)_{23}$. Moreover, since the rotations of the 45 plane are just those along the
one-sphere of the probe worldvolume, the integer $l$ is just the charge under 
$U(1)_{45}$. The  $w$ coordinate parametrizes the 23 plane. Let us denote by 
$W_+^l$ and $W_-^l$ to the two branches of $w$ fluctuations. The $W_{\pm}^l$'s are 
$SO(4)$ singlets and are charged under both $U(1)_{23}$ and $U(1)_{45}$.  As in the D3-D5
case, the fluctuations of types II and III correspond, in the conformal limit, to the
components of a vector field in $AdS_3$ and will be denoted by $V^l$. They are singlets
under $SO(4)$ and $U(1)_{23}$ . Table \ref{tableModesD3D3} is filled in with the
dimensions and quantum numbers of the different fluctuations.

\begin{table}[!h]
\centerline{
\begin{tabular}[b]{|c|c|c|c|c|}   
 \hline
\rule{0mm}{4.5mm} Mode  & $\Delta$  & $SO(4)$ & $U(1)_{23}$ &$U(1)_{45}$ \\ 
\hline 
  \rule{0mm}{5.5mm}$S^l$ & $l+1$   & $(1/2,1/2)$ & $0$ & $l\ge 0$ \\  \hline  
\rule{0mm}{5.5mm}  $W_+^l$ &$l-1$  &  $(0,0)$  & $-1$&$l\ge 1$ \\ \hline  
\rule{0mm}{5.5mm} $W_-^l$ & $3-l$  & $(0, 0)$ &$-1$&$l\le 1$ \\  \hline   
\rule{0mm}{5.5mm} $V^l$ & $l+1$  & $(0, 0)$ &$0$&$l\ge 0$ \\ \hline  
\end{tabular}
}
\caption{Quantum numbers and dimensions of the modes of the D3-D3 intersection.}
\label{tableModesD3D3} 
\end{table}

The mass spectrum $M_S(n,l)$ of the fluctuations $S^l$ has been written in eq.
(\ref{MSd3d3}). The masses of the other modes can
be written in terms of $M_S(n,l)$ as:
\bear
&&M_{W_+}(n,l)\,=\,M_S(n,l-1)\,\,,\qquad(l\ge 2),\rc
&&M_{W_-}(n,l)\,=\,M_S(n,1-l)\,\,,\qquad(l\le 1),\rc
&&M_{V}(n,l)\,=\,M_S(n,l)\,\,,\qquad\qquad\,\,\,(l\ge 0).
\label{D3massrelations}
\eear
Notice that the relation between $M_S$ and $M_{W_-}$ is consistent with the absence of
the
$n=0$ mode in the $S^0$ and $W_-^1$ fluctuations.

In order to relate the different fluctuations to composite operators of the defect theory
let us define following ref. \cite{CEGK} the operator ${\cal B}^l$ for $l\ge 1$ as:
\beq
{\cal B}^l\equiv \tilde b q^{l-1}\,b\,\,.
\eeq
Notice that ${\cal B}^l$ has the same dimension and quantum numbers as the fluctuation 
$W_+^l$. Similarly, the operator ${\cal G}^l$ for $l\le 1$, defined as:
\beq
{\cal G}^l\equiv D_-\tilde b q^{\dagger 1-l}\, D_+b\,+\,
D_+\tilde b q^{\dagger 1-l}\, D_-b\,\,,
\eeq
where $D_\pm=D_+\pm D_-$, has the right properties to be identified with the dual of the
fluctuations $W_-^l$. Moreover, to define the operator dual to $S^l$ we have to build a
vector of $SO(4)$. The natural objects to build  an operator of this sort
are the spinor fields $\psi^{\pm}$. Indeed, such an operator can written as:
\beq
{\cal C}^{\mu l}\equiv \sigma_{ij}^{\mu}\,\big(\,\epsilon_{ik}\bar\psi_k^{+}\,q^l
\psi_j^{-}\,+\,
\epsilon_{jk}\bar\psi_k^{-}\,q^l \psi_i^{-}\,\big)\,\,,
\eeq
where $l\ge 0$ and 
$\mu$ is a $SO(4)$ index. Therefore, according to the proposal of ref. \cite{CEGK}, we have
\beq
S^l\sim {\cal C}^{\mu l}\,\,,\qquad
W_+^l\sim {\cal B}^l\,\,,\qquad
W_-^l\sim {\cal G}^l\,\,.
\eeq
As argued in ref. \cite{CEGK}, the fluctuations $W_+^l$ for $l=1$ are dual to the field 
${\cal B}^1=\tilde b b$, which parametrizes the classical Higgs branch of the theory,
whereas for higher values of $l$ these fluctuations correspond to other holomorphic
curves. Moreover, the field ${\cal C}^{\mu l}$ is a BPS primary and 
${\cal G}^{1-l}$ is a two supercharge descendant of this primary. Notice that this is
consistent with our relation (\ref{D3massrelations}) between the masses of the $S^l$ and
$W_-^l$ fluctuations. Lastly, the fluctuations $V^l$ are dual to two-dimensional vector
currents. The dual operator at the bottom of the Kaluza-Klein tower is a global $U(1)$
current ${\cal J}_B^{M}$:
\beq
V^0\sim {\cal J}_B^{M}\,\,.
\eeq
The expression of ${\cal J}_B^{M}$ has been given in ref. \cite{CEGK}, namely:
\def\lrD{\buildrel \leftrightarrow\over D}
\beq
{\cal J}_B^{M}\equiv \psi_i^{\alpha}\,\rho_{\alpha\beta}^M\,\psi_i^{\beta}\,+\,
i \bar b \lrD{}^M b\,+\,i \tilde b\,\lrD{}^M\,\bar{\tilde b}\,\,,
\qquad (M=0,1)\,\,,
\eeq
where $\alpha,\beta=+, -$ and $\rho^M$ are Dirac matrices in two dimensions.

\setcounter{equation}{0}
\section{Concluding Remarks}
\medskip
\label{Conclusion}

In this paper we have studied the fluctuation spectra of brane probes in the
near-horizon background created by a stack of other branes. In the context of the
generalization of the gauge/gravity correspondence proposed in refs. \cite{KR,KKW} these
fluctuations are dual to open strings stretching between the two types of branes of the
intersection and can be identified, on the field theory side of the correspondence, with
composite operators made up from hypermultiplets in the fundamental representation of
the gauge group. We have mainly studied the cases of D5- and D3-brane probes moving in
the $AdS_5\times S^5$ geometry. In these two cases, if the brane reaches the origin of
the holographic coordinate,  it wraps an  $AdS_{d+2}\times S^d$ ($d=2,1$) submanifold of
the  $AdS_5\times S^5$ background and the corresponding field theory duals are defect
conformal field theories with a fundamental hypermultiplet localized at the defect. The 
spectra of conformal dimensions and the precise mapping between probe fluctuations and
operators of the dual defect theory for the D3-D5 and D3-D3 intersections were obtained
in refs. \cite{WFO} and \cite{CEGK} respectively.

By
allowing a finite separation between the probe and the origin of the holographic
direction we introduce an explicit mass scale in the problem, which is related to the
mass of the hypermultiplet. The system is no longer conformal and develops a mass gap.
By studying the fluctuations of the probes we have obtained the mass spectrum of the
corresponding open string degrees of freedom. Remarkably, for the D3-D5 and D3-D3 systems
the full set of differential equations for the fluctuations can be solved in terms of the
hypergeometric function and the corresponding mass spectra can be obtained analytically. 
These mass spectra display some degeneracies which are consistent with the structure of
the supermultiplets found in refs. \cite{WFO,CEGK} for the corresponding dual operators. 

In the appendices  we have also studied the case of supersymmetric D-brane probes in the 
background of a Dp-brane for $p\not=3$.  In these cases the fluctuation equations can
also be decoupled although one needs to use numerical methods or WKB estimates in order
to get the mass spectra. The fluctuation/operator dictionary for these intersections has
not been worked out in detail in the literature. However, we have obtained relations
between the masses of the different modes which closely resemble the ones found for the 
$AdS_5\times S^5$ background. Moreover, one has to keep in mind \cite{jm} that, in order
to trust these supergravity solutions, both the curvature in string units and the
dilaton must be small. This fact introduces restrictions in the range of the
holographic coordinate for which the correspondence between the supergravity and gauge
theory descriptions is valid \cite{IMSY}. It is interesting to point out that in some
cases we can avoid that our fluctuations enter the ``bad" region by selecting
appropriately the value of the distance $L$.

This paper is a contribution to the program which aims at the extension of the
gauge/gravity correspondence to the case in which the field theory dual contains matter
in the fundamental representation. This program is still in progress and at the present
time it has some unsolved problems, which are also reflected in our results. Let us
comment on some of them. First of all, notice that all mass spectra we have found grow
linearly with the excitation number $n$ for large $n$. This behaviour is common to all
the cases in which the  masses are extracted from the analysis of the fluctuations of
D-branes, both in non-confining and confining backgrounds. This same type of spectrum is
obtained in the effective holography models \cite{Effective}.
It has been argued in ref.
\cite{Schreiber} that the spectrum of highly excited mesons in confining gauge theories
should be a linear function of $\sqrt{n}$ for large $n$. This discrepancy lead us to
think that somehow the approach based on the small fluctuations of brane probes needs to
be corrected. Notice that, for large spin, the open string can be treated
semiclassically. For the D3-D7 intersection this analysis was performed in ref.
\cite{KMMW}, while open spinning strings for the defect conformal field theory were
studied in refs.
\cite{OTY} in connection with their relation with integrable  spin chains 
\cite{LP,DWM}.

Another criticism that one could make to the approach followed here is the fact that we
have employed the probe approximation and we have neglected the backreaction  of
the branes on the geometry. Indeed, in some cases, one can construct a supergravity
background representing the localized intersection \cite{Beyond}. However, these
backgrounds are rather complicated and it is not  easy to extract information about the
gauge theory dual. Maybe, in order to go beyond the probe approximation, it would be
more fruitful to follow the approach recently proposed in \cite{CNP} for the ${\cal
N}=1$ theories, where the supergravity action is supplemented with the action of the
brane, which has been conveniently smeared,  and  a solution of the equations of motion
for the supergravity plus brane system is found. In the approach of \cite{CNP} the
adjoint (color) degrees of freedom are represented by fluxes, whereas the fundamental
(flavor) fields are generated by branes. 

Despite the limitations of our approach just discussed we think that it provides a
non-trivial realization of the holographic idea (see \cite{KOBS} for a rigorous
treatment)  and it would be  interesting in the future to look at some generalizations
of our results. For example, it has been argued in \cite{Higgs} that the Higgs branch of
the D3-D7 system can be generated by turning on a non-zero instanton field on the
D7-brane probe. This result could be clearly generalized for any Dp-D(p+4) intersection.
Similarly, in the Dp-D(p+2) system one can switch a non-zero flux of the worldvolume
gauge field  across the  $S^2$. As checked in ref. \cite{ST}, in order to preserve
some fraction of supersymmetry one has to introduce some bending on the defect boundary.
Clearly, the analysis of the fluctuations around such configurations could shed light to
understand the nature of the deformation induced by the flux on the field theory side. 

It is also of great interest to look at defect theories with reduced supersymmetry. One
of such theories is obtained by embedding a D5-brane in the $AdS_5\times T^{1,1}$
geometry. The precise form of the embedding in this case can be found in
\cite{conifold}. Actually, the $T^{1,1}$ space can be substituted by any Sasaki-Einstein
space \cite{Yamaguchi}, as illustrated in \cite{CEPRV} for the $Y^{p,q}$ manifold. The
supersymmetric defects in the Maldacena-Nu\~nez background have been obtained in ref. 
\cite{CPR}.

We hope to make some progress along these lines in the future.

\medskip
\section*{Acknowledgments}
\medskip

We are grateful to J. D. Edelstein, J. Erdmenger, J. Mas, P. Merlatti, C. N\'u\~nez, A.
Paredes and 
 M.~Schvellinger  for discussions and useful comments.
This work  was
supported in part by MCyT, FEDER and Xunta de Galicia under grant
FPA2005-00188 and by  the EC Commission under  
grants HPRN-CT-2002-00325 and MRTN-CT-2004-005104.

\vskip 1cm
\renewcommand{\theequation}{\rm{A}.\arabic{equation}}
\setcounter{equation}{0}
\medskip
\appendix

\setcounter{equation}{0}
\section{Change of variables for the exact spectra}
\medskip
\label{hyperchange}
Let us consider an equation of the type:
\beq
z(1-z)\,\phi''\,+\,(\alpha\,+\,\beta z)\,\phi'\,+\,
[\,\gamma\,+\,\delta\,z^{-1}\,+\,\epsilon\,(1-z)^{-1}\,]\,\phi=0\,\,,
\label{hypereq}
\eeq
where $\phi=\phi(z)$, the prime denotes derivative with respect to $z$ and $\alpha, \beta,
\gamma,\delta$ and $\epsilon$ are constants. The solution of this equation can be written
as
\beq
\phi(z)\,=\,z^{f}\,(1-z)^{g}\,P(z)\,\,,
\eeq
where $P(z)$ satisfies the hypergeometric equation
\beq
z(1-z)\,P''\,+\,[c-(a+b+1)z]\,P'\,-\,ab\,P\,=\,0\,\,.
\eeq

The values of the exponents $f$ and $g$ are
\bear
&&f\,=\,{1\,-\,\alpha\,+\,\lambda_1\,\sqrt{(\alpha-1)^2\,-\,4\delta}\,\over 2}\,\,,\rc\rc
&&g\,=\,{\alpha+\beta+1\,+\,\lambda_2\,
\sqrt{(\alpha+\beta+1)^2\,-\,4\epsilon}\over 2}\,\,,
\label{fyg}
\eear
where $\lambda_1$ and $\lambda_2$ are signs which can be chosen by convenience. Moreover,
$a$, $b$ and $c$ are given by:
\bear
&&a\,=\,f+g\,-\,{1+\beta+\sqrt{4\gamma\,+\,(\beta+1)^2}\over 2}\,\,,\rc\rc
&&b\,=\,f+g\,-\,{1+\beta-\sqrt{4\gamma\,+\,(\beta+1)^2}\over 2}\,\,,\rc\rc
&&c\,=\,\alpha+2f\,\,.
\eear
There are two  solutions for $P(z)$ in terms of the hypergeometric function. The first
one is:
\beq
P(z)\,=\,F(a,b;c;z)\,\,.
\eeq
The second solution is:
\beq
P(z)\,=\,z^{1-c}\,F(a-c+1, b-c+1; 2-c;z)\,\,.
\eeq

\vskip 1cm
\renewcommand{\theequation}{\rm{B}.\arabic{equation}}
\setcounter{equation}{0}
\medskip

\setcounter{equation}{0}
\section{Fluctuations of the Dp-D(p+2) system}
\medskip
\label{DpD(p+2)}
In this section we analyze the full set of fluctuations of the 
$(p-1|Dp\perp D(p+2))$ intersections for $p<5$. Many of the results are direct
generalizations of those corresponding to the D3-D5 system studied in section
\ref{D3D5}.  Our starting point is the Dirac-Born-Infeld lagrangian density of the
D(p+2)-brane  probe, which is the sum of the Born-Infeld part ${\cal L}_{BI}$ and the
Wess-Zumino part ${\cal L}_{WZ}$, which for the case at hand are given by:
\beq
{\cal L}_{BI}\,=\,-e^{-\phi}\,\sqrt{-\det(\,g+F\,)}\,\,,\qquad\qquad
{\cal L}_{WZ}\,=\,P\big[\,C^{(p+1)}\,\big]\,\wedge F\,\,,
\label{LDpDp+2}
\eeq
where, again, we are taking the D(p+2)-brane tension equal to one and $F=dA$ is the
worldvolume gauge field. We shall  expand the action around the configuration in which
both types of branes are separated a fixed distance $L$. Let $\chi$ be the scalars
transverse to both types of branes and let $X$ denote the scalar which is transverse to
the probe and that is directed along the worldvolume of the Dp-branes of the background
(\ie\  $X\equiv x^p$). At quadratic order, the Born-Infeld lagrangian becomes:
\beq
{\cal L}_{BI}\,=\,-\rho^2\,\sqrt{\tilde g}\,\Bigg[\,{1\over 2}\,
\Bigg[{R^2\over \rho^2+L^2}\Bigg]^{{7-p\over 4}}{\cal G}^{ab}\,
\partial_{a}\chi\partial_{b}\chi\,+\,
{1\over 2}\Bigg[{\rho^2+L^2\over R^2}\Bigg]^{{7-p\over 4}}\,\,{\cal G}^{ab}
\partial_{a}X\partial_{b}X\,+\,{1\over 4}F_{ab}F^{ab}
\Bigg]\,\,,
\label{LBIDpDp+2}
\eeq
where ${\cal G}_{ab}$ is the induced metric of the unperturbed static configuration, 
given by:
\beq
{\cal G}_{ab}d\xi^a d\xi^b\,=\,
\Bigg[{\rho^2+L^2\over R^2}\Bigg]^{{7-p\over 4}}\,\,
dx^2_{1,p-1}\,+\,
\Bigg[{R^2\over \rho^2+L^2}\Bigg]^{{7-p\over 4}}\Big[\,
d\rho^2\,+\,\rho^2\,d\Omega_2^2\,\Big]\,\,.
\eeq

Let us now write the explicit form of the Wess-Zumino term ${\cal L}_{WZ}$. Recall that
the Ramond-Ramond (p+1)-form potential $C^{(p+1)}$ has components along the Minkowski
directions $x^0\cdots x^p$ (see eq. (\ref{CRR})). Therefore, it is clear from the
expression of ${\cal L}_{WZ}$ in (\ref{LDpDp+2}) that the potential $C^{(p+1)}$ induces
a coupling between the scalar $X$ and the components of the worldvolume field $F$ along
the radial coordinate
$\rho$ and along the angular directions of the two-sphere (which we denote by $\theta$
and $\varphi$). Actually, one can prove that:
\beq
{\cal L}_{WZ}\,=\,\Bigg[{\rho^2+L^2\over R^2}\Bigg]^{{7-p\over 2}}\,
\big[\partial_{\rho}X\,F_{\theta\varphi}\,+\,
\partial_{\theta}X\,F_{\varphi\rho}\,+\,
\partial_{\varphi}X\,F_{\rho\theta}\,\big]\,\,.
\eeq
Integrating by parts, one can rewrite ${\cal L}_{WZ}$ as:
\beq
{\cal L}_{WZ}\,=\,-{7-p\over 2}\,{\rho\over R^{7-p}}\,
\Big[\rho^2+L^2\Big]^{{5-p\over 2}}\,
X\epsilon^{ij}F_{ij}\,\,,
\label{LWZDpDp+2}
\eeq
where $i,j$ denote components along the $S^2$ and $\epsilon^{ij}=\pm 1$. As a check of
the expressions (\ref{LBIDpDp+2}) and (\ref{LWZDpDp+2}) for ${\cal L}_{BI}$ and ${\cal
L}_{WZ}$ notice that, by taking $p=3$, they reduce to the ones found in section
\ref{D3D5} for the D3-D5 system.

Let us study first  fluctuations of the scalars $\chi$ which, as is evident
from eqs.  (\ref{LBIDpDp+2}) and (\ref{LWZDpDp+2}), are decoupled from the other modes.
The corresponding equations of motion can be obtained from the ones of a general
intersection with
$p_2=p+2$ and $d=p-1$ and with the $\gamma_i$ written
in eq. (\ref{Dpgammas}). The resulting WKB spectra is obtained by plugging these
parameters in the general expression (\ref{WKBM}). One gets: 
\beq
M^{WKB}_{S}(n,l)\,=\,2\sqrt{\pi}\,{L^{{5-p\over 2}}\over R^{{7-p\over 2}}}\,\,
{\Gamma\Big({7-p\over 4}\Big)\over \Gamma\Big({5-p\over 4}\Big)}\,\,
\sqrt{(n+1)\,\Big(n\,+\,{7-p\over 5-p}\,\big(l+{1\over 2}\big)\,\Big)}\,\,,
\label{DpDp+4-WKBscalarmass}
\eeq
where $R$ is given by eq. (\ref{RDp}).
Moreover, the behavior of the fluctuation $\xi$ when $\varrho\to
0$ is of the form $\xi\sim\varrho^{\gamma}$, where $\gamma$ satisfies a quadratic equation
whose two roots are $\gamma=l, -l-1$. Clearly, the regular solution should correspond to
the root
$\gamma=l$, \ie\ $\xi$ must behave as:
\beq
\xi\sim\varrho^{l}\,\,,
\,\,\,\,\,\,\,\,\,\,\,\,\,\,\,\,\,\,\,\,\,\,\,\,\,\,\,\,
{\rm as}\,\,\,\,\,\,\,\,\,\,\,\varrho\to 0\,\,.
\eeq

In table \ref{MassDpDp+2-Scalar} we compare the numerical results for $\bar M$ with the
WKB  formulas for some $(p-1|Dp\perp D(p+2))$ intersections.

\begin{table}[!h]
\begin{tabular}[b]{|c|c|c|}   
\hline  
\multicolumn{3}{|c|}{$(0|D1\perp D3)$ with $l=0$}\\
\hline
 $n$  & WKB  & Numerical \\ 
\hline   
\ \ 0 & $7.40$  & $5.68$  \\   
\ \ 1 & $34.54$  & $32.40$  \\   
\ \ 2 & $81.42$  & $79.06$  \\   
\ \ 3 & $148.04$  & $145.52$  \\     
\ \ 4 & $234.40$  & $231.76$  \\
\ \ 5 & $340.50$  & $337.76$  \\ 
\hline
\end{tabular}
\qquad
\begin{tabular}[b]{|c|c|c|}   
\hline  
\multicolumn{3}{|c|}{$(1|D2\perp D4)$ with $l=0$}\\
\hline
 $n$  & WKB  & Numerical \\ 
\hline   
\ \ 0 & $5.73$  & $4.34$  \\   
\ \ 1 & $25.21$  & $23.66$  \\   
\ \ 2 & $58.44$  & $56.80$  \\   
\ \ 3 & $105.42$  & $103.74$  \\     
\ \ 4 & $166.15$  & $164.44$  \\
\ \ 5 & $240.63$  & $238.92$  \\ 
\hline
\end{tabular}
\qquad
\begin{tabular}[b]{|c|c|c|}   
\hline  
\multicolumn{3}{|c|}{$(3|D4\perp D6)$ with $l=0$}\\
\hline
 $n$  & WKB  & Numerical \\ 
\hline   
\ \ 0 & $2.15$  & $1.68$  \\   
\ \ 1 & $7.18$  & $6.78$  \\   
\ \ 2 & $15.07$  & $14.72$  \\   
\ \ 3 & $25.84$  & $25.58$  \\     
\ \ 4 & $39.48$  & $39.34$  \\
\ \ 5 & $55.99$  & $56.02$  \\ 
\hline
\end{tabular}
\caption{Numerical and WKB values of $M^2$ (in units of $R^{7-p}L^{p-5}$) for the $S^l$
modes of the Dp-D(p+2) intersection for $p=1,2,4$ and $l=0$.}
\label{MassDpDp+2-Scalar}
\end{table}

\bigskip

We now  want to address the analysis of the remaining fluctuation modes. First of all,
from the expressions of ${\cal L}_{BI}$ and ${\cal L}_{WZ}$ (eqs. (\ref{LBIDpDp+2}) and 
(\ref{LWZDpDp+2})) it is straightforward to find the equation of motion of the scalar
$X$, namely: 
\beq
R^{{7-p\over 2}}\partial_{a}\,\Bigg(\,\rho^2\sqrt{\tilde g}\,
\Big[\rho^2+L^2\Big]^{{7-p\over 4}}
{\cal G}^{ab}\partial_b X\,\Bigg)\,-\,
{7-p\over 2}\,\rho\,\,
\Big[\rho^2+L^2\Big]^{{5-p\over 2}}\,
\epsilon^{ij}F_{ij}\,=\,0\,\,,
\label{DpDp+2Xeq}
\eeq
while those of the gauge field are:
\beq
R^{7-p}\,\partial_a\,\Big[\,\rho^2\sqrt{\tilde g}\,F^{ab}\,\Big]\,-\,(7-p)\,
\rho\,
\Big[\rho^2+L^2\Big]^{{5-p\over 2}}\,
\epsilon^{bi}\partial_iX\,=\,0\,\,,
\label{DpDp+2Aeq}
\eeq
where $\epsilon^{bi}$ is zero unless $b$ is an index along the two-sphere.
As in the D3-D5 case, the scalar $X$ is coupled to the gauge field strength $F_{ij}$
along the two-sphere. To decouple these two fields we will follow the same strategy as
in the exactly solvable case of section \ref{D3D5}. First of all, we introduce the two
types of vector spherical harmonics on $S^2$, namely $Y_i^l$ and $\hat Y_i^l$. These
functions were defined in eqs. (\ref{S2har}) and (\ref{hatS2har}) respectively.
Subsequently, we will define the three types of modes, namely I, II and III, exactly as
in the D3-D5 case and we will be able to decouple the corresponding equations of motion.
For the general $(p-1|Dp\perp D(p+2))$  intersection these equations cannot be solved
analytically, although they are easy to solve numerically and WKB expressions for the
mass levels can be readily found.

\subsection{Type I modes}
Let us adopt  an ansatz  for the gauge field and the scalar $X$ as in eqs. (\ref{typeIA})
and (\ref{typeIX}). The equation of motion of $X$ (eq. (\ref{DpDp+2Xeq})) becomes:
\bear
&&\rho^2\,R^{7-p}\partial_{\mu}\partial^{\mu}\Lambda\,+\,
\partial_{\rho}\Bigg[\rho^2 \Big[\rho^2+L^2\Big]^{{7-p\over 2}}
\partial_{\rho}\Lambda\,\Bigg]\,-\,\rc\rc
&&-\,
l(l+1)\,\Big[\rho^2+L^2\Big]^{{7-p\over 2}}\Lambda\,-\,
(7-p)l(l+1)\rho \Big[\rho^2+L^2\Big]^{{5-p\over 2}}\,\phi\,=\,0\,\,,
\eear
while that of the gauge field (eq. (\ref{DpDp+2Aeq})) reduces to:
\bear
&&R^{7-p}\partial_{\mu}\partial^{\mu}\phi\,+\,
\partial_{\rho}\Bigg[ \Big[\rho^2+L^2\Big]^{{7-p\over 2}}
\partial_{\rho}\phi\,\Bigg]\,-\,\rc\rc
&&-{l(l+1)\over \rho^2}\,\Big[\rho^2+L^2\Big]^{{7-p\over 2}}\phi\,-\,
(7-p)\rho \Big[\rho^2+L^2\Big]^{{5-p\over 2}}\,\Lambda\,=\,0\,\,.
\eear
Let us next define  $V=\rho\Lambda$ as in the D3-D5 case and the following differential
operator:
\beq
\Delta^2_{(p+1)}\,\Psi\,\equiv\,{1\over (\rho^2+L^2)^{{5-p\over 2}}}\,\Bigg[\,
R^{7-p}\,\partial_{\mu}\partial^{\mu}\,\Psi\,+\,
\partial_{\rho}\Big[\,
\,(\,\rho^2\,+\,L^2\,)^{{7-p\over 2}}\,\partial_{\rho}\,\Psi\Big]\,\Bigg]\,\,.
\eeq
With these definitions the equations for the fluctuations become the following system of
coupled differential equations:
\bear
&&\Delta^2_{(p+1)}\,V\,=\,\Bigg[\,l(l+1)\,+\,7-p\,+\, l(l+1)\,{L^2\over
\rho^2}\,\Bigg]V\,+\, (7-p)l(l+1)\,\phi\,\,,\rc\rc
&&\Delta^2_{(p+1)}\,\phi\,=\,\Bigg[\,l(l+1)\,+\, l(l+1)\,{L^2\over \rho^2}\,\Bigg]\phi\,+\,
(7-p)\,V\,\,.
\label{DpDp+2system}
\eear
Let us decouple the equations (\ref{DpDp+2system}) by following a procedure similar to
the one employed in \cite{WFO} for the $L=0$ case. With this purpose we consider first
the system (\ref{DpDp+2system}) for $\rho\to\infty$. In this UV limit
(\ref{DpDp+2system}) reduces to:
\beq
\Delta^2_{(p+1)}\,\pmatrix{V\cr\phi}\,=\,{\cal M}\,\pmatrix{V\cr\phi}\,\,,
\label{UVsystem}
\eeq
with ${\cal M}$ being the following constant matrix:
\beq
{\cal M}\,=\,\pmatrix{l(l+1)+7-p&&(7-p)l(l+1)\cr\cr
7-p&&l(l+1)}\,\,.
\eeq
The UV system (\ref{UVsystem}) is readily decoupled by finding the linear combinations
of the functions $V$ and $\phi$ that diagonalize the matrix ${\cal M}$. Interestingly,
the eigenvalues of ${\cal M}$ are integers and given by:
\beq
(l+1)(l+7-p)\,\,,\qquad\qquad
l(l+p-6)\,\,,
\label{M-eigenvalues}
\eeq
while the corresponding decoupled functions are:
\bear
&&Z^+\,=\,V\,+\, l\phi\,\,,\rc\rc
&&Z^-\,=\,V\,-\,(l+1)\,\phi\,\,.
\label{M-eigenfunctions}
\eear
Let us now write the equations for $Z^{\pm}$ for finite $\rho$. Remarkably, by
substituting the definitions of $Z^{\pm}$ in (\ref{DpDp+2system}), one can verify that
these equations are still decoupled for finite $\rho$ and take the form:
\bear
&&\Delta^2_{(p+1)}\,Z^+\,=\,(l+1)\bigg(\,l+7-p+l{L^2\over \rho^2}\,\bigg)\,Z^+\,\,,\rc\rc
&&\Delta^2_{(p+1)}\,Z^-\,=\,l\bigg(\,l+p-6+(l+1){L^2\over \rho^2}\,\bigg)\,Z^-\,\,.
\label{DpDp+2deco-system}
\eear
As a check, let us notice that when $\rho\to\infty$ the right-hand sides of the two
equations in  (\ref{DpDp+2deco-system}) are just the eigenvalues of ${\cal M}$,
displayed in (\ref{M-eigenvalues}),  multiplying the functions $Z^{\pm}$. To analyze
these equations let us proceed as in the D3-D5 case and
separate variables as:
\beq
Z^{\pm}\,=\,e^{ikx}\,\xi^{\pm}(\rho)\,\,.
\eeq
Except for the $p=3$ case, the resulting decoupled ordinary differential equations for
$\xi^{+}(\rho)$ and $\xi^{-}(\rho)$ cannot be solved in an analytic form. However, we
can extract some important qualitative information on the behaviour of their solutions by
rewriting them  as Schr\"odinger equations. In order to perform this analysis, let us
introduce the reduced quantities $\varrho$ and $\bar M$ as:
\beq
\varrho\,=\,{\rho\over L}\,\,,
\,\,\,\,\,\,\,\,\,\,\,\,\,\,
\bar M^2\,=\,-R^{7-p}\,L^{p-5}\,k^2\,\,.
\label{DpDp+4-newvariables}
\eeq
Moreover, by changing variables as
\beq
e^{y}\,=\,\varrho\,\,,\qquad\qquad
\psi^{\pm}\,=\,{\Big[1+\varrho^2\Big]^{{7-p\over 4}}\over \sqrt{\varrho}}
\,\,\xi^{\pm}\,\,,
\eeq
we can convert the fluctuation equations (\ref{DpDp+2deco-system})
into the Schr\"odinger equation
\beq
\partial_{y}^2\,\psi^{\pm}\,-\,V_{\pm}(y)\,\psi^{\pm}\,=\,0\,\,,
\eeq
where the potentials $V_{\pm}(y)$ are given by:
\bear
&&V_{\pm}(y)=\bigg(l+{1\over 2}\bigg)^2\pm (7-p) \bigg(l+{1\over 2}\pm 1\bigg)
{e^{2y}\over 1+e^{2y}}+\rc\rc
&&\qquad\qquad+{1\over 4}\,(7-p) (3-p) \,
{e^{4y}\over (1+e^{2y})^2}-\bar M^2
{e^{2y}\over ( 1+e^{2y})^{{7-p\over 2}}}\,\,.
\label{DpDp+4potentials}
\eear
By looking at the asymptotic values of the potentials $V_{\pm}$ at $y\to\pm\infty$ we
can get the behaviour of the fluctuations $\xi^{\pm}$ at $\varrho\approx 0, \infty$.
Indeed, from the  potentials (\ref{DpDp+4potentials}) we obtain:
\beq
\lim_{y\to -\infty} V_{\pm}(y)\,=\,\bigg(l+{1\over 2}\bigg)^2\,\,,
\qquad\qquad
\lim_{y\to +\infty} V_{\pm}(y)\,=\,
\bigg(l+{1\over 2}\pm {7-p\over 2}\bigg)^2\,\,.
\eeq
From these values one can prove that  for  $\varrho\approx 0$ the functions $\xi^{\pm}$
behave as:
\beq
\xi^{\pm}\approx c_1\varrho^{l+1}\,+\,c_2\varrho^{-l}\,\,,\qquad\qquad 
(\rho\approx 0)\,\,,
\label{DpDp+2-smallrho}
\eeq
while for $\varrho\to\infty$ one gets:
\bear
&&\xi^{+}\approx d_1^+\varrho^{-(l+7-p)}\,+\,d_2^+\varrho^{l+1}\,\,,\rc\rc
&&\xi^{-}\approx d_1^-\varrho^{-l}\,+\,d_2^-\varrho^{l+p-6}\,\,,
\qquad\qquad (\varrho\to\infty)\,\,.
\label{DpDp+2-largerho}
\eear
Obviously, the regular solutions at $\varrho\to 0$ for any $l\ge 0$ are those which
behave as $\varrho^{l+1}$ for small $\varrho$, \ie\ those with $c_2=0$ in eq.
(\ref{DpDp+2-smallrho}). Moreover, we should also impose that $\xi^{\pm}$ vanish when 
$\varrho\to \infty$. From eq. (\ref{DpDp+2-largerho}) we notice that the behaviour of
the  $\xi^{+}$ modes is different from that of the $\xi^{-}$'s and, therefore, both
cases must be analyzed separately. Clearly,  only those  $\xi^{+}$ modes for which 
$d_2^+=0$ are normalizable. In analogy with the D3-D5 case studied in section \ref{D3D5},
let us call $I_+^l$ to  these modes. On the contrary, the two exponents of $\varrho$ in
the expression giving the $\varrho\to\infty$ behaviour of  $\xi^{-}$ in
(\ref{DpDp+2-largerho}) could be negative and, thus, we  have two types of modes.
Generalizing the case of the D3-D5 system, the modes with $d_2^-=0$ ($d_1^-=0$) will be
denoted by $I_-^l$ ($\tilde I_-^l$). Therefore, the different behaviours at $\varrho\to
\infty$ are:
\bear
&&I_+^l\qquad\Longrightarrow \qquad \xi^{+}\sim \varrho^{-(l+7-p)}\,\,,
\qquad (l\ge 0)\,\,,\rc\rc
&&I_-^l\qquad\Longrightarrow \qquad \xi^{-}\sim \varrho^{-l}\,\,,
\qquad\qquad \,\,\, (l\ge 1)\,\,,\rc\rc
&&\tilde I_-^l\qquad\Longrightarrow \qquad \xi^{-}\sim \varrho^{l+p-6}\,\,,
\qquad \,\,\,\,\,\,(1 \le l< 6-p)\,\,,
\label{DpDp+2-behaviours}
\eear
where, we have taken into account that for $l=0$ only the $\xi^{+}$ modes exist (see
section \ref{D3D5}). As in the D3-D5 intersection, notice that the $\tilde I_-^l$ modes
only exist for a finite number of values of the angular quantum number $l$.

The  behaviours  displayed in (\ref{DpDp+2-behaviours})
can be used to characterize the different types of modes in the numerical calculations
of the energy levels (see below). It is also interesting to study these levels in the 
WKB approximation. In particular it is interesting to find out how the WKB approximation
distinguishes between the two types of $\xi^{-}$ modes. The key observation in this
respect is to realize that within the WKB approach the solution of the wave equation
that is selected is that in which the ``wave function'' $\psi^{\pm}$ vanishes when we
move beyond the turning points of the potential. Using the fact that 
$\psi^{\pm}\approx \varrho^{3-{p\over 2}}\, \xi^{\pm}$ for large $\varrho$, one can check
immediately that for the $I_+^l$ modes 
$\psi^{+}\approx \varrho^ {-4-l+{p\over 2}}$, which always vanish for
$\varrho\to\infty$. On the contrary, for the $I_-^l$ ($\tilde I_-^l$) modes 
$\psi^{-}\approx \varrho^{3-l-{p\over 2}}$ 
($\psi^{-}\approx \varrho^{-3+l+{p\over 2}}$), which means that for $l\ge 3-{p\over 2}$
the WKB approximation picks up the $I_-^l$ branch whereas for 
$l\le 3-{p\over 2}$ the $\tilde I_-^l$ modes are selected\footnote{Notice that for 
$l= 3-{p\over 2}$ the two types of modes behave in the same way for $\varrho\to\infty$
and there is no real distinction between them. This case  can only happen for even
$p$.}.  The sign of $l-3+{p\over 2}$ is relevant when one computes   the quantity
$\beta_2$, defined in (\ref{alphabeta2}),  which in turn is needed to apply  the WKB
energy formula (\ref{generalWKBlevels}). Applying these ideas to the case at hand, 
the WKB energy levels for the $I_\pm$ modes are given by:
\beq
M^{WKB}_{I_\pm}(n,l)\,=\,
2\sqrt{\pi}\,\,{L^{{5-p\over 2}}\over R^{{7-p\over 2}}}\,\,
{\Gamma\Big({7-p\over 4}\Big)\over \Gamma\Big({5-p\over 4}\Big)}\,\,
\sqrt{\bigg(n+1\bigg)\,\bigg(n\,+\,{7-p\over 5-p}\,(l+ {1\over 2}\pm 1)\,\bigg)}\,\,,
\label{WKBIpm}
\eeq
whereas the WKB energy levels for the $\tilde I_-$ modes are:
\beq
M^{WKB}_{\tilde I_-}(n,l)\,=\,
2\sqrt{\pi}\,\,{L^{{5-p\over 2}}\over R^{{7-p\over 2}}}\,\,
{\Gamma\Big({7-p\over 4}\Big)\over \Gamma\Big({5-p\over 4}\Big)}\,\,
\sqrt{\bigg(n+1\bigg)\,\bigg(n\,+\,{7-p\over 5-p}\,
+\,{3-p\over 5-p}\,\Big(\,l+ {1\over 2}\,\Big)\,\bigg)}\,\,.
\label{WKB-tildeI-}
\eeq
Let us use the WKB mass formulae (\ref{WKBIpm}) and (\ref{WKB-tildeI-}) to notice some
regularities, which  can be subsequently checked with the numerical calculations. First
of all, by direct comparison of (\ref{DpDp+4-WKBscalarmass}) and (\ref{WKBIpm}) one gets
that the masses of the  $I_\pm$ modes are related to those of the scalar fluctuations as:
\beq
M^{WKB}_{I_\pm}(n,l)\,=\,M^{WKB}_{S}(n,l\pm 1)\,\,.
\label{WKBrelations}
\eeq
We  have also verified that this relation holds  numerically. In the case of
the D3-D5 intersection the analogue of  (\ref{WKBrelations}), namely  
(\ref{massrelations}), was crucial to organize the different fluctuations in massive
supermultiplets. Notice also that the WKB formula (\ref{WKB-tildeI-}) for the masses of
the $\tilde I_-$ modes gives a result which is independent of $l$ only for $p=3$ and,
actually, it coincides with the exact result (\ref{Itildelevels}) in the case of the
D3-D5 system. In  table \ref{MassDpDp+2-I} we list the mass levels obtained
numerically for the $I_+$ fluctuations for $l=0$ and for the $\tilde I_-$ modes for
$l=1$. As mentioned above  the levels for $I_-^{l=1}$ are equal, within the accuracy of
our numerical calculation, to the ones of the $S^{l=0}$ fluctuations, which were given
at the beginning of this section.

\bigskip
\begin{table}[!h]
\qquad\qquad\qquad
\begin{tabular}[b]{|c|c|c|}   
\hline  
\multicolumn{3}{|c|}{$(0|D1\perp D3)$}\\
\hline
\rule{0mm}{5.0mm} $n$  & $I_+^{l=0}$  & $\tilde I_-^{l=1}$ \\ 
\hline   
\ \ 0 & $27.06$  & $22.04$  \\   
\ \ 1 & $69.41$  & $63.75$  \\   
\ \ 2 & $131.38$  & $125.26$  \\   
\ \ 3 & $213.01$  & $206.55$  \\     
\ \ 4 & $314.36$  & $307.60$  \\
\ \ 5 & $435.41$  & $428.41$  \\ 
\hline
\end{tabular}
\qquad\qquad\qquad
\begin{tabular}[b]{|c|c|c|}   
\hline  
\multicolumn{3}{|c|}{$(1|D2\perp D4)$ }\\
\hline
\rule{0mm}{5.0mm} $n$  & $I_+^{l=0}$  & $\tilde I_-^{l=1}$ \\ 
\hline   
\ \ 0 & $21.04$  & $15.00$  \\   
\ \ 1 & $52.13$  & $43.63$  \\   
\ \ 2 & $96.92$  & $86.02$  \\   
\ \ 3 & $155.43$  & $142.17$  \\     
\ \ 4 & $227.68$  & $212.06$  \\
\ \ 5 & $313.67$  & $259.71$  \\ 
\hline
\end{tabular}
\qquad
\caption{$\bar M^2$ for the $I_+^{l=0}$ and $\tilde I_-^{l=1}$ modes of the
D1-D3 intersection (left) and D2-D4 system (right).}
\label{MassDpDp+2-I}
\end{table}

\subsection{Type II modes}
As a generalization of the case of the D3-D5 intersection, let us consider a
configuration in which the scalar field $X$ vanishes and the vector field has only
non-vanishing components along the Minkowski directions $x^{\mu}$, which are given by the
ansatz of eqs. (\ref{D3D5II}) and (\ref{D3D5IIbis}). It is immediate to verify that the
equations for the gauge field components $A_{\rho}$ and $A_{i}$ are identically
satisfied, whereas the equation of motion for $A_{\mu}$, in terms of the reduced
variables $\varrho=\rho/L$ and $\bar M^2\,=\,R^{7-p}\,L^{p-5}\,M^2$,  is equivalent to:
\beq
\partial_{\varrho}\big(\,\varrho^2\,\partial_{\varrho} \chi\,\big)\,+\,
\Bigg[\,\bar M^2\,\,{\varrho^2\over \big(\,\varrho^2+1\,\big)^{{7-p\over 2}}}\,-\,
l(l+1)\,\Bigg]\,\chi\,=\,0\,\,.
\label{DpDp+2II}
\eeq
Eq. (\ref{DpDp+2II}) is exactly the same as the one corresponding to the scalar
fluctuations. Therefore we conclude that, as it happened for the D3-D5 system,  the
spectrum of type II vector modes is identical to the one corresponding to the scalar
modes.

\subsection{Type III modes}
Let us consider the type III modes with the same ansatz (\ref{D3D5IIIansatz})  as in the
D3-D5 system. The equation for $A_{\rho}$ is now:
\beq
\rho^2\,R^{7-p}\partial_{\mu}\partial^{\mu}\phi\,-\,
l(l+1)\,\Big[\rho^2+L^2\Big]^{{7-p\over 2}}\,
(\,\phi\,-\,\partial_{\rho}\tilde\phi\,)\,=\,0\,\,.
\eeq
The equation for $A_i$ is:
\beq
R^{7-p}\partial_{\mu}\partial^{\mu}\tilde\phi\,+\,
\partial_{\rho}\,\Big[\,(\rho^2+L^2)^{{7-p\over 2}}\,
(\partial_{\rho}\tilde\phi\,-\,\phi)\,\Big]\,=\,0\,\,,
\eeq
and the equation for $A_{\mu}$ is exactly the same as in the D3-D5 intersection, namely 
(\ref{D3D5IIIAmu}).
Using these relations, the two equations written above are equivalent. Separating
variables in
$\phi$ as in the D3-D5 case, we obtain the following equation:
\beq
{1\over \varrho^2}\,\partial^2_{\varrho}\,\Big(\varrho^2\zeta\Big)\,+\,
\Bigg[\,{\bar M^2\over (1+\varrho^2)^{{7-p\over 2}}}
\,-\,{l(l+1)\over\varrho^2}\,\Bigg]\zeta\,=\,0\,\,.
\label{DpDp+2typeIII}
\eeq
In order to obtain the spectrum derived from this equation, let us perform a change of
variables to convert (\ref{DpDp+2typeIII}) into the Schr\"odinger equation (\ref{Sch}).
This change of variables is the following:
\beq
e^y\,=\,\varrho\,\,,\qquad\qquad
\psi\,=\,\varrho^{{3\over 2}}\,\,\zeta\,\,.
\eeq
The potential $V$ of  (\ref{Sch}) after this change of variables is:
\beq
V\,=\,\Big(\,l+{1\over 2}\,\Big)^{2}\,-\,\bar M^2\,\,{e^{2y}\over 
(1\,+\,e^{2y})^{{7-p\over 2}}}\,\,,
\eeq
which is just the same as the one corresponding to the scalar excitations.  Therefore,
we conclude that the mass levels of these  modes of  type III are the same as the
ones corresponding to the scalar fluctuations. Putting together the results corresponding
to the modes of types II and III, we conclude that the vector fluctuations have a mass
spectrum which coincides with that of the scalar modes, namely:
\beq
M_{V}(n,l)\,=\,M_{S}(n,l)\,\,,
\eeq
a result which generalizes the one for the D3-D5 intersection.

\vskip 1cm
\renewcommand{\theequation}{\rm{C}.\arabic{equation}}
\setcounter{equation}{0}
\medskip

\section{Fluctuations of the Dp-Dp system}
\medskip
The lagrangian density of a Dp-brane probe in the background generated by  a stack of
Dp-branes is the sum of the Born-Infeld and Wess-Zumino terms, which are given by:
\beq
{\cal L}_{BI}\,=\,-e^{-\phi}\,\sqrt{-\det(\,g+F\,)}\,\,,\qquad\qquad
{\cal L}_{WZ}\,=\,P\big[\,C^{(p+1)}\,\big]\,\,.
\label{LDpDp}
\eeq
In this section we will analyze the $(p-2|Dp\perp Dp)$ intersection for $2\le p< 5$. The
induced metric on the worldvolume of the probe for a static configuration of such
intersection in which the branes are separated a constant distance $L$ is given by:
\beq
{\cal G}_{ab}d\xi^a d\xi^b\,=\,
\Bigg[{\rho^2+L^2\over R^2}\Bigg]^{{7-p\over 4}}\,
dx^2_{1,p-2}\,+\,
\Bigg[{R^2\over \rho^2+L^2}\Bigg]^{{7-p\over 4}}\Big[\,
d\rho^2\,+\,\rho^2\,d\Omega_1^2\,\Big]\,\,.
\eeq
Let us now study the quadratic fluctuations around the static configuration. As in
previous sections, let us denote by $\chi$  the scalars transverse to both types of
branes. Moreover, we will assume that the Dp-brane probe intersects the Dp-branes
of the background along  $x^0\cdots x^{p-2}$ and we will denote the remaining Minkowski
coordinates as:
\beq
\lambda_1=x^{p-1}\,\,,\qquad\qquad
\lambda_2=x^p\,\,.
\eeq 
The lagrangian for the quadratic fluctuations  can be obtained straightforwardly by
expanding  (\ref{LDpDp}). Indeed, the contribution from the Born-Infeld lagrangian is:
\beq
{\cal L}_{BI}\,=\,-\rho\,\Bigg[\,{1\over 2}\,
\Bigg[{R^2\over \rho^2+L^2}\Bigg]^{{7-p\over 4}}{\cal G}^{ab}\,
\partial_{a}\chi\partial_{b}\chi\,+\,
{1\over 2}\Bigg[{\rho^2+L^2\over R^2}\Bigg]^{{7-p\over 4}}\,\,{\cal G}^{ab}
\partial_{a}\lambda_i\partial_{b}\lambda_i\,+\,{1\over 4}F_{ab}F^{ab}
\,\,\Bigg]\,\,,
\eeq
while the  Wess-Zumino term  becomes:
\beq
{\cal L}_{WZ}\,=\,\Bigg[{\rho^2+L^2\over R^2}\Bigg]^{{7-p\over 2}}
\epsilon^{ij}\partial_{\rho} \lambda_i\,\partial_{\varphi} \lambda_j\,\,.
\eeq

The analysis of the $\chi$ fluctuations reduces to the general case of section 
\ref{general} (see eq. (\ref{fluc})) with $\gamma_1$ and $\gamma_2$ as in eq.
(\ref{Dpgammas}), $p_2=p$ and $d=p-2$.  Notice that in this case the equation for 
the fluctuations  depends on $l^2$. As argued in the analysis of the 
$(1|D3\perp D3)$ intersection, for this real field $\chi$ we can restrict ourselves to
the case $l\ge 0$. By using the asymptotic limits of the equivalent Schr\"odinger
potential  (eq. (\ref{generalV-limit})) one readily gets that 
$\chi\sim \varrho ^{\pm l}$ both for $\varrho\to\infty$ and $\varrho\to 0$. The $S^l$
modes in this case will be defined as those modes which behave as $\varrho^l$ for
$\varrho\approx 0$ and as $\varrho^{-l}$ for $\varrho\to\infty$. 
The WKB mass formula for these modes can be obtained from
(\ref{WKBM}), namely:
\beq
M^{WKB}_S(n,l)\,=\,2\sqrt{\pi}\,{L^{{5-p\over 2}}\over R^{{7-p\over 2}}}\,\,
{\Gamma\Big({7-p\over 4}\Big)\over \Gamma\Big({5-p\over 4}\Big)}\,\,
\sqrt{(n+1)\,\Big(n\,+\,{7-p\over 5-p}\,l\,\Big)}\,\,.
\eeq
Notice that this result coincides with the exact one  (\ref{MSd3d3}) for $p=3$ and $l=1$.
Moreover, the quadratic and linear terms in $n$  for $M^2$ are also reproduced for $p=3$
and arbitrary
$l$. 
In table \ref{MassDpDp} we give the numerical results and the WKB estimates for
$\bar M^2$  for the intersections of the type $(p-2|Dp\perp Dp)$ for $p=2,4$ and $l=1$.

\begin{table}[!h]
\qquad\qquad
\begin{tabular}[b]{|c|c|c|}   
\hline  
\multicolumn{3}{|c|}{$(0|D2\perp D2)$ with $l=1$}\\
\hline
 $n$  & WKB  & Numerical \\ 
\hline   
\ \ 0 & $11.46$  & $11.34$  \\   
\ \ 1 & $36.67$  & $36.54$  \\   
\ \ 2 & $75.63$  & $75.50$  \\   
\ \ 3 & $128.34$  & $128.20$  \\     
\ \ 4 & $194.80$  & $194.66$  \\
\ \ 5 & $275.01$  & $274.88$  \\ 
\hline
\end{tabular}
\qquad\qquad\qquad
\begin{tabular}[b]{|c|c|c|}   
\hline  
\multicolumn{3}{|c|}{$(2|D4\perp D4)$ with $l=1$}\\
\hline
 $n$  & WKB  & Numerical \\ 
\hline   
\ \ 0 & $4.31$  & $4.68$  \\   
\ \ 1 & $11.48$  & $11.88$  \\   
\ \ 2 & $21.53$  & $21.94$  \\   
\ \ 3 & $34.45$  & $34.86$  \\     
\ \ 4 & $50.24$  & $50.66$  \\
\ \ 5 & $68.91$  & $69.34$  \\ 
\hline
\end{tabular}
\caption{Numerical and WKB values for the $S^l$ modes of the D2-D2 and D4-D4
intersections for $l=1$.}
\label{MassDpDp}
\end{table}

\subsection{Scalar fluctuations}

The equations of motion of the $\lambda_i$'s derived from 
${\cal L}_{BI}+{\cal L}_{WZ}$ are:
\beq
R^{{7-p\over 2}}\,\,\partial_a\,\Bigg[\,\rho\,(\rho^2+L^2)^{{7-p\over 4}}\,
{\cal G}^{ab}\partial_b \lambda_i\,\Bigg]\,-\,(7-p)
\rho\,
( \rho^2+L^2)^{{5-p\over 2}}\,\epsilon^{ij}
\partial_{\varphi} \lambda_j\,=\,0\,\,.
\eeq
Let us separate variables and define the complex combination $w$ as in the
D3-D3 case (eqs. (\ref{D3D3separation}) and (\ref{D3D3w-def})). The decoupled equations
become:
\beq
\partial_{\varrho}\Bigg[\,\varrho\,(\varrho^2+1)^{{7-p\over 2}}\,
\partial_{\varrho}\,w\Bigg]\,+\,
\Bigg[\,\bar M^2\varrho\,-\,l^2\,{(\varrho^2+1)^{{7-p\over 2}}\over \varrho}+
(7-p)l\,\varrho (\varrho^2+1)^{{5-p\over 2}}\,\Bigg]w\,=\,0\,\,.
\label{DpDp-w-eq}
\eeq
In order to transform this equation into the Schr\"odinger equation (\ref{Sch}), let us
change variables as:
\beq
e^{y}\,=\,\varrho\,\,,\qquad\qquad
\psi\,=\,\Big[1+\varrho^2\Big]^{{7-p\over 4}}
\,\,w\,\,.
\eeq
The potential in this case becomes:
\bear
&&V(y)=l^2+(7-p) \bigg(1-l\bigg)
{e^{2y}\over 1+e^{2y}}+\rc\rc
&&\qquad\qquad+{1\over 4}\,(7-p) (3-p) \,
{e^{4y}\over (1+e^{2y})^2}-\bar M^2
{e^{2y}\over ( 1+e^{2y})^{{7-p\over 2}}}\,\,.
\eear
The asymptotic values of $V$ can be readily computed, with the result:
\beq
\lim_{y\to -\infty} V(y)\,=\,l^2\,\,,
\qquad\qquad
\lim_{y\to +\infty} V(y)\,=\,
\bigg(l-{7-p\over 2}\bigg)^2\,\,.
\eeq
From these values and the relation between $\psi$ and $w$ we obtain that 
for $\varrho\approx 0$:
\beq
w\approx c_1\varrho^{l}\,+\,c_2\varrho^{-l}\,\,,\qquad\qquad 
(\varrho\approx 0)\,\,,
\eeq
whereas for large $\rho$ we have:
\beq
w\approx d_1\varrho^{-l}\,+\,d_2\varrho^{l+p-7}\,\,,\rc\rc
\qquad\qquad (\varrho\to\infty)\,\,.
\eeq
Following the steps of the analysis of the scalar fluctuations in the D3-D3 system, 
the different modes are characterized by their behaviour at $\varrho\to\infty$, as
follows: 
\bear
&&W_+^l\qquad\Longrightarrow \qquad w\sim \varrho^{-l}\,\,,
\rc\rc
&&W_-^l\qquad\Longrightarrow \qquad w\sim \varrho^{l+p-7}\,\,.
\label{DpDp-behaviours}
\eear
In addition the modes should not blow up at $\varrho=0$, \ie\  they should behave
as $\varrho^{|l|}$ near $\varrho=0$. Interestingly, the WKB approximation selects the 
$W_+^l$ or $W_-^l$ modes, depending on the value of $l$. Indeed, for large $\varrho$
the functions $w$ and $\psi$ are related as
$w\sim \varrho^{{p-7\over 2}}\psi$ and, within the WKB approach, the wave function
$\psi$ vanishes as $\psi\sim \varrho^{-|l-{7-p\over 2}|}$ when $\varrho\to\infty$. It
follows that $w$ behaves as in the $W_+^l$ branch for $l\ge {7-p\over 2}$ and as in the 
$W_-^l$ modes for $l\le {7-p\over 2}$. To compute the actual values of the WKB energy
levels we need to evaluate the coefficients $\alpha_i$ and $\beta_i$ defined in eqs. 
(\ref{WKBalphabetauno}) and (\ref{alphabeta2}). In the present case these coefficients
are:
\beq
\alpha_1\,=\,2\,\,,\qquad
\alpha_2\,=\,2|\,l\,|\,\,,\qquad
\beta_1\,=\,5-p\,\,,\qquad
\beta_2\,=\,|\,2l-7+p\,|\,\,.
\label{DpDpalphabeta}
\eeq
The appearance of the absolute value on the expressions of $\alpha_2$ and $\beta_2$
in  (\ref{DpDpalphabeta}) forces us to consider different ranges of $l$. 
For $l\ge {7-p\over 2}$ the WKB method selects the $W_+^l$ branch and the corresponding
energy levels are:
\beq
M^{WKB}_{W_+}(n,l)\,=\,2\sqrt{\pi}\,{L^{{5-p\over 2}}\over R^{{7-p\over 2}}}\,\,
{\Gamma\Big({7-p\over 4}\Big)\over \Gamma\Big({5-p\over 4}\Big)}\,\,
\sqrt{(n+1)\,\Big(n\,+\,{7-p\over 5-p}\,(l-1)\,\Big)}\,\,,
\qquad \bigg(\,l\ge {7-p\over 2}\,\bigg)\,\,.
\label{DpDpWKBW_+}
\eeq
This mass spectrum is related to the one corresponding to the transverse scalar
excitations as follows:
\beq
M^{WKB}_{W_+}(n,l)\,=\,M^{WKB}_{S}(\,n,l-1\,)\,\,,\qquad\qquad
\bigg(\,l\ge {7-p\over 2}\,\bigg)\,\,.
\label{DpDpSW+}
\eeq
When $l\le 0$ the $W_-$ branch is selected by the WKB approach and one gets the following
mass formula:
\beq
M^{WKB}_{W_-}(n,l)\,=\,2\sqrt{\pi}\,{L^{{5-p\over 2}}\over R^{{7-p\over 2}}}\,\,
{\Gamma\Big({7-p\over 4}\Big)\over \Gamma\Big({5-p\over 4}\Big)}\,\,
\sqrt{(n+1)\,\Big(n\,+\,{7-p\over 5-p}\,(1-l)\,\Big)}\,\,,
\qquad \bigg(\,l\le 0\,\bigg)\,\,,
\eeq
which is related to the spectrum of the $S^l$ modes as:
\beq
M^{WKB}_{W_-}(n,l)\,=\,M^{WKB}_{S}(\,n,1-l\,)\,\,,\qquad\qquad
\bigg(\,l\le 0\,\bigg)\,\,.
\label{DpDpSW-}
\eeq
We have checked that the relations (\ref{DpDpSW+}) and (\ref{DpDpSW-}) are well
satisfied by the masses computed by solving numerically the corresponding fluctuation
equations.

For the range $0\le l\le {7-p\over 2}$ the WKB method picks up the $W_-$ branch and
one gets:
\beq
M^{WKB}_{W_-}(n,l)\,=\,2\sqrt{\pi}\,{L^{{5-p\over 2}}\over R^{{7-p\over 2}}}\,\,
{\Gamma\Big({7-p\over 4}\Big)\over \Gamma\Big({5-p\over 4}\Big)}\,\,
\sqrt{(n+1)\,\Big(n+2l+{7-p\over 5-p}\,(1-l)\,\Big)}\,\,,
\qquad \bigg(\,0\le l\le {7-p\over 2}\,\bigg)\,\,.
\label{DpDpW-intermediate}
\eeq
We have checked that (\ref{DpDpW-intermediate}) represents fairly well the results
obtained numerically.

Notice that the expression (\ref{DpDpWKBW_+}) of $ M^{WKB}_{W_+}$ vanishes for $n=0$,
$l=1$. Actually, when the mass
$M$ vanishes, the equation (\ref{DpDp-w-eq}) for the fluctuations can be written as:
\beq
\varrho^2(1+\varrho^2)\,\partial^2_{\varrho}w\,+\,
\big[1+(8-p)\varrho^2\,\big]\,\varrho\,\partial_{\varrho}w\,
-\,l\big[l-(7-p-l)\varrho^2\,\big]\,w\,=\,0\,\,.
\eeq
This equation can be mapped into the hypergeometric equation. Actually, one of its
solutions is just $w=\rho^{-l}$. The second solution can be written in terms of the
hypergeometric function as follows:
\beq
w\,=\,\rho^l\,F(l, {7-p\over 2};1+l;-\rho^2)\,\,.
\eeq
When $p=3$ and $l=1$ this solution reduces to the one written in (\ref{higgsbranch}),
which represents the holomorphic map of the Higgs branch of the D3-D3 intersection.

\subsection{Vector fluctuations}
Let us analyze the configurations of the gauge field which are given by the same
ansatzs as in the D3-D3 intersection. First of all, we consider  a configuration
as the one displayed in eq. (\ref{D3D3typeII}), in which the only non-vanishing
components of the gauge field are $A_{\mu}$, which depends on a function $\phi(\rho)$. 
After a short calculation one can verify that the equations of motion of the gauge field
reduce to the following equation for $\phi$:
\beq
\partial_{\varrho}\,(\varrho\,\partial_{\varrho}\,\phi)\,+\,\Bigg[\,
\bar M^2\,{\varrho\over (1+\varrho^2)^{{7-p\over 2}}}\,-\,
{l^2\over \varrho}\,\Bigg]\,\phi\,=\,0\,\,,
\label{dpdpfluct}
\eeq
where we have already used the reduced variables $\varrho$ and $\bar M$ defined in eq.
(\ref{DpDp+4-newvariables}). Eq.  (\ref{dpdpfluct}) is just the same as the one
satisfied by the transverse scalars and, therefore, the corresponding spectra are
identical. Let us next consider an ansatz as in  eq. (\ref{D3D3typeIII}), which depends
on two functions $\phi$ and
$\tilde\phi$.  The equation for $A_{\rho}$ in this case becomes:
\beq
il\partial_{\rho}\tilde\phi\,-\,l^2\phi\,+\,
M^2 R^{7-p}{\rho^2\over (\rho^2+L^2)^{{7-p\over 2}}}\,\phi\,=\,0\,\,.
\eeq
Moreover, the equation for $A_\varphi$ is:
\beq
\partial_{\rho}\Bigg[\,{(\rho^2+L^2)^{{7-p\over 2}}\over \rho}\,
\Big(\,\partial_{\rho}\tilde\phi\,+\,il\phi\,\Big)\,\Bigg]\,+\,
{M^2R^{7-p}\over \rho}\,\tilde\phi\,=\,0\,\,.
\eeq
The equation for $A_{\mu}$ gives a relation between $\phi$ and $\tilde\phi$, which is the
same as in the D3-D3 intersection, namely eq. (\ref{D3D3phi-tildephi}). By using this
relation the remaining equations reduce to the following equation for $\phi$:
\beq
\partial_{\rho}\,\Big[\,\rho\,\partial_{\rho}(\,\rho \phi\,)\,\Big]\,+\,
\Bigg[\,{\rho^2\over (\rho^2+L^2)^{{7-p\over 2}}}\,M^2\,-\,l^2\,\Bigg]\,\phi\,=\,0\,\,.
\eeq
This equation becomes again just the same as the one corresponding to the transverse
scalars if we redefine the fluctuation as $\hat\phi=\rho\phi$. Then, the spectrum of
these modes coincides again with that of the transverse scalars. The conclusion we
arrive at is that, also in this case, the scalar and vector modes are degenerate in
mass, \ie:
\beq
M_V(n,l)\,=\,M_S(n,l)\,\,.
\eeq

\renewcommand{\theequation}{\rm{D}.\arabic{equation}}
\setcounter{equation}{0}
\medskip

\section{Fluctuations of the Dp-D(p+4) system}
\medskip
Let us consider a D(p+4)-brane  probe embedded in the background created by a Dp-brane in
such a way that the probe fills the (p+1)-dimensional worldvolume of the background
brane for $1\le p <5$. If $L$ denotes the distance between both types of branes, the
induced metric  on the D(p+4) worldvolume is:
\beq
{\cal G}_{ab}d\xi^a d\xi^b\,=\,
\Bigg[{\rho^2+L^2\over R^2}\Bigg]^{{7-p\over 4}}\,\
dx^2_{1,p}\,+\,
\Bigg[{R^2\over \rho^2+L^2}\Bigg]^{{7-p\over 4}}\Big[\,
d\rho^2\,+\,\rho^2\,d\Omega_3^2\,\Big]\,.
\label{indp4}
\eeq
The dynamics of the probe is governed by a lagrangian density which is the sum of the
Born-Infeld term ${\cal L}_{BI}$ and the Wess-Zumino term ${\cal L}_{WZ}$. The
expression of ${\cal L}_{BI}$ is just the same as the one appearing in 
(\ref{LDpDp+2}) and (\ref{LDpDp}). By expanding it around the static configuration, one
gets the following expression:
\beq
{\cal L}_{BI}\,=\,-\rho^3\,\sqrt{\tilde g}\,\Bigg[\,{1\over 2}\,
\Bigg[{R^2\over \rho^2+L^2}\Bigg]^{{7-p\over 4}}{\cal G}^{ab}\,
\partial_{a}\chi\partial_{b}\chi\,+\,{1\over 4}F_{ab}F^{ab}
\,\,\Bigg]\,,
\label{bip4}
\eeq
where $\chi$ are the fluctuations of the scalars transverse to the probe and $F_{ab}$
is the strength of the wordlvolume gauge field. The Wess-Zumino lagrangian in this case
is:
\beq
{\cal L}_{WZ}\,=\,{1\over2}\,P\big[\,C^{(p+1)}\,\big]\,\wedge F\,\wedge F\,.
\label{wzgnp4}
\eeq
Using the expression of the Ramond-Ramond  potential $C^{(p+1)}$  given in eq.
(\ref{CRR}) and  dropping terms that do not contribute to the equations of motion we get:
\beq
{\cal L}_{WZ}\,=\,\Bigg[{\rho^2+L^2\over R^2}\Bigg]^{{7-p\over 2}}\,
\epsilon_{ijk}\,\partial_\rho\,A_i\,\partial_j\,A_k\,.
\label{wzp4}
\eeq
\subsection{Scalar fluctuations}
The equations of motion for the scalar fields $\chi$ are a particular case of those
studied in section \ref{general}. Indeed, in this case one must take $\gamma_1$ and
$\gamma_2$ as given in eq. (\ref{Dpgammas}), $p_2=p+4$ and $d=p$. Using these values in
the WKB estimate (\ref{WKBM}), we arrive at the expression:
 \beq
M^{WKB}_S(n,l)\,=\,2\sqrt{\pi}\,{L^{{5-p\over 2}}\over R^{{7-p\over 2}}}\,\,
{\Gamma\Big({7-p\over 4}\Big)\over \Gamma\Big({5-p\over 4}\Big)}\,\,
\sqrt{(n+1)\,\Big(n\,+\,{7-p\over 5-p}\,(l+1)\,\Big)}\,\,.
\label{MSDpDp+4}
\eeq

In   table  \ref{MassDpDp+4} we compare 
the numerical values of the masses for the scalar fluctuations with the corresponding
WKB estimates.

\begin{table}[!h]
\begin{tabular}[b]{|c|c|c|}   
\hline  
\multicolumn{3}{|c|}{$(1|D1\perp D5)$ with $l=0$}\\
\hline
 $n$  & WKB  & Numerical \\ 
\hline   
\ \ 0 & $14.80$  & $14.70$  \\   
\ \ 1 & $49.35$  & $49.22$  \\   
\ \ 2 & $103.63$  & $103.50$  \\   
\ \ 3 & $177.65$  & $177.54$  \\     
\ \ 4 & $271.41$  & $271.30$  \\
\ \ 5 & $384.91$  & $384.80$  \\ 
\hline
\end{tabular}
\qquad
\begin{tabular}[b]{|c|c|c|}   
\hline  
\multicolumn{3}{|c|}{$(2|D2\perp D6)$ with $l=0$}\\
\hline
 $n$  & WKB  & Numerical \\ 
\hline   
\ \ 0 & $11.46$  & $11.34$  \\   
\ \ 1 & $36.67$  & $36.54$  \\   
\ \ 2 & $75.63$  & $75.50$  \\   
\ \ 3 & $128.34$  & $128.20$  \\     
\ \ 4 & $194.80$  & $194.66$  \\
\ \ 5 & $275.01$  & $274.88$  \\ 
\hline
\end{tabular}
\qquad
\begin{tabular}[b]{|c|c|c|}   
\hline  
\multicolumn{3}{|c|}{$(4|D4\perp D8)$ with $l=0$}\\
\hline
 $n$  & WKB  & Numerical \\ 
\hline   
\ \ 0 & $4.31$  & $4.68$  \\   
\ \ 1 & $11.48$  & $11.88$  \\   
\ \ 2 & $21.53$  & $21.94$  \\   
\ \ 3 & $34.45$  & $34.86$  \\     
\ \ 4 & $50.24$  & $50.66$  \\
\ \ 5 & $68.91$  & $69.34$  \\ 
\hline
\end{tabular}
\caption{Values of $\bar M^2$ obtained numerically and with the WKB method for the
scalar modes of the D1-D5, D2-D6 and D4-D8 intersections for l=0.}
\label{MassDpDp+4}
\end{table}

\subsection{Vector fluctuations}

The equation of motion of the gauge field derived from ${\cal L}_{BI}+{\cal L}_{WZ}$ of
eqs. (\ref{bip4}) and (\ref{wzp4}) takes the form:
\beq
\partial_a\,\Big[\,\rho^3\sqrt{\tilde g}\,F^{ab}\,\Big]\,-\,(7-p)\,
{\rho\over R^{7-p}}\,
\Big[\rho^2+L^2\Big]^{{5-p\over 2}}\,
\epsilon_{bjk}\,\partial_jA_k\,=\,0\,\,,
\label{vectoreomdp4}
\eeq
where $\epsilon_{bjk}$ is non-vanishing only when $b$ is an index along the $S^3$. 

Let us study the solutions of this equation. Following \cite{KMMW}, we can expand $A_\mu$
and $A_\rho$ in (scalar) spherical harmonics on the $S^3$, and $A_i$ in vector spherical
harmonics. We can construct three classes of vector spherical harmonics on the $S^3$.
One is simply given by $Y^l_i\equiv\nabla_iY^l$, and the other two, denoted by
$Y_i^{l,\pm}$ for 
$l\ge 1$, transform in the $({l\mp 1\over 2}, {l\pm 1\over 2})$ representation of 
$SO(4)$ and satisfy:
\bear
&&\nabla^i\,\nabla_i\,Y_j^{l,\pm}-R^k_j\,Y_k^{l,\pm}=-(l+1)^2\,Y_j^{l,\pm}\,,\rc
&&\epsilon_{ijk}\,\nabla_j\,Y_k^{l,\pm}=\pm(l+1)\,Y_i^{l,\pm}\,\,,\rc
&&\nabla^i\,Y_i^{l,\pm}=0\,,
\label{harmp4}
\eear
where $R^i_j=2\delta^i_j$ is the Ricci tensor of a three-sphere of unit radius.

\subsubsection{Type I modes}
As argued in \cite{KMMW}, the modes containing $Y_i^{l,\pm}$ do not mix with the others
due to the fact that they are in different representations  of $SO(4)$. Accordingly, 
let us take the ansatz:
\beq
A_{\mu}=0\,\,,\qquad\qquad
A_{\rho}=0\,\,,\qquad\qquad
A_i\,=\,\Lambda^{\pm}(x,\rho)\,Y_i^{l,\pm}(S^3)\,.
\label{typeIAp4}
\eeq
The equation of motion (\ref{vectoreomdp4}) reduces to the following equation for
$\Lambda^{\pm}(x,\rho)$:
\bear
&&\rho\,R^{7-p}\,\partial_{\mu}\partial^{\mu}\Lambda^{\pm}\,+\,
\partial_{\rho}\Big[\,\rho
\,(\,\rho^2\,+\,L^2\,)^{7-p\over2}\,\partial_{\rho}\,\Lambda^{\pm}\,\Big]\,-\,
(l+1)^2\,{(\,\rho^2\,+\,L^2\,)^{7-p\over2}\over
\rho}\,\Lambda^{\pm}\mp\rc\rc
&&\qquad\qquad\mp\,(7-p)\,(l+1)\,\rho\,(\,\rho^2\,+\,L^2\,)^{5-p\over2}
\Lambda^\pm\,=\,0\,.
\label{typeIeomp4}
\eear
Let  us separate variables in (\ref{typeIeomp4}) as in previous 
cases, namely:
\beq
\Lambda^{\pm}(x,\rho)\,=\,e^{ikx}\, \xi^{\pm}(\rho)\,.
\label{typeIchvbp4}
\eeq
Moreover, we shall introduce the reduced quantities
$\varrho$ and  $\bar M$, defined as in (\ref{DpDp+4-newvariables}). 
By changing variables as:
\beq
e^{y}\,=\,\varrho\,\,,\qquad\qquad
\psi^{\pm}\,=\,\Big[1+\varrho^2\Big]^{{7-p\over 4}}
\,\, \xi^{\pm}\,\,,
\label{typeIschvp4}
\eeq
we can convert the fluctuation equation (\ref{typeIeomp4}) into a Schr\"odinger equation
for
$\psi^{\pm}$, with the potential $V_{\pm}$  given by:
\bear
&&V_{\pm}(y)=\bigg(l+1\bigg)^2\pm (7-p) \bigg(l+1\pm 1\bigg)
{e^{2y}\over 1+e^{2y}}+\rc\rc
&&\qquad\qquad+{1\over 4}\,(7-p) (3-p) \,
{e^{4y}\over (1+e^{2y})^2}-\bar M^2
{e^{2y}\over ( 1+e^{2y})^{{7-p\over 2}}}\,.
\label{typeIschpp4}
\eear
By looking at the asymptotic value of the potential $V_{\pm}$ at 
$y\to\pm\infty$ we
can get the behaviour of the fluctuations $\xi^{\pm}$ at 
$\varrho\approx 0, \infty$.
Indeed, from the above potentials we obtain:
\beq
\lim_{y\to -\infty} V_{\pm}(y)\,=\,\bigg(l+1\bigg)^2\,\,,
\qquad\qquad
\lim_{y\to +\infty} V_{\pm}(y)\,=\,
\bigg(l+1\pm {7-p\over 2}\bigg)^2\,.
\label{typeIschplmp4}
\eeq
 From these values one can prove that  for  $\varrho\approx 0$:
\beq
\xi^{\pm}\approx c_1\varrho^{l+1}\,+\,c_2\varrho^{-(l+1)}\,\,,\qquad\qquad
(\varrho\approx 0)\,\,,
\label{typeItozrp4}
\eeq
whereas for $\varrho\to\infty$ one gets:
\bear
&& \xi^{+}\approx d_1^+\varrho^{-(l+8-p)}\,+\,d_2^+\varrho^{l+1}\,\,,\rc\rc
&& \xi^{-}\approx d_1^-\varrho^{-(l+1)}\,+\,d_2^-\varrho^{l+p-6}\,\,,
\qquad\qquad (\varrho\to\infty)\,.
\label{typeItoinftyp4}
\eear
Obviously, the regular solutions should behave as $\varrho^{l+1}$ as 
$\varrho\to 0$. The regularity at the UV requires also the vanishing of 
$\xi^{\pm}$ when $\varrho\to\infty$. For the $\xi^+$ fluctuation this requirement is
only satisfied when $d_2^+=0$ in (\ref{typeItoinftyp4}). This condition defines the
so-called $I^l_+$ modes. On the other hand, for the $\xi^-$ fluctuation we have clearly
two possibilities. The modes with $d_2^-=0$ for $l\ge 1$ will be denoted by $I^l_-$,
whereas those with $d_1^-=0$ for $1\le l < 6-p$ will be called $\tilde I^l_-$.
Summarizing, the different behaviours at $\varrho\to\infty$ are:
\bear
&&I^l_+\quad\Longrightarrow\quad\xi^+\sim\varrho^{-(l+8-p)}\,,
\qquad\qquad (l\ge 1)\,\,,\rc\rc
&&I^l_-\quad\Longrightarrow\quad\xi^-\sim\varrho^{-(l+1)}\,,
\qquad\qquad (l\ge 1)\,\,,\rc\rc
&&\tilde I^l_-\quad\Longrightarrow\quad\xi^-\sim\varrho^{l+p-6}\,\,,
\qquad\qquad (1\le l < 6-p)\,\,.
\label{typeIdef}
\eear
In order to get information about the mass levels for these  modes, let us compute
the  different spectra in the WKB approximation. One can readily verify that the WKB
method selects always the $I^l_+$ modes of the $\xi^+$ fluctuation, whereas it picks up
one of the two branches of the $\xi^-$ modes, depending on the value of $l$. Indeed, if 
$l\ge {5-p\over 2}$, the $I^l_-$ branch is selected, while the $\tilde I^l_-$ modes are
picked up otherwise. The corresponding WKB mass levels are given by:
\bear
&& M^{WKB}_{I_\pm}(n,l)\,=\,2\sqrt{\pi}\,\,{L^{{5-p\over 2}}\over R^{{7-p\over 2}}}\,
{\Gamma\Big({7-p\over 4}\Big)\over \Gamma\Big({5-p\over 4}\Big)}\,\,
\sqrt{(n+1)\,\left(n\,+\,{7-p\over 5-p}\,(l+1\pm1)\,\right)}\,,
\qquad\qquad\rc\rc
&& M^{WKB}_{\tilde I_-}(n,l)\,=\,2\sqrt{\pi}\,\,{L^{{5-p\over 2}}\over R^{{7-p\over
2}}}\, {\Gamma\Big({7-p\over 4}\Big)\over \Gamma\Big({5-p\over 4}\Big)}\,\,
\sqrt{(n+1)\,\left(n\,+\,{7-p\over 5-p}\,+\,{3-p\over
5-p}\,(l+1)\,\right)}\,\,,\qquad\qquad
\label{typeIwkbp4}
\eear
where we have assumed that for each case $l$ varies in the range just discussed. By
comparing eqs. (\ref{typeIwkbp4}) and (\ref{MSDpDp+4})  we conclude that:
\beq
M_{I_{\pm}}(n,l)\,=\,M_{S}(n,l\pm 1)\,\,.
\label{I-S-relationDp4}
\eeq
Actually, the relation (\ref{I-S-relationDp4}) is satisfied by  the masses found
numerically with large accuracy and, therefore it seems to hold beyond the WKB
approximation\footnote{For $p=3$  the relation (\ref{I-S-relationDp4}) is exactly
satisfied by the analytical spectra found in
\cite{KMMW}.}. Moreover, the WKB formula in (\ref{typeIwkbp4}) for the masses of the 
$\tilde I_-^l$ modes reproduces reasonably well the values found in the numerical
calculations.

\subsubsection{Type II modes}
As before we shall take the ansatz:
\beq
A_{\mu}\,=\,\phi_{\mu}(x,\rho)\,Y^l(S^2)\,\,,\qquad\qquad
A_{\rho}\,=\,0\,\,,\qquad\qquad
A_i\,=\,0\,\,,
\label{IImd}
\eeq
where $\partial^{\mu}\phi_{\mu}=0$. 
The equations of motion for $A_{\rho}$ and $A_i$ are
automatically satisfied. Let us, expanding as before in a plane wave
basis, represent $\phi_\mu$ as:
\beq
\phi_\mu=\xi_\mu\,e^{ikx}\,\chi(\rho)\,,
\label{phiwave}
\eeq
with $\xi_\mu$ being a constant vector satisfying $k^{\mu}\xi_\mu\,=\,0$. 
The equation for $A_\mu$ yields:
\beq
\partial_{\varrho}\big(\varrho^3\partial_\varrho\chi\big)\,+\,
\bar M^2\,{\varrho^3\over
(1+\varrho^2)^{7-p\over2}}\,\chi\,-\,
l\,(l+2)\,\chi\,=\,0\,,
\eeq
which is the same equation as for the transverse scalar modes. Therefore, we conclude
that
\beq
M_{II}(n,l)\,=\,M_{S}(n,l)\,\,.
\eeq

\subsubsection{Type III modes}
Let us take the following form for the gauge field:
\beq
A_{\mu}\,=\,0\,\,,\qquad\qquad
A_{\rho}\,=\,\phi(x,\rho)\,Y^l(S^3)\,\,,\qquad\qquad
A_i\,=\,\tilde \phi(x,\rho)\,Y^l_i(S^3)\,\,.
\eeq
The equation for $A_{\rho}$ becomes:
\beq
\rho^3\,R^{7-p}\,\partial_{\mu}\,\partial^{\mu} \phi\,
+\,l(l+2)\,\rho\,(\rho^2+L^2)^{7-p\over 2}\,
(\partial_{\rho}\tilde \phi-\phi\,)\,=\,0\,.
\eeq
The equation for $A_{\mu}$ can be written as:
\beq
\partial_{\mu}\,\Big[\,l(l+2)\,\rho\,\tilde \phi\,-\,\partial_{\rho}(\rho^3 \phi)\,\Big]\,=\,0\,.
\eeq
Expanding $\phi$ and $\tilde\phi$ in a plane wave basis one can  write:
\beq
l(l+2)\tilde\phi\,=\,{1\over\rho}\partial_{\rho}(\rho^3\phi)\,\,.
\eeq
For $l\not= 0$, one can use this relation to eliminate $\tilde\phi$ 
in favor of $\phi$.
The equation of motion of $A_{\rho}$ becomes:
\beq
\partial_{\rho}\,\left({1\over\rho}\,\partial_\rho(\rho^3\,\phi)\right)\,-\,l(l+2)\,\phi\,+\,R^{7-p}{\rho^2\over 
(\rho^2+L^2)^{7-p\over 2}}\,\partial_{\mu}\,\partial^{\mu}\phi\,=\,0\,.
\eeq
The equation for $A_i$ results equivalent to the above one.

Let  us separate variables as in the Dp-D(p+2) case, namely:
\beq
\phi(x,\rho)\,=\,e^{ikx}\, \zeta(\rho)\,.
\label{type3chvbp4}
\eeq
In terms of the reduced quantities $\varrho$ and $\bar M$ introduced before in
(\ref{DpDp+4-newvariables}), and  by changing variables as:
\beq
e^{y}\,=\,\varrho\,\,,\qquad\qquad
\psi\,=\,\varrho^2\,\zeta\,,
\label{type3schvp4}
\eeq
we can convert the fluctuation equation into a Schr\"odinger equation, where the
potential
$V(y)$ is given by:
\beq
V(y)=\bigg(l+1\bigg)^2-\bar M^2
{e^{2y}\over ( 1+e^{2y})^{{7-p\over 2}}}\,.
\label{type3schpp4}
\eeq
This potential is just the same as the one corresponding to the scalar fluctuations. It
follows that the masses of these fluctuations are the same as those corresponding to the
scalar modes, \ie:
\beq
M_{III}(n,l)\,=\,M_{S}(n,l)\,\,.
\eeq

\renewcommand{\theequation}{\rm{E}.\arabic{equation}}
\setcounter{equation}{0}
\medskip

\section{Fluctuations of the F1-Dp systems}
\medskip

Let us now consider the intersection $(0|F1\perp Dp)$. We will treat the fundamental
string as background and the Dp-brane as a probe. The corresponding near-horizon
supergravity solution is:
\bear
&&ds^2\,=\,{r^6\over R^6}\,\,
\big(\,-dt^2\,+\,(dx^1)^2\,\big)\,\,+\,
d\vec y\cdot d\vec y\,\,,\rc\rc
&&e^{-\phi}\,=\,{R^3\over r^3}\,\,,
\label{F1metric}
\eear
where $R$ is given in (\ref{F1back-parameters}), 
 $\vec y\,=\,(y^1,\cdots,y^8)$ and $r^2=\vec y^{\,2}$. We shall place now a
Dp-brane at a constant value of $x^1$ and take the following set of worldvolume
coordinates:
\beq
\xi^a\,=\,(t,y^1,\cdots,y^p)\,\,.
\eeq
As before, we shall denote by $\vec z$ the vector formed by the $8-p$ coordinates 
$(y^{p+1},\cdots, y^8)$. If $\rho^2=(y^1)^2+\cdots +(y^p)^2$ and if the Dp-brane is 
located at
$|\vec z|=L$, the induced metric is:
\beq
ds^2_I\,=\,-
\Bigg[{\rho^2\,+\,L^2\over R^2}\Bigg]^3\,
dt^2\,+\,
d\rho^2\,+\,\rho^2\,d\Omega_{p-1}^2\,\,,
\label{F1Dpmetric}
\eeq
where we have assumed that $p>1$. We shall limit ourselves here to  study the
fluctuations of the scalars $\chi$ transverse to both the F1 and the Dp-brane. These 
fluctuations are governed by the lagrangian (\ref{fluct-lag-general}) for $p_2=p$, $d=0$
and for the $\gamma_i$ exponents written in eq. (\ref{F1back-parameters}). 
By changing variables as
$e^y=\varrho$ and $\psi=\varrho^{{p-2\over 2}}\xi$, we can convert the fluctuation
into the Schr\"odinger equation (\ref{Sch}) with potential:
\beq
V(y)\,=\,\Big(l+{p-2\over 2}\Big)^2\,-\,\bar M^2\,\,
{e^{2y}\over (e^{2y}\,+\,1)^{3}}\,\,.
\label{F1Dp-potential}
\eeq
Notice that
the potential (\ref{F1Dp-potential}) is invariant if we simultaneously change $l\to l+1$
and $p\to p-2$. This means that the spectrum of the $(0|F1\perp Dp)$ intersection at
angular quantum number $l$ is equivalent to that of $(0|F1\perp D(p-2))$ at quantum
number $l+1$. For this system  the corresponding WKB mass levels are:
\beq
\bar M_{WKB}\,=\,\pi\,
\sqrt{(n+1)\,\Bigg(n\,+\,{3\over 4}\,(p-2)\,+\,{3l\over 2}\,\Bigg)}\,\,.
\eeq

As discussed in section \ref{generalnumeric}, 
one can prove from the asymptotic values of the potential (\ref{F1Dp-potential}) that,
for both $\varrho\to 0$ and $\varrho\to \infty$, the fluctuation $\xi$ behaves as
$\xi\sim \varrho^{\gamma}$ with  $\gamma=l,-(l+p-2)$. We will select numerically the
regular solutions as those which behave as $\varrho^l$ for small 
$\varrho$ and as $\varrho^{-(l+p-2)}$ for large $\varrho$. 
In table \ref{MassF1Dp} we collect some numerical results and the corresponding WKB
estimates for some $(0|F1\perp Dp)$ intersections. By looking at the potential of the
equivalent Schr\"odinger problems, we notice
that  the transverse fluctuations of the 
$(0|F1\perp D3)$ intersection are equivalent to those of the 
$(0|D1\perp D3)$ configuration, while the $(0|F1\perp D4)$ intersection is equivalent to
$(0|D1\perp D5)$.

\medskip
\begin{table}[!h]
\qquad\qquad
\begin{tabular}[b]{|c|c|c|}   
\hline  
\multicolumn{3}{|c|}{$(0|F1\perp D2)$ with $l=1$}\\
\hline
 $n$  & WKB  & Numerical \\ 
\hline   
\ \ 0 & $14.80$  & $14.70$  \\   
\ \ 1 & $49.35$  & $49.22$  \\   
\ \ 2 & $103.63$  & $103.50$  \\   
\ \ 3 & $177.65$  & $177.54$  \\     
\ \ 4 & $271.41$  & $271.30$  \\
\ \ 5 & $384.91$  & $384.80$  \\ 
\hline
\end{tabular}
\qquad\qquad\qquad
\begin{tabular}[b]{|c|c|c|}   
\hline  
\multicolumn{3}{|c|}{$(0|F1\perp D5)$ with $l=0$}\\
\hline
 $n$  & WKB  & Numerical \\ 
\hline   
\ \ 0 & $22.21$  & $27.06$  \\   
\ \ 1 & $64.15$  & $69.40$  \\   
\ \ 2 & $125.84$  & $131.36$  \\   
\ \ 3 & $207.26$  & $213.02$  \\     
\ \ 4 & $308.43$  & $314.36$  \\
\ \ 5 & $429.33$  & $435.40$  \\ 
\hline
\end{tabular}
\caption{$\bar M^2$ for the transverse scalar fluctuations of F1-D2 ($l=1$) and
F1-D5 ($l=0$) obtained numerically and with the WKB approximation.}
\label{MassF1Dp}
\end{table}

\renewcommand{\theequation}{\rm{F}.\arabic{equation}}
\setcounter{equation}{0}
\medskip

\section{Fluctuations of the M-theory intersections}
\medskip
According to the analysis performed in section \ref{general}
the basic supersymmetric orthogonal intersections of M-theory are 
$(1|M2\perp M5)$, $(3|M5\perp M5)$ and $(0|M2\perp M2)$.
The M2 and M5 eleven-dimensional near-horizon metrics are:
\bear
&&ds^2_{M2}\,=\,{r^{4}\over R^{4}}\,\,
(\,-dt^2\,+\,(dx^1)^2\,+\,(dx^2)^2\,)\,+\,
{R^{2}\over r^{2}}\,\,
d\vec y\cdot d\vec y\,\,,\rc\rc
&&ds^2_{M5}\,=\,{r\over R}\,\,
(\,-dt^2\,+\,(dx^1)^2\,+\,\cdots+(dx^5)^2)\,+\,
{R^{2}\over r^{2}}\,\,
d\vec y\cdot d\vec y\,\,,
\eear
where the radii $R$ are given in eqs. (\ref{M2back-parameters}) and
(\ref{M5back-parameters}). In this appendix we look at the fluctuations of the
transverse scalars for the three BPS intersections listed above. We will verify that the
corresponding differential equations for these M-theory systems are identical to some of
the ones already studied for the type II theory, as expected naturally from the relation
between these two theories.

\subsection{$\bf {(1|M2\perp M5)}$ intersection}
Let us consider a M5-brane probe in the M2-brane background written above and let us take
the worldvolume coordinates as $\xi^{a}\,=\,(t, x^1,y^1,y^2,y^3,y^4)$. We shall assemble
the orthogonal coordinates in the vector $\vec z\,=\,(y^5,y^6,y^7,y^8)$. For an embedding
with $x^2$ constant and $|\vec z|=L$, we have the following induced metric:
\beq
ds^2_I\,=\,
\Bigg[{\rho^2\,+\,L^2\over R^2}\Bigg]^{2}\,
\big(\,-dt^2\,+\,(dx^1)^2\,\big)\,+\,
{R^2\over \rho^2\,+\,L^2 }\,
(\,d\rho^2\,+\,\rho^2\,d\Omega_3^2\,)\,\,,
\label{m2m5inmetric}
\eeq
where we have used spherical coordinates to parametrize the $(y^1,y^2,y^3,y^4)$ variables.
Notice that for $L=0$, the above metric corresponds to an $AdS_2\times S^3$ defect of
the 
$AdS_4\times S^7$ geometry. The equation for the fluctuations around this configuration
is:
\beq
{R^{6}\over (\rho^2+L^2)^{3}}\,\,\partial^{\mu}\partial_{\mu}\,\chi\,+\,
{1\over \rho^{3}}\,\partial_{\rho}\,(\rho^{3}\partial_{\rho}\chi)\,+\,
{1\over \rho^2}\,\nabla^i\nabla_i\,\chi=0\,\,,
\eeq
which is just the same equation as that of the transverse scalars in the D1-D5 system.
Therefore, the corresponding mass levels are exactly the same as in the D1-D5
intersection.

\subsection{$\bf {(3|M5\perp M5)}$ intersection}
We now consider a M5-brane probe in the M5-brane geometry. The worldvolume coordinates are
$(t, x^1,x^2,x^3,y^1,y^2)$ and the orthogonal space is parametrized by the vector 
$\vec z\,=\,(y^3,y^4,y^5)$. For $|\vec z|=L$ and constant $x^4$ and $x^5$, the induced
metric is:
\beq
ds^2_I\,=\,
\Bigg[{\rho^2\,+\,L^2\over R^2}\Bigg]^{{1\over 2}}\,
\big(\,-dt^2\,+\,(dx^1)^2\,+\cdots +(dx^3)^2\big)\,+\,
{R^2\over \rho^2\,+\,L^2 }\,
(\,d\rho^2\,+\,\rho^2\,d\Omega_1^2\,)\,\,.
\label{m5m5inmetric}
\eeq
For $L=0$ this metric corresponds to an $AdS_5\times S^1$ defect on the 
$AdS_7\times S^4$ background geometry. The equation for the fluctuations is:
\beq
{R^{2}\over (\rho^2+L^2)^{{3\over 2}}}\,\,\partial^{\mu}\partial_{\mu}\,\chi\,+\,
{1\over \rho}\,\partial_{\rho}\,(\rho\,\partial_{\rho}\chi)\,+\,
{1\over \rho^2}\,\nabla^i\nabla_i\,\chi=0\,\,,
\eeq
which is identical to the one corresponding to the transverse scalars of the D4-D4
system. 
\subsection{$\bf {(0|M2\perp M2)}$ intersection}
Let us put a M2-brane in the M2 geometry and take $(t, y^1,y^2)$ as worldvolume
coordinates. Now $\vec z\,=\,(y^3,\cdots,y^8)$ and we consider an embedding at $x^1$ and
$x^2$ constant and $|\vec z|=L$. the induced metric is:
\beq
ds^2_I\,=\,
-\Bigg[{\rho^2\,+\,L^2\over R^2}\Bigg]^{2}\,
\,dt^2\,+\,
{R^2\over \rho^2\,+\,L^2 }\,
(\,d\rho^2\,+\,\rho^2\,d\Omega_1^2\,)\,\,,
\label{m2m2inmetric}
\eeq
which for $L=0$ is just $AdS_2\times S^1$. The equation for the fluctuations becomes:
\beq
{R^{6}\over (\rho^2+L^2)^{3}}\,\,\partial^{0}\partial_{0}\,\chi\,+\,
{1\over \rho}\,\partial_{\rho}\,(\rho\,\partial_{\rho}\chi)\,+\,
{1\over \rho^2}\,\nabla^i\nabla_i\,\chi=0\,\,,
\eeq
and is identical to the one for the transverse scalars of the F1-D2 system.


\bigskip\bigskip\bigskip


















\begin{thebibliography}{99}



\bibitem{jm} J.~M.~Maldacena, ``The large $N$ limit of superconformal field
theories and supergravity'', {\it Adv.\ Theor.\ Math.\ Phys.}\  {\bf 
2} (1998) 231,
{\rm hep-th/9711200}.



\bibitem{MAGOO}For a review see, O. Aharony, S. Gubser, J. Maldacena, 
H. Ooguri and Y. Oz,
``Large $N$ field theories, string theory and gravity",
{\sl Phys.  Rept. } {\bf 323} (2000) 183, {\rm hep-th/9905111}.


\bibitem{Wilson}J. M. Maldacena, ``Wilson loops in large N field theories",
{\sl \prl} {\bf 80 }(1998) 4859, {\rm hep-th/9803002};\\
S. J. Rey and J. Yee, ``Macroscopic strings as heavy quarks in large 
N theory and anti-de Sitter
supergravity", {\sl Eur. Phys. J.} {\bf C22 }(2001) 379, {\rm hep-th/9803001}.




\bibitem{KR}A. Karch and L. Randall, ``Locally localized gravity", 
{\sl \jhep} {\bf 0105 } (2001) 008, {\rm hep-th/0011156}; ``Open and closed string
interpretation of SUSY CFT's on branes with boundaries", 
{\sl \jhep} {\bf 0106 } (2001) 063.



\bibitem{KKW} A. Karch and E. Katz, ``Adding flavor to AdS/CFT",
{\sl \jhep} {\bf 0206 }(2002) 043, {\rm hep-th/0205236};\\
A. Karch, E. Katz and N. Weiner,
``Hadron masses and screening from AdS Wilson loops",
{\sl \prl} {\bf 90 }(2003) 091601, {\rm hep-th/0211107}.






\bibitem{D3D7}
A. Fayyazuddin and M. Spalinski, ``Large N superconformal gauge 
theories and supergravity
orientifolds",
{\sl \np} {\bf B535 }(1998) 219, {\rm hep-th/9805096};\\
O. Aharony, A. Fayyazuddin and J. M. Maldacena, ``The large N limit 
of ${\cal N}=1,2$ field
theories from threebranes in F-theory",
{\sl \jhep} {\bf 9807 }(1998) 013, {\rm hep-th/9806159};\\
M. Gra\~na and J. Polchinski, ``Gauge/gravity duals with holomorphic dilaton",
{\sl \pr} {\bf D65 }(2002) 126005, {\rm hep-th/0106014};\\
M. Bertolini, P. Di Vecchia, M. Frau, A. Lerda and R. Marotta,
``N=2 gauge theories on systems of fractional D3/D7 branes",
{\sl \np} {\bf B621 }(2002) 157, {\rm hep-th/0107057};\\
M. Bertolini, P. Di Vecchia, G. Ferretti and R. Marotta,
``Fractional branes and N=1 gauge theories",
{\sl \np} {\bf B630 }(2002) 222, {\rm hep-th/0112187}.


\bibitem{KMMW}M. Kruczenski, D. Mateos, R. Myers and D. Winters,
``Meson spectroscopy in AdS/CFT with flavour",
{\sl \jhep} {\bf 0307 }(2003) 049, {\rm hep-th/0304032}.







\bibitem{Sonnen} T. Sakai and J. Sonnenschein, ``Probing flavored 
mesons of confining
gauge theories by supergravity",
{\sl \jhep} {\bf 0309 }(2003) 047, {\rm hep-th/0305049}.



\bibitem{Johana} J. Babington, J. Erdmenger, N. Evans, Z. Guralnik 
and I. Kirsch,
``Chiral symmetry breaking and pions in non-supersymmetric 
gauge/gravity duals", 
{\sl \pr} {\bf D69} (2004) 066007, {\rm hep-th/0306018};\\
R. Apreda, J. Erdmenger and N. Evans, ``Scalar effective potential for
D7-brane probes which break chiral symmetry", {\rm hep-th/0509219};\\
R. Apreda, J. Erdmenger, N. Evans, J. Grosse and Z. Guralnik,
``Instantons on D7 brane probes and AdS/CFT with flavour", 
{\rm hep-th/0601130}.



\bibitem{KMMW-two}M. Kruczenski, D. Mateos, R. Myers and D. Winters,
`Towards a holographic dual of large-$N_c$ QCD",
{\sl \jhep} {\bf 0405 }(2004) 041, {\rm hep-th/0311270};\\
J. L. F. Barbon, C. Hoyos, D. Mateos and R. C. Myers,
``The holographic life of the eta'", 
{\sl \jhep} {\bf 0410 }(2004) 029, {\rm hep-th/0404260};\\
A.~Armoni, ``Witten-Veneziano from Green-Schwarz'',
{\sl \jhep} {\bf 0406}, 019 (2004), 
{\rm hep-th/0404248};\\
 J.~L.~Hovdebo, M.~Kruczenski, D.~Mateos, R.~C.~Myers and D.~J.~Winters,
``Holographic mesons: Adding flavor to the AdS/CFT duality,''
{\sl Int.\ J.\ Mod.\ Phys.}\ A {\bf 20} (2005) 3428.



\bibitem{Ouyang}P. Ouyang,
``Holomorphic D7-branes and flavored N=1 gauge dynamics", 
{\sl \np} {\bf B699 }(2004) 207, {\rm hep-th/0311084};\\
T.~S.~Levi and P.~Ouyang, 
``Mesons and flavor on the conifold'', {\rm hep-th/0506021}.

\bibitem{WH}X.-J. Wang and S. Hu, ``Intersecting branes and 
adding flavors to the Maldacena-N\'u\~nez background",
{\sl \jhep} {\bf 0309 }(2003) 017  {\rm hep-th/0307218}.



\bibitem{flavoring}C. N\'u\~nez, A. Paredes and A. V. Ramallo,
``Flavoring the gravity dual of ${\cal N}=1$ Yang-Mills with probes",
{\sl \jhep} {\bf 0312 }(2003) 024, {\rm hep-th/0311201}.


\bibitem{Hong}S. Hong, S. Yoon, M. J. Strassler, 
``Quarkonium from the fifth dimension", 
{\sl \jhep} {\bf 0404 }(2004) 046, {\rm hep-th/0312071}.




\bibitem{Evans}N. Evans, J. P. Shock, ``Chiral dynamics from AdS space",
{\sl \pr} {\bf D70} (2004) 046002,{\rm hep-th/0403279};\\
N. Evans, J. P. Shock and T. Waterson, ``D7 brane embeddings and chiral
symmetry breaking", {\sl \jhep} {\bf 0503 }(2005) 005, {\rm hep-th/0502091};\\
J. P. ~Shock, ``Canonical coordinates and meson spectra for scalar deformed
${\cal N}=4$ SYM from the AdS/CFT correspondence", {\rm hep-th/0601025}.






\bibitem{Ghoroku} K. Ghoroku, M. Yahiro, ``Chiral symmetry breaking driven by
the dilaton", {\sl \pl} {\bf B604 }(2004) 235, {\rm hep-th/0408040};
``Holographic models for mesons at finite temperature", 
{\rm hep-ph/0512289};\\
K. Ghoroku, T. Sakaguchi, N. Uekusa and M. Yahiro, ``Flavor quark at high
temperature from a holographic model", 
{\sl \pr} {\bf D71} (2005) 106002, {\rm hep-th/0502088};\\
I. Brevik, K. Ghoroku and  A. Nakamura, ``Meson mass and confinement force
driven by the dilaton", {\rm hep-th/0505057}. 


\bibitem{conifold}D. Arean, D. Crooks and A. V. Ramallo, 
``Supersymmetric probes on the conifold", 
{\sl \jhep} {\bf 0411 }(2004) 035, {\rm hep-th/0408210}.



\bibitem{Kuper} S. Kuperstein, ``Meson spectroscopy from holomorphic probes
on the warped deformed conifold", 
{\sl \jhep} {\bf 0503 }(2005) 014, {\rm hep-th/0411097}.






\bibitem{Sakai}T. Sakai and S. Sugimoto, ``Low energy hadron physics in
holographic QCD", {\sl \ptp} {\bf 113 }(2005) 843, {\rm hep-th/0412141};
``More on a holographic dual of QCD", 
{\sl \ptp} {\bf 114 }(2006) 1083, {\rm hep-th/0507073};

\bibitem{APR}
  D.~Arean, A.~Paredes and A.~V.~Ramallo,
  ``Adding flavor to the gravity dual of non-commutative gauge theories,''
{\sl \jhep} {\bf 0508} (2005) 017, {\rm hep-th/0505181}.




\bibitem{WFO}O. DeWolfe, D. Z. Freedman and H. Ooguri, 
``Holography and defect conformal field theories", 
{\sl \pr} {\bf D66} (2002) 025009, {\rm hep-th/0111135}.


\bibitem{EGK}J. Erdmenger, Z. Guralnik and I. Kirsch,
``Four-dimensional superconformal theories with interacting boundaries or defects",
{\sl \pr} {\bf D66} (2002) 025020, {\rm hep-th/0203020}.

\bibitem{ST}K. Skenderis and M. Taylor, ``Branes in AdS and pp-wave spacetimes",
{\sl \jhep} {\bf 0206 } (2002) 025,
{\rm hep-th/0204054}.


\bibitem{CEGK} N. R. Constable, J. Erdmenger, Z. Guralnik and I. Kirsch,
``Intersecting D3-branes and holography", 
{\sl \pr} {\bf D68}, 106007 (2003), {\rm hep-th/0211222}.


\bibitem{Kirsch}I. Kirsch, ``Generalizations of the AdS/CFT correspondence", 
{\sl Fortsch. Phys. } {\bf 52} (2004)724. 




\bibitem{EGHK}J. Erdmenger, Z. Guralnik, R. Helling and I. Kirsch,
``A worldvolume prespective on the recombination of intersecting branes",
{\sl \jhep} {\bf 0404 } (2004) 64, {\rm hep-th/0309043}.










\bibitem{MInahan}
J. A. Minahan, ``Glueball mass spectra and other issues for 
supergravity duals of QCD models",
{\sl \jhep} {\bf 9901 }(1999) 020, {\rm hep-th/9811156}.


\bibitem{RS}J. G. Russo and K. Sfetsos,
``Rotating D3-branes and QCD in three dimensions",
{\sl \atmp} {\bf 3}(1999) 131, {\rm hep-th/9901056}.



\bibitem{glueball}
C. Csaki, H. Ooguri, Y. Oz and J. Terning, ``Glueball mass spectrum 
from supergravity",
{\sl \jhep} {\bf 9901 }(1999) 017, {\rm hep-th/9806021};\\
R. de Mello Koch, A. Jevicki, M. Mihailescu and J. P. Nunes, 
``Evaluation of glueball masses
from supergravity", {\sl \pr} {\bf D 58}(1998) 105009, {\rm hep-th/9806125}.




\bibitem{IMSY} N. Itzhaki, J. M. Maldacena, J. Sonnenschein and S.
Yankielowicz, ``Supergravity and the large-N limit of theories with sixteen
supercharges", {\sl\pr} {\bf D58} (1998)046004,  {\rm hep-th/9802042}.







\bibitem{Effective}
 J.~Erlich, E.~Katz, D.~T.~Son and M.~A.~Stephanov,
``QCD and a holographic model of hadrons'',
{\sl \prl} {\bf 95} (2005) 261602, {\rm hep-ph/0501128};\\
 L.~Da Rold and A.~Pomarol,
``Chiral symmetry breaking from five dimensional spaces'',
{\sl \np} {\bf B 721}(2005) 79, {\rm hep-ph/0501218};\\
 G.~F.~de Teramond and S.~J.~Brodsky,
``The hadronic spectrum of a holographic dual of QCD'',
 {\sl \prl} {\bf 94}, 201601 (2005), {\rm hep-th/0501022};\\
 G.~Siopsis,``Correspondence principle for a brane in Minkowski space and vector 
 mesons'', {\sl \jhep} {\bf 0510}, 059 (2005)
{\rm hep-th/0503245};\\
 N.~Mahajan,
``Revisiting 5D chiral symmetry breaking and holographic QCD models'',
 {\sl \pl} {\bf B623}, 119 (2005), {\rm hep-ph/0506098};\\
J.~Hirn and V.~Sanz,
``Interpolating between low and high energy QCD via a 5D Yang-Mills model'',
{\sl \jhep} {\bf 0512}, 030 (2005),  {\rm hep-ph/0507049};\\
 T.~Hambye, B.~Hassanain, J.~March-Russell and M.~Schvellinger,
``On the $\Delta I = 1/2$ rule in holographic QCD'',
 {\rm  hep-ph/0512089}
 




\bibitem{Schreiber} E. Schreiber, ``Excited mesons and quantization of strings
endpoints", {\rm   hep-th/0403226}.



\bibitem{OTY}K. Okamura, Y. Takayama and K. Yoshida, 
``Open semiclassical strings and long defect operators in AdS/dCFT correspondence", 
{\sl \pr} {\bf D 71}(2005) 126006, {\rm hep-th/0410139};
``Integrability and higher loops in AdS/dCFT correspondence", 
{\sl \pl} {\bf B624}(2005) 115, {\rm hep-th/0504209};
``Open spinning strings and AdS/CFT duality", {\rm hep-th/0511139}.



\bibitem{LP}P. Lee and J. Park, ``Open strings in pp-wave backgroud from defect
conformal field theory", {\sl \pr} {\bf D 67}(2003) 026002, {\rm hep-th/0203257}.


\bibitem{DWM}O. De Wolfe and N. Mann,
``Integrable spin chains in defect conformal field theory", 
{\sl \jhep} {\bf 0404 } (2004) 035, {\rm hep-th/0401041}.




\bibitem{Beyond}
  S.~A.~Cherkis and A.~Hashimoto,
  ``Supergravity solution of intersecting branes and AdS/CFT with flavor",
{\sl \jhep} {\bf 0211 }(2002) 036, {\rm hep-th/0210105};\\
 H.~Nastase,
  ``On Dp-Dp+4 systems, QCD dual and phenomenology", 
{\rm hep-th/0305069};\\
 B.~A.~Burrington, J.~T.~Liu, L.~A.~Pando Zayas and D.~Vaman,
  ``Holographic duals of flavored N = 1 super Yang-Mills: Beyond the probe
  approximation'',
{\sl \jhep} {\bf 0502 }(2005) 022, {\rm hep-th/0406207};\\
 J.~Erdmenger and I.~Kirsch,
 ``Mesons in gauge/gravity dual with large number of fundamental fields",
{\sl \jhep} {\bf 0412 }(2004) 025, {\rm hep-th/0408113};\\  
I.~Kirsch and D.~Vaman,
  ``The D3/D7 background and flavor dependence of Regge trajectories",
{\sl \pr} {\bf D 72}(2005) 026007, {\rm hep-th/0505164}.
  

  
\bibitem{CNP}R. Casero, C. N\'u\~nez and A. Paredes, ``Towards the string dual of
${\cal N}=1$ SQCD-like theories", {\rm hep-th/0602027}.









\bibitem{KOBS}A. Karch, A. O'Bannon and K. Skenderis, ``Holographic renormalization of
probe D-branes in AdS/CFT", {\rm   hep-th/0512125}.

\bibitem{Higgs}  J. Erdmenger, J. Grosse and Z. Guralnik, 
``Spectral flow on the Higgs branch and AdS/CFT duality", 
{\sl \jhep} {\bf 0506 } (2005) 052, {\rm hep-th/0502224};\\
R. Apreda, J. Erdmenger, N. Evans and Z. Guralnik, 
``Strong coupling effective Higgs potential and a first order thermal phase
transition from AdS/CFT duality", 
{\sl \pr} {\bf D 71}(2005) 126002, {\rm hep-th/0504151}.



\bibitem{Yamaguchi}S. Yamaguchi, ``AdS branes corresponding to superconformal defects", 
{\sl \jhep} {\bf 0306 } (2003) 002,
{\rm hep-th/0505007}.




\bibitem{CEPRV}F. Canoura, J. D. Edelstein, L. A. Pando Zayas, A. V. Ramallo and
D. Vaman, ``Supersymmetric branes on $AdS_5\times Y^{p,q}$ and their field theory
duals", {\rm hep-th/0512087}.



\bibitem{CPR}
  F.~Canoura, A.~Paredes and A.~V.~Ramallo,
 ``Supersymmetric defects in the Maldacena-Nunez background",
  {\sl \jhep} {\bf 0509 } (2005) 032, {\rm hep-th/0507155}.












\end{thebibliography}
\end{document}